\documentclass[12pt,preprint]{aastex}

\newcommand\diamondrule{\line{$\m@th
   \leaders\hrule\hfill\rlap{$\m@th\bracerd\braceld$}
   \braceru\bracelu\leaders\hrule\hfill$}}

\newcommand\upskip{\vskip-9pt}

\newcommand\bupskip{\vskip-17pt}
\newcommand\smallneg{\kern-.0800em}
\newcommand\negskip{\kern-.5em}

\newcommand\lsim{\rlap{\raise.4ex\hbox{$<$}}\lower.55ex\hbox{$\sim$}\,}
\newcommand\gsim{\rlap{\raise.4ex\hbox{$>$}}\lower.55ex\hbox{$\sim$}\,}
\newcommand\implies{{\bf=\kern-0.45em>}}

\newcommand\unit{\,\rm}
\newcommand\back{\negthinspace\negthinspace}
\newcommand\kms{\rm\, km\cdot s^{-1}}
\newcommand\Kkms{\rm\, K\cdot\kms}

\newcommand\um{\unit\mu m}

\newcommand\CO{\rm {}^{12}\smallneg CO}
\newcommand\COex{\rm {}^{12}\smallneg C{}^{16}\smallneg O}

\newcommand\cO{\rm {}^{13}\smallneg CO}
\newcommand\cOex{\rm {}^{13}\smallneg C{}^{16}\smallneg O}
\newcommand\cObf{\bf {}^{13}\smallneg CO}

\newcommand\Co{\rm C{}^{18}\smallneg O}
\newcommand\Coex{\rm {}^{12}\smallneg C{}^{18}\smallneg O}

\newcommand\Jthree{\rm J=3\rightarrow 2}

\newcommand\Jtwo{\rm J=2\rightarrow 1}

\newcommand\Jone{\rm J=1\rightarrow 0}

\newcommand\COone{\CO\ \Jone}
\newcommand\cOthree{\cO\ \Jthree}

\newcommand\cOtwo{\cO\ \Jtwo}

\newcommand\Cotwo{\Co\ \Jtwo}
\newcommand\Coone{\Co\ \Jone}
\newcommand\cOone{\cO\ \Jone}

\newcommand\nH{\rm n(H_2)}

\newcommand\Tk{\rm T_{{}_K}}
\newcommand\Tkz{\rm T_{{}_{K0}}}
\newcommand\Tkone{\rm T_{{}_{K1}}}

\newcommand\Tr{\rm T_{{}_R}}

\newcommand\Tbg{\rm T_{{}_{BG}}}
\newcommand\Trz{\rm T_{{}_{R0}}}
\newcommand\Trone{\rm T_{{}_{R1}}}

\newcommand\cKkms{\unit cm^{-2}\cdot (K\cdot km\cdot s^{-1})^{-1}}
\newcommand\ckms{\unit cm^{-2}\cdot (km\cdot s^{-1})^{-1}}

\newcommand\too{\rightarrow}
\newcommand\Td{\rm T_{d}}
\newcommand\Tdz{\rm T_{d0}}
\newcommand\Tdo{\rm T_{d1}}
\newcommand\Tdc{\rm T_{dc}}
\newcommand\DT{\rm \Delta T}
\newcommand\rd{\rm r_{_{240}}}

\newcommand\Ia{\rm I_\nu(140\um)}
\newcommand\Ib{\rm I_\nu(240\um)}
\newcommand\Ic{\rm I(\cO)}

\newcommand\MJsr{\unit MJy\cdot sr^{-1}}
\newcommand\MJkk{\unit MJy\cdot sr^{-1}\cdot (K\cdot km\cdot s^{-1})^{-1}}
\newcommand\NHd{\rm N_d(H)}
\newcommand\NHdo{\rm N_{d1}(H)}

\newcommand\NHth{\rm N_{13}(H_2)}
\newcommand\Ngth{\rm N_{13}(H\,I+2H_2)}

\newcommand\filf{\eta_{_f}(\nu)}
\newcommand\Dv{\rm\Delta v}
\newcommand\Dvc{\rm\Delta v_c}
\newcommand\nvco{\rm {N_{c1}\over\Dvc}}
\newcommand\nvcz{\rm {N_{c0}\over\Dvc}}
\newcommand\nvtco{\rm {N_{c1}(\cO)\over\Dvc}}
\newcommand\nvtcz{\rm {N_{c0}(\cO)\over\Dvc}}
\newcommand\nvhco{\rm {N_{c1}(H_2)\over\Dvc}}
\newcommand\nvhcz{\rm {N_{c0}(H_2)\over\Dvc}}
\newcommand\NDv{\rm N(\cO)/\Delta v}

\newcommand\Xdvdr{\rm X(\cO)/(dv/dr)}

\newcommand\degree{\rlap{$^\circ$}\kern.06em} 

\newcommand{\mymail}{wwall@inaoep.mx}

\slugcomment{Version --  19/May/2006}

\shorttitle{Diagnostic of Dust/Molecular Gas Conditions}
\shortauthors{Wall}

\begin{document}


\title{Comparison of $\cObf$ Line and Far-Infrared Continuum\\
	Emission as a Diagnostic of Dust and Molecular Gas\\
	Physical Conditions:\\
	I. Motivation and Modeling}


\author{W. F. Wall}
\affil{Instituto Nacional de Astrof\'{\i}sica, \'Optica, y Electr\'onica,
Apdo. Postal 51 y 216, Puebla, Pue., M\'exico}
\email{\mymail}







\begin{abstract}

Determining temperatures in molecular clouds from ratios of CO rotational lines or from ratios
of continuum emission in different wavelength bands suffers from reduced temperature sensitivity 
in the high-temperature limit.  In theory, the ratio of far-IR, submillimeter, or millimeter 
continuum to that of a $\cO$ (or $\Co$) rotational line can place reliable upper limits on the 
temperature of the dust and molecular gas.  Consequently,  far-infrared continuum data from the 
{\it COBE}/{\it DIRBE} instrument and Nagoya 4-m $\cOone$ spectral line data were used to plot
240$\um$/$\cOone$ intensity ratios against 140$\um$/240$\um$ dust color temperatures, allowing
us to constrain the multiparsec-scale physical conditions in the Orion$\,$A and B molecular 
clouds.

The best-fitting models to the Orion clouds consist of two components: a component near the
surface of the clouds that is heated primarily by a very large-scale (i.e. $\sim 1\,$kpc) 
interstellar radiation field and a component deeper within the clouds.  The former has a fixed
temperature and the latter has a range of temperatures that varies from one sightline to another.
The models require a dust-gas temperature difference of 0$\pm 2\,$K and suggest that 40-50\% of
the Orion clouds are in the form of dust and gas with temperatures between 3 and 10$\,$K.  These
results have a number implications that are discussed in detail in later papers.   These include 
stronger dust-gas thermal coupling and higher Galactic-scale molecular gas temperatures than are 
usually accepted, an improved explanation for the N(H$_2$)/I(CO) conversion factor, and ruling out 
one dust grain alignment mechanism.

\end{abstract}




\section{Introduction\label{sec1}}

Development of a complete theory of star formation requires understanding the 
physical conditions of the molecular gas in which stars form, both on sub-parsec 
and multi-parsec scales.  On the latter size scales, observations of molecular clouds 
can provide insights into large-scale patterns of star formation such as in 
spiral arms, rings, or in starburst regions of external galaxies \citep[e.g., see][]
{Rand92, Vogel88, Aalto99, Kuno95, Wong00, Sheth00, Seaq98, Kiku98, Pag97, Gus93, 
Rieu92, Wild92, Harris91, Tilanus91, Turner91, Verter89, Lo87, Sorai00, GB00, Harr99,
W91, Sage91, Scov85, Nish01, Nish01a, Kennicutt89, Martin01, Rownd99, Bryant99, 
Mulchaey97, Aalto95, W93, Puxley90, Verter87, Verter88}.  The diagnostics
of molecular cloud physical, or excitation, conditions are usually obtained 
from observed molecular spectral lines.  Despite being composed almost 
entirely of molecular hydrogen, only very warm ($\gsim few\times 100\unit K$) gas in 
interstellar molecular clouds is sampled {\it in emission\/} with the purely rotational 
lines of $\rm H_2$ \citep{Rodriguez01, Rosenthal00, Wright99, Parmar91} and the 
ro-vibrational lines sample even warmer ($\gsim 2000\unit K$) or fluorescently excited 
$\rm H_2$ \citep[e.g.,][]{Takami00, Fuente99, Moorwood90, Moorwood88, Black87}.  This 
is because the purely rotational transitions are quadrupolar and only occur between widely 
spaced ($\rm h\nu/k\geq 510\,K$) levels \citep[for a more detailed discussion, see]
[and references therein]{W99}.   Consequently, estimating the 
properties of the bulk of the molecular gas, which is cold \citep[kinetic temperatures 
of $\sim 5$--20$\,$K, see, e.g.,][]{Sanders85}, requires observing lines of trace 
molecules with dipole transitions and closely spaced levels.  The molecule CO and its 
isotopologues, like $\cO$ and $\Co$, are the usual molecules of choice for estimating the 
large-scale properties of molecular clouds, because the most common isotopologue of 
CO --- $\COex$ --- is the most abundant molecule after $\rm H_2$, the levels are 
closely spaced ($\rm h\nu/k = 5.5\,K$ for $\Jone$ of $\COex$), and the critical 
density of the lowest transition is low (i.e. $\nH\simeq 3\times 10^3\,cm^{-3}$ for 
$\Jone$).  These characteristics of CO ensure that the lines are often strong --- with 
radiation temperatures or $\Tr$ (i.e., Rayleigh-Jeans brightness temperatures as 
opposed to Planck brightness temperatures) of a few Kelvins --- for $\COone$ 
observations of clouds in our Galaxy, \citep[e.g.,][]{Sanders85} and occur throughout 
most of the molecular ISM.  Observations of lines of other molecules are also important 
for coming to a holistic picture of molecular clouds.  For example, the rotational 
lines of CS sample the dense gas ($\nH\gsim 10^5\, cm^{-3}$) most directly associated 
with star formation \citep{ELada92}.  Nevertheless, because the rotational lines of 
CO sample the excitation conditions in the bulk of the molecular gas, these lines 
provide insights into the environment in which the star-forming dense gas exists and, 
consequently, providing insights into star formation on the scales of many parsecs.  
Multi-parsec-scale observations of the lines of CO and its isotopologues, $\cO$ and
$\Co$, are carried out on external galaxies (e.g., see the references listed near 
the beginning of this paragraph) or by low-resolution (few arcminutes), wide-coverage 
(many degrees) mapping of Galactic molecular clouds \citep[e.g., see][]{Lee01, Dame01,
Dame87, Dame94, Grabelsky88, Bronfman88, Cohen86, Sakamoto95, Robinson88, Sanders84,
Tachihara00, Mizuno98, Kawamura98, Heyer98, Boulanger98, Onishi96, Oka98, Nagahama98, 
Sakamoto97a, Sakamoto94, Bally87, Maddalena86}.  

Besides the usually observed main isotopologue $\CO$ (written this way to 
explicitly distinguish it from the rarer isotopologues), the lines of 
the rarer isotopologues $\cO$ (short for $\cOex$) and $\Co$ (short for $\Coex$) 
furnish very important additional information on molecular cloud structure.  The 
lower-J lines of $\CO$ are often optically thick and cannot probe deeply into the 
substructures, such as clumps or filaments, of molecular clouds.  Also, the 
opacity of the $\CO$ lines reduces their sensitivity to the effects of changing 
density and changing radiative trapping.  Therefore, the lines of $\cO$ or $\Co$ 
are crucial for realistically constraining the molecular cloud excitation
conditions, because these lines, being from lower abundance isotopologues, are 
often optically thin on multi-parsec scales.  This greater sensitivity of the 
$\cO$ and $\Co$ lines to cloud structure and physical properties makes them
capable of identifying interesting regions within
molecular clouds.  Of particular interest is locating regions of warm (e.g., 
$\gsim 50$--$100\unit K$) molecular gas, because energetic events 
and massive stars associated with recent star formation heat the 
surrounding molecular gas.  Such gas is sometimes identified from ratios of 
strengths of CO rotational lines \citep[e.g.,][]{Wilson01, Plume00, Howe93,
Graf93, Graf90, Boreiko89, Fixsen99, Harris85, Harr99, W91, Gus93, Wild92, 
Harris91}.  

\begin{figure}
\epsscale{0.65}
\plotone{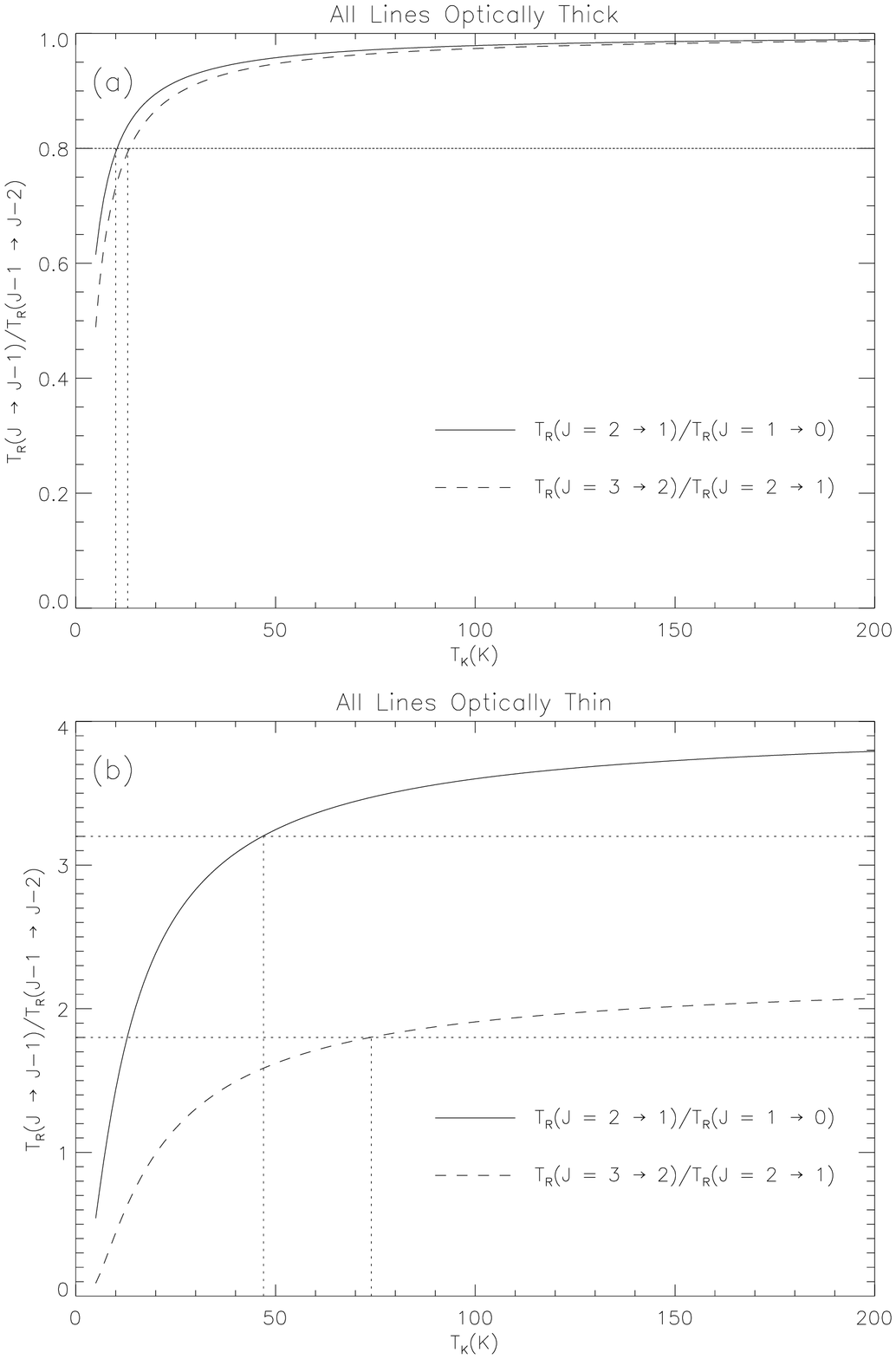}
\caption{The panels above show CO rotational line ratios as a function of kinetic 
temperature, $\Tk$, in the LTE limit.  The solid curves show the ratio of the 
radiation temperature, $\Tr$, of the $\Jtwo$ line to that of the $\Jone$ line.  
The dashed curves show the $\Tr(\Jthree)/\Tr(\Jtwo)$ ratio.  The horizontal dotted 
lines mark the levels of the ratios at 80\% of their asymptotic values in the
high-$\Tk$ limit.  The vertical dotted lines mark the $\Tk$ values for which the 
ratios achieve this 80\% level.  Panel (a) shows the behavior of these ratios in 
the case where all the lines are optically thick.  Panel (b) shows this behavior 
in the optically thin case. 
\label{fig1}}
\end{figure}

Examples of how two of the line ratios 
involving the three lowest rotational lines of CO vary with gas kinetic 
temperature, $\Tk$, in local thermodynamic equilibrium or LTE are shown in 
Figure~\ref{fig1}.  All the curves in Figure~\ref{fig1} show the same trend: 
the line ratios lose their sensitivity to $\Tk$ in the high-$\Tk$, or 
Rayleigh-Jeans, limit. This is a consequence of the line ratios depending on 
ratios of two Planck functions at different frequencies.  The maximum $\Tk$ 
to which a given line ratio is sensitive can be quantified by adopting a 
calibration uncertainty for the line ratios.  If we adopt the optimistic 
value of 20\% for the uncertainties in the ratios, then a ratio that has 
an observed value $\simeq 80\%$ of its asymptotic value in the Rayleigh-Jeans
limit is observationally indistinguishable from this asymptotic value.  
Similarly, the value of the kinetic temperature at this 80\% level is the 
maximum kinetic temperature for which the given ratio is sensitive.  Two 
important conclusions can be drawn from the plots of Figure~\ref{fig1}:
\begin{enumerate}
\item Optically thin lines, like those of $\cO$, have a greater range of 
sensitivity to $\Tk$ than do optically thick lines, like those of $\CO$.
Specifically, $\Tr(\Jtwo)/\Tr(\Jone)$ loses
$\Tk$ sensitivity above 46$\,$K in the optically thin case and above 10$\,$K
in the optically thick case.  
\item Ratios involving higher-J lines have a greater range of temperature
sensitivity than those with lower-J lines. For $\Tr(\Jthree)/\Tr(\Jtwo)$ the 
maximum $\Tk$ values are 73$\,$K and 13$\,$K for the optically thin and thick
cases respectively.  These values are higher than the corresponding values for
the $\Tr(\Jtwo)/\Tr(\Jone)$ ratio, especially for the optically thin case. 
\end{enumerate} 
It is clear then that unambiguous identification of warm molecular gas requires 
observations of high-J lines of $\CO$ and, sometimes, the optically thin lines
of $\cO$ (see references cited at the end of the previous paragraph).  (It 
should be mentioned that simply observing the optically thick lines of $\CO$ 
does not automatically give the kinetic temperature, because the observed lines 
are not necessarily thermalized and the gas does not necessarily fill the beam 
in the velocity interval about the line peak.)  
      
However, identifying warm gas in this manner can be complicated.  One complication 
is that observing higher and higher lines up the rotational ladder becomes increasingly 
difficult with ground-based observations, because the atmosphere becomes increasingly 
opaque, on average, and because sub-millimeter receivers become increasingly noisier
with decreasing observational wavelength.  Another complication is that non-LTE 
effects become more important higher in the rotational ladder: distinguishing between 
the density effects on the line ratios and kinetic temperature effects becomes 
difficult.  One way to separate such effects is to observe many CO rotational lines, 
but this can be observationally intensive --- especially if high-J $\cO$ lines are also 
observed.  Another approach to finding warm gas is using the $\CO/\cO$ intensity ratio.  
This ratio is very sensitive to the molecular gas kinetic temperature in
the LTE, high-$\Tk$ limit.  For example, the $\COone/\cOone$ intensity ratio
in LTE goes like $\Tk^2$ to within 25\% for $\Tk\geq 20\,K$.  While the advantage of
this ratio is that it only requires two lines with very
similar frequencies, the considerable disadvantage is that the $\CO$
and $\cO$ lines occur in very different optical depth regimes.  This results
in three problems.  One is that the $\CO$ line will be effectively probing a 
different volume of gas from that of the $\cO$ line.  Another is that the 
levels of radiative trapping will be very different between the $\CO$ and
$\cO$ lines, resulting in some sensitivity to density for the $\CO/\cO$ 
intensity ratio.  The third problem caused by the very different opacities
in the $\CO$ and $\cO$ lines is that the $\CO/\cO$ intensity ratio will
have some dependence on gas column density; indeed, in LTE this ratio is the
inverse of the optical depth of the $\cO$ line (for the $\CO$ line in the
optically thick limit and $\cO$ in the optically thin limit).  While the
problems above are not necessarily prohibitively difficult, it would be 
helpful nonetheless to have two optically thin tracers with very different
dependences on temperature, with relative insensitivity to gas density, and 
that are ubiquitous in the molecular ISM.  The use of such tracers would 
complement the use of lines of CO and of other molecules in diagnosing the 
presence of warm molecular gas simply and unambiguously.

Another way to trace molecular gas, and interstellar gas in general, is to 
observe the continuum emission from the dust grains found throughout most of
the ISM.  This continuum emission is usually observed at mid-IR, far-IR, 
submillimeter, and millimeter wavelengths.  On parsec scales, dust continuum 
emission is usually optically thin for wavelengths $\gsim 15\um$.  All-sky 
surveys like IRAS and {\it COBE}/{\it DIRBE} cover specific IR wavelengths from the
mid-IR to far-IR and near-IR to far-IR, respectively \citep{irasx, dirbex}.  
Surveys at such wavelengths in the continuum are valuable for probing the structure 
and excitation of the ISM \citep[see, for example,][]{Dupac01, W96, Bally91, Zhang89,
Werner76, Heiles00, Reach98, Boulanger98, Lagache98, Goldsmith97, Sodroski94,
Boulanger90, Sellgren90, Scoville89, Sodroski89, Leisawitz88}.  
Ratios of continuum intensities at different wavelengths, for instance,
can constrain estimates of dust grain temperatures.  A physical model that includes 
dust-gas thermal coupling could then, in theory, constrain gas temperatures 
\citep[e.g., ][]{Weingartner01, Mochizuki00, Hollenbach89, Tielens85, Burke83}.  
Optically thin dust continuum intensity is $\tau_\nu B_\nu(T_d)$, where $\tau_\nu$ 
is the optical depth at frequency $\nu$, and $\rm B_\nu(\Td)$ is the Planck function
evaluated at frequency $\nu$ and dust temperature, $\Td$. In the far-infrared, the 
optical depth is often approximated as a power-law $\tau_\nu\propto\nu^\beta$, where 
the emissivity index $\beta$ depends on the size and optical properties of the grains.
Consequently, as with CO rotational lines, the ratio of continuum intensities at 
different wavelengths loses sensitivity to dust temperature in the Rayleigh-Jeans 
limit.  As an example, we will consider the ratio of continuum intensities (per 
unit frequency) at the wavelength of 450$\um$ to that at 850$\um$, i.e. 
$\rm I_\nu(450\um)/I_\nu(850\um)$, and adopt a 20\% calibration uncertainty for this 
ratio.  We will also adopt $\beta=2$, which is a reasonable number for wavelengths 
longer than about 100$\um$ \citep{Andriesse74, Seki80, W96}.  These adopted values 
imply that the $\rm I_\nu(450\um)/I_\nu(850\um)$ ratio loses sensitivity to dust 
temperature for $\Td>37\, K$ (again from adopting 80\% of the asymptotic value in 
the Rayleigh-Jeans limit).  This maximum $\Td$ can be improved by roughly a factor 
of 2 by combining the 450$\um$ observations with observations at wavelengths longer 
than 850$\um$, but real improvement over this 37$\,$K limit requires observations at 
wavelengths even shorter than 450$\um$.   Observations at such short wavelengths must 
deal with increased atmospheric opacity and sometimes must be space-based (e.g., IRAS 
and {\it COBE}). 

A more direct ground-based approach is suggested by the work of \citet{Schloerb87}, 
who show that a direct comparison of optically thin CO line emission with submillimeter 
continuum can provide temperature estimates of the dust and molecular gas in molecular 
clouds \citep[see also][]{Swartz89}.  \citet{Schloerb87} derive an expression for 
the millimeter continuum to integrated $\Cotwo$ line strength ratio, 
$\rm I_\nu(submm)/I(\Cotwo)$, which goes like $\Tk\Td$ in the high-temperature
limit in LTE.  For similar gas and dust temperatures, this means that the 
$\rm I_\nu(submm)/I(\Cotwo)$ ratio and, similarly, the $\rm I_\nu(submm)/I(\cOone)$
ratio are actually {\it more sensitive to temperature as that temperature increases.\/} 
This is in stark contrast to ratios of rotational lines of a given isotopologue of CO
and to ratios of continuum intensities at different frequencies, which, as discussed
above, {\it lose\/} sensitivity to temperature in the high-temperature limit.  In
addition, comparing the continuum intensity with that of a low-J line like the
$\Jone$ line of $\cO$ (or of $\Co$) reduces the sensitivity of the continuum-to-line
ratio to gas density.  In short, the intensity ratio of the submillimeter continuum to 
an optically thin CO line may be the sought-for alternative diagnostic of molecular
gas temperature that was mentioned above.

Despite the potential diagnostic advantages of the continuum-to-line ratio \hfil\break
[e.g., $\rm I_\nu(submm)/I(\cOone)$], interpretation of this ratio and its 
spatial variations assumes the following:
\begin{itemize}
\item {\it The ratio of the mass of $\cO$ to that of dust does not vary spatially.\/}  
Even though the gas-to-dust mass ratio may vary, the $\cO$-to-dust mass ratio may 
not vary significantly within a single source, but may vary by up to an order of
magnitude from source to source \citep{Swartz89}.
\item {\it The gas along a given line of sight is predominantly molecular.\/}  
Appreciable quantities of atomic or ionized gas and their associated dust would
complicate the interpretation of $\rm I_\nu(submm)/I(\cOone)$.  This is because 
the submillimeter continuum emission would {\it not\/} originate in dust {\it 
only\/} associated with molecular gas, but the $\cO$ emission {\it would\/} 
originate only from the molecular gas.  Consulting H$\,$I 21-cm and radio continuum 
data can check this assumption.  
\item {\it The molecular gas density does not vary spatially.\/}  If this 
assumption is false, then spatial gradients in the $\rm I_\nu(submm)/I(\cOone)$ 
ratio may be probing spatial gradients in the molecular gas density rather than
in the molecular gas temperature.  Again, observing the lower-J rotational lines 
of $\cO$, especially the $\Jone$, line can minimize this problem.
\item {\it The gas and dust temperatures have spatial gradients in the same
direction.\/}  Even if $\Tk\ne\Td$, only assuming that the spatial gradients are 
in the same direction for both $\Tk$ and $\Td$ is necessary.  This is reasonable 
since both gas and dust are cooler with greater distance from heating sources, 
especially on multi-parsec scales. 
\item {\it Dust grain properties, like the dust mass absorption coefficient 
($\rm\kappa_{d\nu}$) and emissivity index ($\beta$), do not vary spatially.\/}
Observations of Orion, for example, show that such variations do occur for
$\beta$ on sub-parsec scales \citep[e.g.,][]{Johnstone99}, but, to first 
order, adopting a single value on multi-parsec scales is possible. 
\item {\it The $\cO$ line used in the comparison is truly optically thin.\/} 
If the $\cO$ line opacity varies from optically thin to optically thick, the 
$\rm I_\nu(submm)/\Ic$ ratio would again show gradients. The cause of the 
opacity change could still be due to a change in temperature, but it could also 
be due to a change in column density (and even volume density).  Observing the 
corresponding $\Co$ line can test this.
\item {\it There is no temperature gradient along the line of sight.\/}  Such 
a gradient would result in the submillimeter emission being biased toward the 
warmer regions along the line of sight and the $\cO$ emission much less so 
biased (and even biased toward the colder regions if LTE applies). Even in the
hypothetical case of $\Tk=\Td$ at every point along the line of sight, these
biases could lead the observer, armed with observations at additional frequencies,
to believe that $\Tk\ne\Td$.  This could also lead to the erroneous result that
comparisons from one line of sight to another show no coupling between gas
and dust temperatures.  The effects of gradients along the line of sight can
be minimized by observing clouds on multi-parsec scales, where such gradients
would be less extreme. 
\item {\it Thermal and non-thermal gas emission make a negligible contribution 
to the submillimeter continuum\/}.  This should indeed be the case. In M$\,$82, for 
example, \citet{Thronson89} estimate that $\sim 30\%$ of the 
1300$\um$ flux in the central 30-45$''$ is due to gas emission. This would be 
even less at shorter wavelengths and in regions with no dominant supernova
remnants nor HII regions. 
\end{itemize}
These assumptions may be reasonable first approximations for
observations on multi-parsec scales or, depending on the assumption, 
can be tested for each source observed.

Testing the overall validity of the above assumptions and, as a result, 
the behavior and effectiveness of the $\rm I_\nu(submm)/\Ic$ ratio 
as a diagnostic of molecular cloud physical conditions requires 
maps of a molecular cloud in the submillimeter or far-infrared (FIR) continuum, 
in the $\cOone$ or $\Coone$ spectral lines, and of dust or gas temperatures, 
all with resolutions at multi-parsec scales.  The most reliable maps of
temperatures associated with molecular clouds would be from ratios of
FIR maps.  Ignoring certain complications for the moment, such as 
line-of-sight gradients in dust properties (e.g., $\Td$), uncertainties 
in those properties (e.g., $\beta$), and dust grains {\it not\/} in thermal 
equilibrium, the ratio of the intensities of FIR maps at two different 
wavelengths gives reliable dust temperatures because such intensities are 
in the Wien limit of the Planck function.  In this limit, the dust 
temperature derived from the observed intensity ratio is insensitive to that 
ratio.  For example, the observed continuum intensity ratio of 
$\Ia/\Ib=1.66$ would imply $\Td=20\,K$.  A 10\% 
uncertainty in that intensity ratio would imply only a 5\% uncertainty in 
the derived dust temperature (or $\pm 1\unit K$).  One complication, that of 
stochastically-heated dust grains (i.e., grains 
{\it not\/} in thermal equilibrium), can be effectively avoided by using
two FIR maps at wavelengths of $\gsim 100\um$ \citep[for example, see][]{Desert90,
W96}.  This, therefore, 
rules out using IRAS maps, whose nomimal wavelengths do not go beyond 100$\um$. 
On the other hand, maps from the {\it DIRBE\/} instrument aboard the {\it COBE\/}
spacecraft cover the whole sky in ten IR bands from 1.25$\um$ out
to 240$\um$ \citep{dirbex}.  Two maps are at nominal wavelengths of 
longer than 100$\um$ (i.e., 140$\um$ and 240$\um$), making them virtually free 
of emission from stochastically-heated grains.  Thus the {\it DIRBE\/} maps at 
140$\um$ and 240$\um$ and their ratio can be compared with a map of $\cOone$ 
(or $\Coone$ if available) of a given molecular cloud.  Maps of entire 
clouds in our Galaxy in $\cOone$ have recently become available \citep[e.g., see]
[]{Lee01, Mizuno98, Kawamura98, Nagahama98}.  Therefore, the observational data 
exist to test the sensitivity and reliability to molecular cloud physical 
conditions of the ratio of the submillimeter or FIR continuum to $\cO$ line 
intensities. 

The appropriate molecular cloud, or cloud{\it s\/}, for which this testing
should be applied must satisfy a number of criteria:
\begin{itemize}
\item[---] The test clouds should be bright in the desired tracers (i.e., FIR and
$\cO$) and completely or nearly completely mapped in these tracers,
\item[---] largely mapped in tracers of atomic and ionized gas, such as
HI 21-cm and radio continuum, 
\item[---] out of the Galactic plane to reduce confusion with foreground 
and background material, 
\item[---] several degrees in extent so as to accommodate many beams of the 
large {\it DIRBE\/} beam ($0\degree.7$),
\item[---] and should have as large a range of dust temperatures as possible 
at the {\it DIRBE\/} resolution. 
\end{itemize}
The clouds of Orion most closely satisfy these criteria \citep[see, e.g.,][]{Bally91,
W96, Nagahama98, Zhang91, Green91, Chromey89, Heiles74, Reich78, Berkhuijsen72,
Haslam70}.  While there are other clouds that {\it may seem\/} to satisfy these 
criteria as well \citep[e.g., Chamaeleon-Musca,][]{Mizuno98}, the Orion clouds 
have the largest range of dust temperatures at the {\it DIRBE\/} resolution 
\citep[see Figure~2 of][]{Lagache98}.  The Orion$\,$A and B molecular clouds are 
bright in $\cOone$ \citep{Nagahama98}, and in the FIR continuum at 140$\um$ and 
240$\um$ \citep{W96}, and has been completely or largely mapped in these tracers.  
Much of the Orion constellation has been mapped in H$\,$I \citep{Zhang91, Green91, 
Chromey89, Heiles74} and radio continuum \citep{Reich78, Berkhuijsen72, Haslam70}, 
thereby permitting tests of whether or not molecular gas dominates the cloud surface,
or column, density for each line of sight.  Much of Orion has also been mapped
in $\CO$ lines \citep{Wilson05, Lang00, Ikeda99, Sakamoto97a, Sakamoto94, Maddalena86}, 
which can also provide crude estimates of the molecular gas column densities.  Note that
only \citet{Maddalena86} have mapped the Orion all three major clouds: Orion$\,$A, 
Orion$\,$B and the $\lambda\,$Orionis ring.  The Orion clouds are about 15-20$^\circ$ 
out of the Galactic plane and about $6^\circ\times 3^\circ$ (or about $50\unit pc\times 
25\, pc$ for an adopted distance of 450$\,$pc) in size, down to some ill-defined zero 
level.  An additional advantage to the clouds of Orion is that these clouds have been 
surveyed at many other wavelengths on the scales of many parsecs, thereby providing 
further checks on models.  These other wavelengths include, for example, the optical 
surveys of the Orion OB1 and $\lambda\,$Ori OB associations \citep{Brown95, Brown94, 
Warren78, Warren77, Murdin77, Blaauw64}, the near-IR to far-IR all-sky survey of 
{\it COBE}/{\it DIRBE\/} \citep{W96}, the mid-IR to far-IR all-sky survey of IRAS 
\citep{Bally91}, and submillimeter maps of small sections of Ori$\,$A and B 
\citep[e.g.,][]{Johnstone01, Mitchell01, Johnstone99}.  Therefore, the Orion$\,$A 
and B molecular clouds were chosen for testing the behavior of the 
$\rm I_\nu(submm)/\Ic$ ratio, using the {\it DIRBE\/} maps at 140$\um$ and 240$\um$ 
and the $\cOone$ data of \citet{Nagahama98}.

In addition to testing the diagnostic abilities of the continuum-to-line ratio, the
current data may also permit us to test theoretical models describing interactions
between dust and gas.   Specifically, models of dust-gas thermal coupling use very
simple assumptions, such as large, single-size, spherical dust grains with smooth
surfaces \citep[e.g., see][]{Goldsmith01, Burke83}.   Are these simplistic assumptions
sufficient for explaining the observations?   This will be examined. 

The current paper, or Paper~I, describes the one- and two-component modeling of the data.
Paper~II \citep{W05} in this series confirms and, in some cases, modifies the model results 
using simulated data.  Paper~III \citep{W05a} examines systematic effects not explicitly 
considered in the models and also discusses the scientific implications of the results.

\section{Data Processing\label{sec2}}

\subsection{DIRBE Data\label{ssec21}}

Details of the {\it COBE\/} mission and {\it DIRBE\/} instrument can be found in \citet{Boggess92}, 
\citet{Silverberg93}, \citet{Hauser98}, and the \citet{dirbex}.  Much in this section has been 
discussed in \citet{W96} (hereafter referred to as W96) in Section~2.1 of that paper.  The data in 
the current paper are from the 1998 release, whereas the data of W96 were of the 1993 
release.  Also, the current paper concentrates on Bands 9 (i.e. 
$\lambda = 140\um$) and 10 ($\lambda = 240\um$), while W96 discussed
all 10 wavelength bands.  Consequently, the data processing
in this section largely concerns the maps in the 140$\um$ and 240$\um$ bands from the
1998-release of the {\it DIRBE\/} data. 

While a number of {\it DIRBE\/} data products are available, the ZSMA (Zodi-Subtracted Mission
Averaged) maps were the most relevant for the current research. \citep[See the][for 
more details of the {\it DIRBE\/} data products described in this paragraph.]{dirbex} These maps 
are the average sky brightnesses determined from averaging over the full 40 weeks of the 
{\it COBE\/} mission, thereby maximizing the signal-to-noise ratio of the data.  Maximizing the
signal-to-noise ratio is particularly important for the 140$\um$ and 240$\um$ bands,
because these have the noisiest data.  The average sky brightness in each wavelength 
band was determined after removal of the {\it DIRBE\/} interplanetary dust model for each of 
the ten wavelength bands, which effectively removes the zodiacal emission from the maps.  
The ZSMA maps are available in the form of FITS tables that include the pixel number, 
the intensity in that pixel in MJy/sr, and the noise in that pixel in MJy/sr.  The
FITS tables were converted into maps in the equal-area Mollweide projection using IDL 
routines available from NASA/GSFC through the worldwide web.  The fxbopen and 
fxbread routines were used to open the FITS files and read in the data, respectively. The 
data included the pixel numbers, sky brightnesses, and noise levels. The pix2xy routine 
was used to create maps from the data in the {\it DIRBE\/} sky-cube format.  This was followed 
by converting the sky-cube maps to maps in the Mollweide projection using the reproj 
routine.  Maps of both the sky brightnesses and of the rms noise levels were created 
this way.  The maps of the rms noise levels are hereafter referred to as the sigma
maps. 

After the creation of the Mollweide maps using the IDL routines from NASA/GSFC, the
section of the all-sky maps corresponding to the Orion region was extracted.  As was 
done in W96, a cosecant-law background of the form $a_c csc(|b|)$ was subtracted 
from the sky-brightness maps, so as to remove foreground and background emission 
associated with the Galactic plane.  The fitting procedure was virtually identical
to that used in W96 with a slight change in the algorithm used to determine the
Galactic latitude $b$ from the pixel position.  The earlier paper used a simplistic
linear interpolation for determining $b$ in the cosecant-law fitting routine.  The
current paper uses the exact expression that was used in creating the maps in the 
Mollweide projection.  Nevertheless, over the latitude range of the Orion maps (i.e. 
$b = -5$ to $-31\degree\,$), the improvement in the algorithm only contributes a 4\% 
change to the scale factors, $a_c$.  As stated in W96, the uncertainties in the 
map intensities due to the uncertainty in the background subtraction is about 10\%.
The effects of this subtraction on the results will be discussed in Paper~II \citep{W05}.

These Orion maps were compared with those of
the 1993 release of the {\it DIRBE\/} data.  The brightnesses of the 140 
and 240$\um$ maps were reduced by about 15\% with respect to those of the 1993 maps
made.  Accordingly, luminosities and masses determined from
these maps are also reduced by this amount.  However, the 140$\um$/240$\um$ 
dust color temperatures are virtually unchanged. 

\begin{figure}
\epsscale{0.59}
\plotone{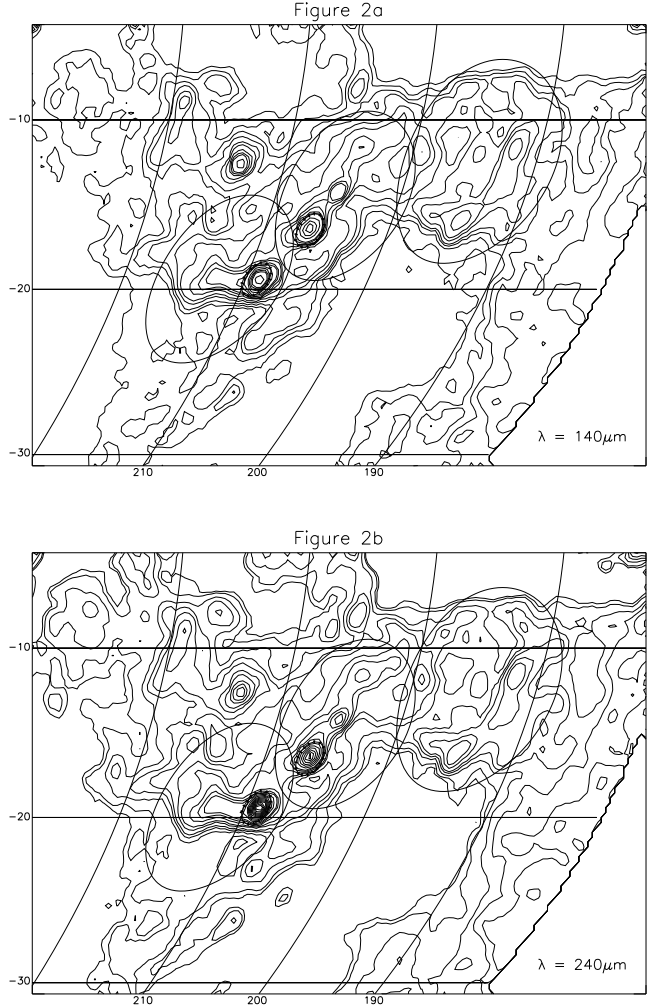}
\caption{The {\it DIRBE\/} Surface Brightness Maps for 
$\lambda = 140$ and 240$\um$ from the 1998-release
are displayed in the equal-area Mollweide projection with a Galactic
longitude-latitude coordinate grid.  The analysis concentrates on the 
three $10^\circ$-$12^\circ$ circular fields that appear in the map as
roughly elliptical loops; circles on the sky appear distorted in this 
projection. The fields, from east to west, are the Orion$\,$A, 
Orion$\,$B, and $\lambda\,$Orionis Fields.  Also shown with dotted
ellipses are the smaller ($2^\circ$ diameter) Orion$\,$Nebula and 
NGC$\,$2024 Fields (which are circular on the real sky).   The absolute 
photometric uncertainties of the maps are 10.6\% for the 140$\um$ map 
and 11.6\% for the 240$\um$ map.  All maps have been ``cleaned" of 
zodiacal light and had $cosecant(|b|)$ backgrounds subtracted.  The 
contour levels for the 140$\um$ map are 5, 10, 15, 20, 30, 40, 60, 80, 
100, 150, 200, 400, 600, 800, 1000, 1500, 1900$\MJsr$ and for the 
240$\um$ map are  3, 5, 10, 15, 20, 30, 40, 60, 80, 100, 150, 200, 
250, 300, 350, 400, 450 500, 550, 600, 650, 700, 750$\MJsr$.
\label{fig2}}
\end{figure}

Before creating maps of the 140$\um$/240$\um$ dust color temperatures, the 140$\um$ 
and 240$\um$ maps were smoothed, as was done in
W96.  Without smoothing these maps, the resultant dust temperature map
appears to be very choppy: noticeable, large, and random variations that occur
within a beam size.  The maps were smoothed with a square flat-topped box that
was 3$\,$pixels $\times$ 3$\,$pixels in width.  Since the {\it DIRBE\/} beam size is
roughly the same size (about 2.3$\,$pixels $\times$ 2.3$\,$pixels) and shape, the 
equivalent resolution after smoothing is roughly $\sqrt2$ larger than the original 
42$'$ or about 1$\degree$.  The resultant maps of $\Ia$ and
$\Ib$ are shown in Figure~\ref{fig2}. 

As with the intensity maps, the corresponding sigma maps must also be smoothed.
Here, however, the smoothing procedure is not so straightforward.  Suppose the
smoothing box is $n_x\,$pixels $\times$ $n_y\,$pixels and $I_{ij}$ is the 
intensity at pixel $ij$ in the unsmoothed map.  Then the intensity in the
smoothed map at the central pixel of the smoothing box, $I_s$, is 
\begin{equation}
I_s = {\strut 1\over \strut N} \sum_{i=1}^{n_x} \sum_{j=1}^{n_y} I_{ij}\quad ,
\label{dp01}
\end{equation}
where $N=n_x n_y$ is the total number of pixels in the smoothing box.  If 
the pixels are independent, then the rms noise level in the smoothed map
at the central pixel of the smoothing box, $\sigma_s$, is
\begin{equation}
\sigma_s^2 = {\strut 1\over \strut N^2} \sum_{i=1}^{n_x} \sum_{j=1}^{n_y} 
\sigma_{ij}^2\quad .
\label{dp02}
\end{equation}
However, if the $I_{ij}$ are already convolved with a beam the same size and
shape as the smoothing box (i.e. $n_x\,$pixels $\times$ $n_y\,$pixels),  then
$\sigma_s$ can be approximated by
\begin{equation}
\sigma_s^2 = {\strut 1\over \strut 2N} \sum_{i=1}^{n_x} \sum_{j=1}^{n_y} 
\sigma_{ij}^2\quad .
\label{dp03}
\end{equation}
Equation (\ref{dp02}) states that the noise in the smoothed map is a factor
$\sqrt N$ below the quadrature average of the noise levels in the smoothing
box.  This is expected when the pixels are independent because the 
random walk applies.  On the other hand, equation (\ref{dp03}) states that
the noise in the smoothed map is only a factor of $\sqrt2$ below the quadrature 
average of the noise levels in the smoothing box.  This is because the area of
the effective beam is roughly doubled (exactly doubled in the case of Gaussian
smoothing and a Gaussian beam) after convolution with a smoothing box the same 
size as the original beam.  Equation (\ref{dp03}) has been tested with
simulated maps of random noise.  After discarding pixels along the map edges,
it was found that for 99.9\% of the pixels equation (\ref{dp03}) must be either 
corrected downward or upward by up to 30\%.  This is not a large correction to 
random noise estimates.  Therefore, equation (\ref{dp03}) was used 
in smoothing the sigma maps. 

\begin{figure}
\epsscale{0.80}
\plotone{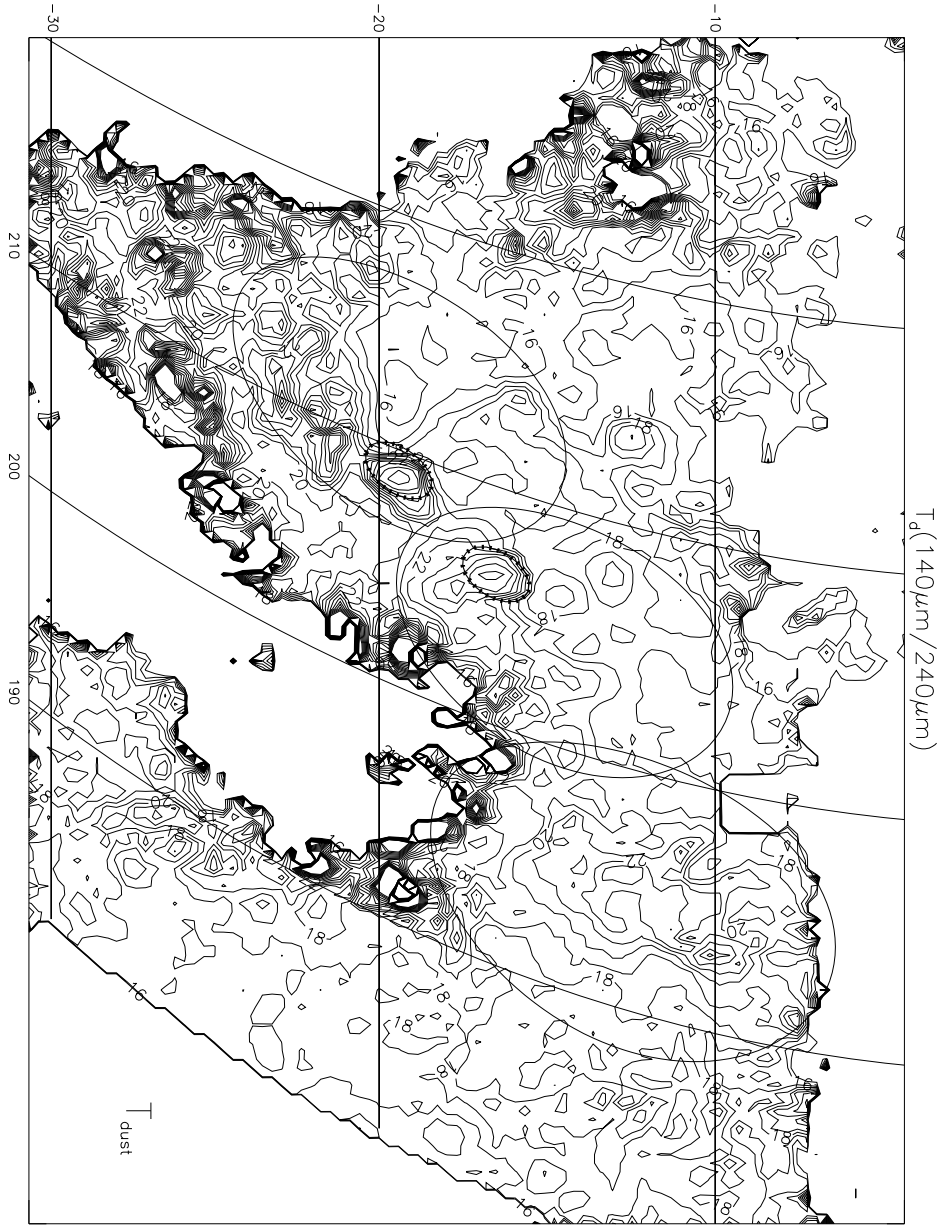}
\caption{Spatial distribution of the $\rm 140\um/240\um$ color
temperature, $\Td$, is illustrated ($\beta=2$ emissivity index unless
otherwise stated).  Contour levels are 15, 16, 17, ..., 30$\,$K.  All 
positions in the $\Td$ map where $\Ia < 1\,\MJsr$ or 
$\Ib < 1\,\MJsr$ were set to zero, resulting in the close 
spacing of contours at the edges of the emitting regions.
\label{fig3}}
\end{figure}

The map of the dust color temperatures (which is the actual dust grain temperature
for the large grains in thermal equilibrium in the case of homogeneous dust 
properties along the line of sight, assuming the spectral emissivity index is 
correct) is created in the same way as in W96.  The dust temperature, $\Td$, is 
derived from inverting
\begin{equation}
\rm {\cal R}={\nu_9^\beta B_{\nu 9}(\Td) 
K_{\nu 9}(\Td,\beta)
\over \nu_{10}^\beta B_{\nu_{10}}(\Td) K_{\nu_{10}}(\Td,\beta)}\quad ,
\label{dp04}
\end{equation}
where $\rm {\cal R}\equiv I_{\nu 9}/I_{\nu_{10}}$, $\rm I_{\nu 9}$ and $\rm I_{\nu_{10}}$ 
are the quoted {\it DIRBE\/} intensities at frequencies $\nu_9$ and $\nu_{10}$,
$\nu_9$ is the frequency at $\lambda_9=140\um$, $\nu_{10}$ is the frequency 
at $\lambda_{10}=240\um$,  $\rm B_\nu(\Td)$ 
is the Planck function, and $\rm K_\nu$ is the {\it DIRBE\/} band color correction.
Even though these color corrections never vary by more than 8\% from 1.0, their
ratio, $K_{\nu 9}/K_{\nu_{10}}$, can vary by up to 14\% from 1.0 in the temperature
range, $\Td=15$--$30\unit K$.  Consequently, the color corrections were necessary
for obtaining accurate dust color temperatures.  Although there is evidence that
$\beta$ can vary spatially by 10--20\% on sub-parsec scales \citep[e.g., see][]
{Johnstone99}, adopting $\beta=2.0$ is still valid on the scale of a few parsecs 
for the reasons given in W96.  Equation (\ref{dp04}) was solved for $\Td$
by plotting $\Td$ as a function of ${\cal R}$ over the temperature range 2.1$\,$K to
210$\,$K, and fitting polynomials over three subranges within the full range.
The polynomial fits have a typical accuracy 0.08$\,$K and never deviate 
by more than 0.4$\,$K, which occurs at $\Td = 2.1\,K$ and 45$\,$K.  The polynomial
fits quickly determine $\Td$ for each pixel.  The $\Td$ map is
shown in Figure~\ref{fig3}. 

The 1998 release of the {\it DIRBE\/} data includes the sigma maps and this allows 
the computation of the map of random errors in $\Td$, i.e., a $\sigma(\Td)$
map.  To compute this map rapidly and easily, the Wien approximation is applied
to the Planck functions in equation (\ref{dp04}).  The Wien version of (\ref{dp04})
is differentiated and rearranged to give the following simple expression:
\begin{equation}
{\sigma(\Td)\over\Td} = {\Td\over A}{\sigma({\cal R})\over {\cal R}}\quad ,
\label{dp05}
\end{equation}
where it was assumed that the color corrections are only weak functions of
$\Td$ and where $\rm A = h(\nu_9 - \nu_{10})/k = 42.8\, K$.  The constants h 
and k are Planck's constant and the Maxwell-Boltzmann constant, respectively.  
The relative error in ${\cal R}$, $\sigma({\cal R})/{\cal R}$, was computed from the 
quadrature sum of the relative errors of the intensities in the 140$\um$ and 
240$\um$ maps. It must be emphasized that even though the {\it uncertainties\/} 
in $\Td$ are computed from an expression that is based on the Wien approximation 
--- i.e., (\ref{dp05}) --- {\it no\/} such approximation was used in determining 
the $\Td$ itself.  Therefore, the $\Td$ that appears in (\ref{dp05}) was computed 
from (\ref{dp04}) (or an accurate approximation of it), which includes the full
Planck function and the color corrections.  Using the accurately determined
$\Td$ improves the accuracy of the derived $\sigma(\Td)$ values.  A detailed
comparison with $\sigma(\Td)$ determined from a numerical differentiation of
(\ref{dp04}) shows that the $\sigma(\Td)$ derived from (\ref{dp05}) only needs
to be corrected upwards by 20--25\% in the temperature range 15--30$\,$K.  Such
a correction was not applied and makes little difference to the results. (This
is discussed later in more detail.)

Besides the random uncertainties due to noise, there is the overall
calibration or photometric uncertainty of the map in each wavelength band.  
These are 10.6\% and 11.6\% in the 140$\um$ and 240$\um$ bands, respectively 
\citep{Hauser98}.  How these systematic uncertainties affect the derived
color temperatures, $\Td$, depends on the relative calibration uncertainty
of the 140$\um$ band with respect to the 240$\um$ band.  In W96 the 
calibration uncertainty in the $\Ia/\Ib$ ratio was
specified as 6\%.  Adopting that value for the current paper gives an
uncertainty of $\pm 0.4\unit K$ for $\Td=15\, K$ and $\pm 1.6\unit K$
for $\Td=30\, K$.  These uncertainties are small compared to the relevant
range of values of $\Td$ (i.e., 15--30$\,$K) itself.  Consequently, the 
systematic uncertainty in the 140$\um$/240$\um$ color temperature is 
roughly equivalent to an upward or downward shift of about 1$\,$K to the 
entire $\Td$ map.

\subsection{$\cObf$ Data\label{ssec22}}

The $\cO$ data were obtained with the Nagoya 4-m antenna and cover both the
Orion$\,$A and Orion$\,$B clouds.  The beamsize is 2\rlap{$'$}.7 and Orion$\,$A
was mapped with a 2$'$ grid spacing.  Full details of the Orion$\,$A observations
are given in \citet{Nagahama98}.  The Orion$\,$B cloud was mapped with an 8$'$
grid spacing.  The data consist of velocity-integrated intensities of the
$\cOone$ line and were gridded to the {\it DIRBE\/} resolution in the Mollweide
projection to allow comparison with the {\it DIRBE\/} data.  The regridding procedure
was a straightforward conversion from the galactic longitude and latitude of
the central position of each pointing of the beam on the sky to {\it DIRBE\/} pixel
coordinates in the Mollweide projection.  These pixel coordinates were rounded 
off to the nearest whole number to give the {\it DIRBE\/} pixel that contains the center 
of the beam.  This simple approach does not take into account the cases where a 
given pointing of the 2\rlap{$'$}.7 beam spills over the edges of given {\it DIRBE\/} 
pixel.   However, this is not an important consideration given that the 
{\it DIRBE\/}-gridded map is then smoothed by 3$\,$pixels $\times$ 3$\,$pixels. 
Each {\it DIRBE\/} pixel was ``hit" by a few pointings of the Nagoya beam and the 
corresponding intensities were averaged for each {\it DIRBE\/} pixel.  The 1-$\sigma$ 
noise level of the Nagoya integrated intensity map was 0.7$\Kkms$.  After regridding 
to the {\it DIRBE\/} resolution, the noise-level was computed for each {\it DIRBE\/} 
pixel by dividing the Nagoya map noise-level by the square-root of the number of hits 
of the Nagoya beam in that pixel.  Hence, the regridding process produced both an
integrated intensity map of the $\cOone$ line and a corresponding sigma map. 
As was done with the {\it DIRBE\/} maps, the $\cO$ map was smoothed by 3$\,$pixels 
$\times$ 3$\,$pixels.  The sigma map was smoothed using equation (\ref{dp02}),
since the pixels are independent in this case. 

As mentioned above, the Orion$\,$B section of the original Nagoya map has an
8$'$ grid spacing even though the beam is only 2\rlap{$'$}.7 in size.  This
spacing gives gaps in the coverage.  These gaps do not result in large systematic 
errors in the velocity-integrated intensities in the final map, largely because 
all the data were gridded to the {\it DIRBE\/} resolution.  Systematic errors in the map 
intensities due to incomplete coverage can be estimated by comparing the Orion$\,$A 
observations on the grid with 2$'$ spacing with those same observations on a grid 
with 8$'$ spacing.  Both the 2$'$ grid and 8$'$grid Orion$\,$A data were regridded 
to the {\it DIRBE\/} resolution and then smoothed by 3$\,$pixels $\times$ 3$\,$pixels.  The 
median of the intensities of the map that was originally on an 8$'$ grid were 10--11\% 
lower than that of the map that was originally on a 2$'$ grid.  Therefore, it is 
possible that the Orion$\,$B observations need to be corrected upwards by, on average, 
about 10\%.  Such a correction was not applied, because this is only a small change 
that will not affect the results significantly and because application of a single
scale factor to {\it all\/} the Orion$\,$B intensities is overly simplistic. 

\begin{figure}
\epsscale{0.85}
\plotone{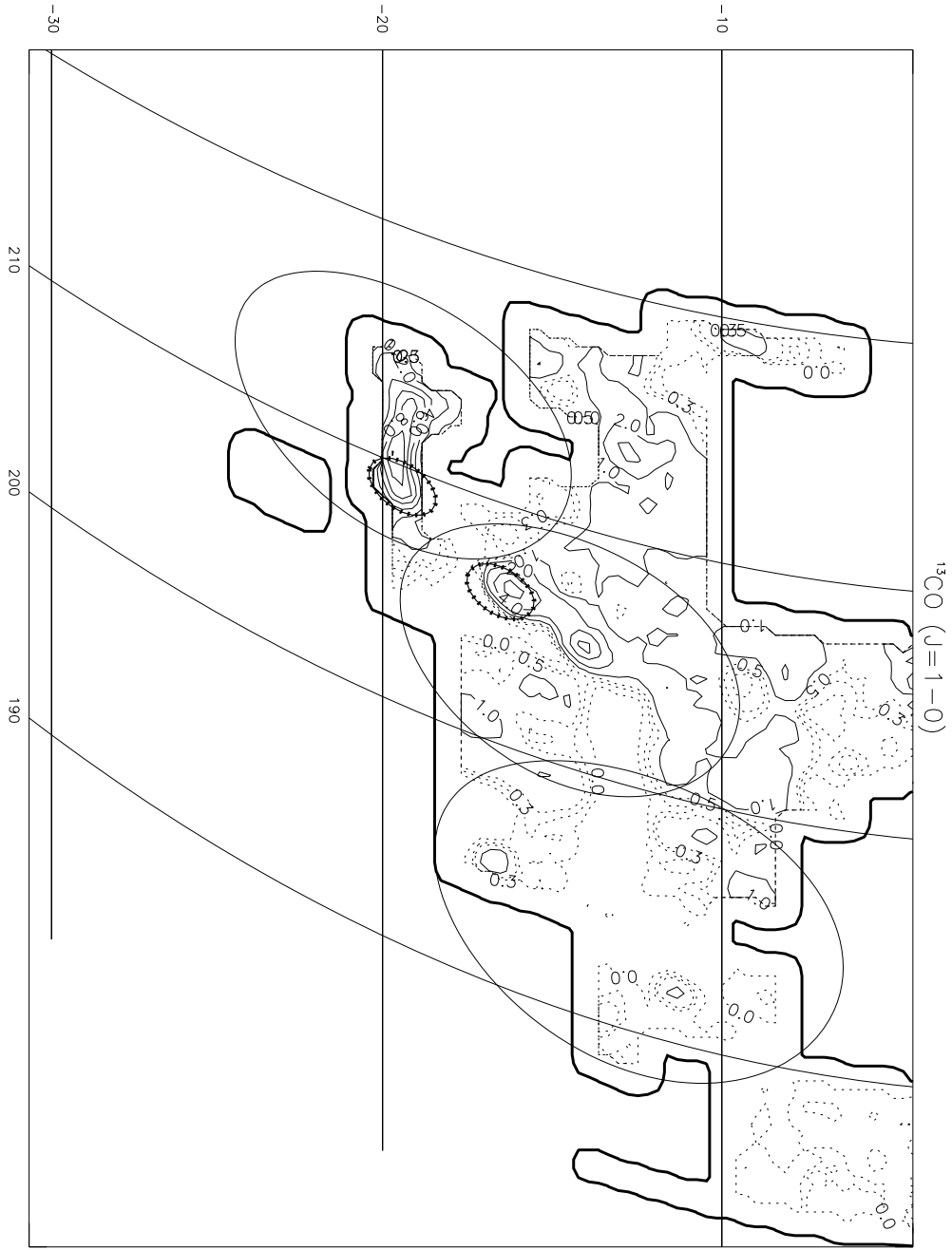}
\caption{Map of the velocity-integrated radiation temperature of the
$\cOone$ line is shown for the Orion region.  Contour levels are 0,
0.3, 0.5, 1, 2, 4, 6,...,12$\Kkms$. The thick contour indicates the 
edge of the mapped area.
\label{fig4}}
\end{figure}

A final consideration for comparison of the $\cO$ data with the {\it DIRBE\/} data is
whether background subtraction, as was done for the {\it DIRBE\/} data, is necessary. 
As was found in W96 for the $\CO$ data, such background subtraction for
molecular line data produced virtually no change in their intensities in the
Orion fields. Consequently, no background subtraction was applied to the $\cO$ 
map.  The map of velocity-integrated radiation temperature of the $\cOone$
line, or $\Ic$, is shown in Figure~\ref{fig4}. 

\subsection{Other Datasets\label{ssec23}}

Other datasets used in this research include the map of the integrated intensities
of the $\COone$ line, the map of the integrated intensities of the H$\,$I 21-cm line, 
and the map of the peak $\COone$ radiation temperatures.  The maps of the integrated 
intensities of the $\COone$ line and of the H$\,$I 21-cm line were described in W96
\citep[see also][]{Dame87, Maddalena86, Heiles74}.  As in W96, an integrated
CO line intensity to molecular gas column density, or X-factor, of $\rm N(H_2)/
\int\Tr dv = 2.6\times 10^{20}\cKkms$ is assumed.  Recent work by \citet{Dame01}
suggests that an X-factor of $\sim 2.2\times 10^{20}\cKkms$ is more appropriate for
the Orion clouds.  As will be discussed in Section~\ref{ssec36}, this is indeed a more 
appropriate value (at least for the one-component models fitted to the data), but the 
value of $2.6\times 10^{20}\cKkms$ is adopted for consistency with W96.  After regridding
the data to the {\it DIRBE\/} resolution and performing the usual 3$\,$pixel $\times$ 3$\,$pixel
smoothing, the noise levels in the CO and H$\,$I integrated
intensity maps were estimated from the negative tails of the histograms.  The 1-$\sigma$
noise levels are 0.7 and 0.2 in units of $10^{20}\unit cm^{-2}$ for the CO and HI,
respectively.  As described in W96, a smooth background representing large-scale
background and foreground Galactic emission was subtracted from the H$\,$I map.  No
such background was subtracted from the CO map because it made little difference
to the resulting intensities.  Both the regridded CO and H$\,$I maps can be seen in
Figure~3a of W96. 

The peak $\COone$ radiation temperatures were provided by R. Maddalena and T. Dame
(priv. comm.).  As with the other datasets, these radiation
temperatures were gridded to the Mollweide projection with the same pixels as in
the {\it DIRBE\/} maps and then smoothed by 3$\,$pixels $\times$ 3$\,$pixels.  The negative
tail of the histogram of the map of peak $\COone$ radiation temperatures suggests
a noise level of 0.15$\,$K.  This is consistent with the noise level in the map
of integrated CO brightnesses for line widths of about 2$\kms$.  As with the 
integrated CO map, background subtraction was not necessary.  This map looks very
similar to the integrated CO map and, consequently, is not shown.

\section{Models and Results\label{sec3}}

\begin{figure}
\epsscale{0.85}
\plotone{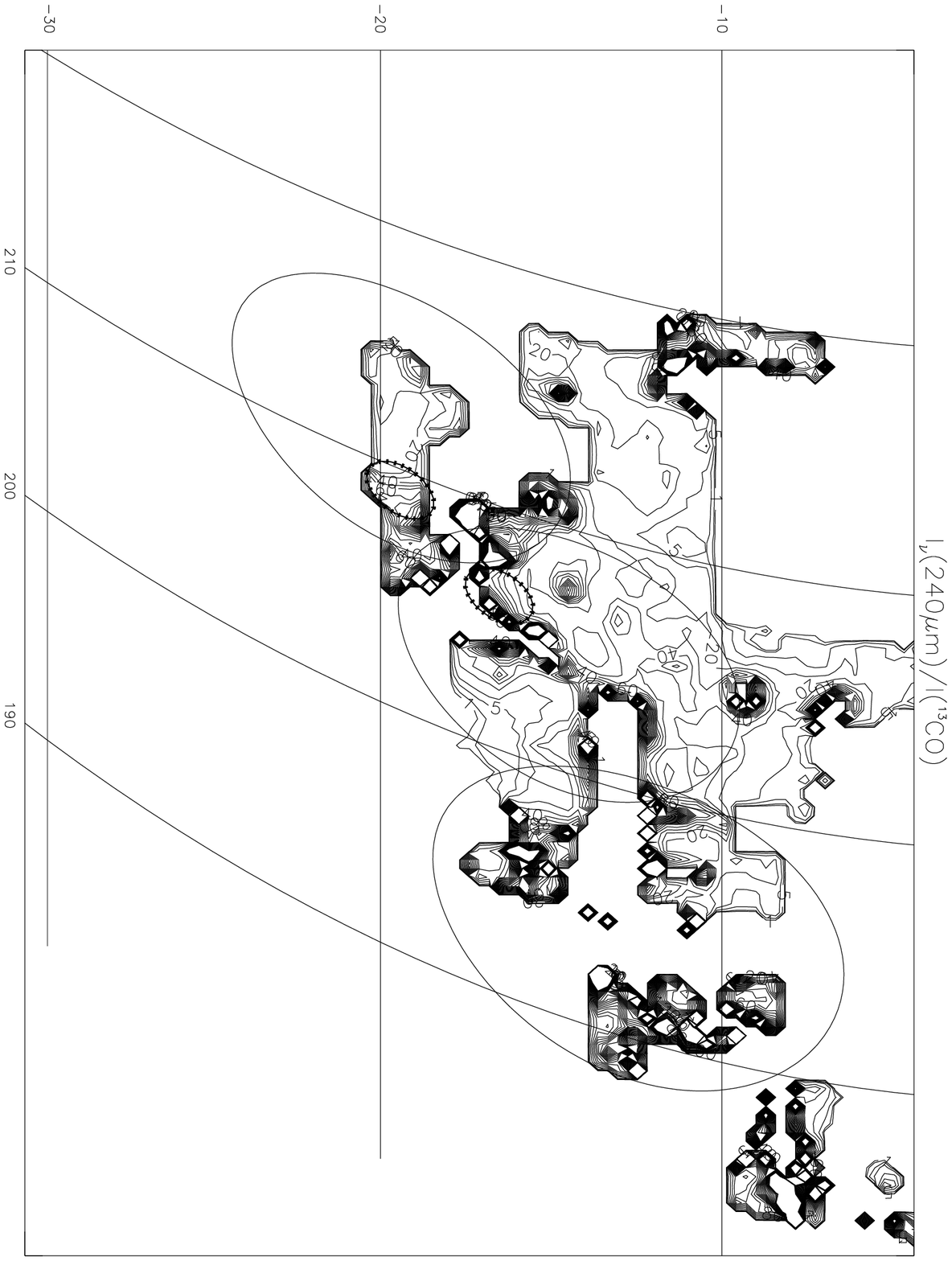}
\caption{The map of $\rd$ or $\Ib/\Ic$ is shown for 
the Orion region.  Contour levels are 1, 2, 5, 10, 20, 30,..., 250$\MJkk$. 
\label{fig5}}
\end{figure}

\begin{figure}
\epsscale{0.73}
\plotone{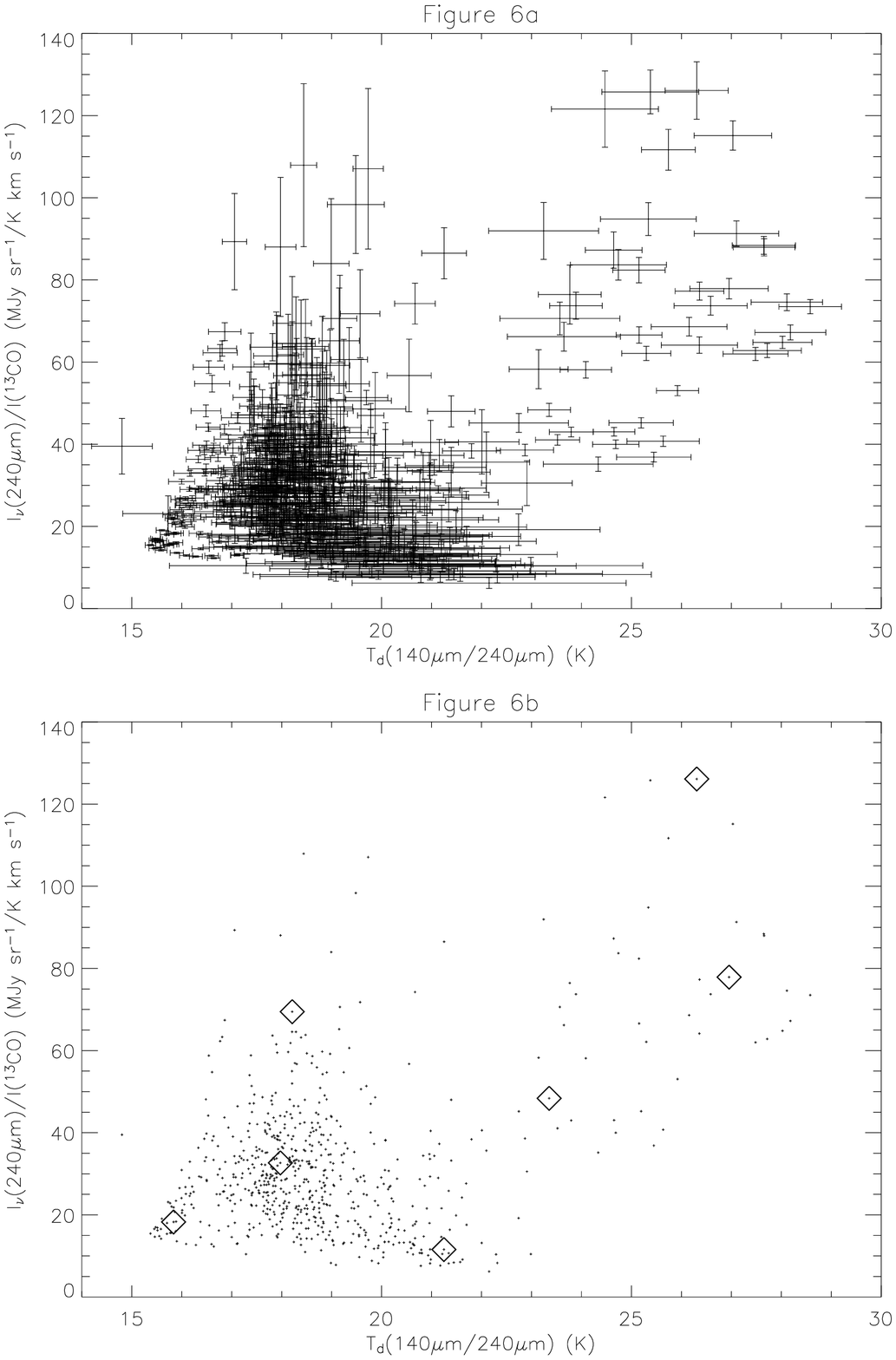}
\caption{Plots of $\rd$ or $\Ib/\Ic$ versus $\Td$ are shown 
for the Orion region.  The plots include points at more than 5-$\sigma$ in
$\Ia$, $\Ib$, and $\Ic$.  Panel a) includes 
the error bars of the random errors (i.e. systematic errors are not included), 
whereas in Panel b) the error bars are excluded to more clearly reveal the 
overall pattern in the plotted points.  The diamonds in this panel are the 
fiducial points discussed in Section~\ref{ssec35} and listed in 
Table~\ref{tbl-2x}. 
\label{fig6}}
\end{figure}

The ratio $\Ib/\Ic$, abbreviated as $\rd$, is shown as a contour map in 
Figure~\ref{fig5} and is plotted against the 140$\um$/240$\um$ color 
temperature, or $\Td$, in Figure~\ref{fig6}.  The points in Figure~\ref{fig6} 
represent the high signal-to-noise positions: those positions with signal at 
or above 5-$\sigma$ in $\Ia$, $\Ib$, and $\Ic$.  The total number of points 
in the plot is 674, but these are not totally independent.  Given that the 
effective beamsize is about 3$\,$pixels $\times$ 3$\,$pixels, the effective 
number of points is about 70. (This is actually a conservative estimate to 
the effective number of points, because, even though the pixels within the 
effective beam are not completely {\it in}dependent, neither are they 
completely {\it de}pendent.)  This number is important for assigning a 
confidence level to the model fits.

The physical models used to interpret Figure~\ref{fig6} include 
single-component LTE models, single-component non-LTE models, and
two-component non-LTE models.  Except where mentioned otherwise,
the models are based on the following assumption:
\smallskip
\hfil\break
{\it The only physical parameters that change from one line of sight
to the next are the dust temperature, $T_{_d}$, and the gas kinetic 
temperature, $T_{_K}$, while maintaining a constant difference,
$\DT \equiv T_{_d} - T_{_K}$. Other physical parameters such as
gas density, dust-to-gas mass ratio, dust mass absorption coefficient,
etcetera are assumed to be constant from position to position.\/}
\smallskip
\hfil\break
This will be referred to as the {\it basic assumption\/}.  Obviously,
this assumption is not strictly correct, but is nevertheless roughly
correct in describing the overall trends in the data.  Thus the
$\chi^2$ statistic that quantifies the goodness of the fit is not
only measuring the scatter due to observational uncertainties, but 
also the scatter due to the real physical variations not accounted
for in the basic assumption.  The best fit of a certain class of
models is found by adjusting a set of model parameters, $a_j$, so
as to minimize the $\chi^2$.  Because there are errors in both 
variables --- i.e. $x$ and $y$ or, more specifically, $\Td$ and 
$\rd$ --- the expression for $\chi^2$ must reflect this:
\begin{eqnarray}
\chi^2 &=&\sum_i d_i^2\quad ,
\label{mr1}\\
\noalign{\noindent where}
d_i^2 &=& \left[{x_{o\,i}-x_{m\,i}(\{a_j\})\over\sigma(x_{o\,i})}\right]^2\ +\
\left[{y_{o\,i}-y_{m\,i}(\{a_j\})\over\sigma(y_{o\,i})}\right]^2\quad ,
\label{mr2}
\end{eqnarray}
and where $(x_{o\,i},y_{o\,i})$ represents observed data point $i$, 
$(x_{m\,i}(\{a_j\}),y_{m\,i}(\{a_j\})$ is the point on the model curve
closest to data point $i$, the $\{a_j\}$ are the parameters that 
characterize the model curve, and $\sigma(x_{o\,i})$ and $\sigma(y_{o\,i})$
are the random errors for data point $i$ in $x$ and $y$, respectively.  The 
closest point on the model curve to data point $i$ is defined as that 
point on the curve that gives the minimum $d_i^2$.  Expressions (\ref{mr1}) 
and (\ref{mr2}) represent orthogonal regression, in which the fit seeks to
minimize the perpendicular distance between each point and the model curve.


    In the following sections, we examine the following cases:
LTE, non-LTE one component, and non-LTE two components.

\subsection{LTE Models and Results\label{ssec31}}

\begin{figure}
\epsscale{0.8}
\plotone{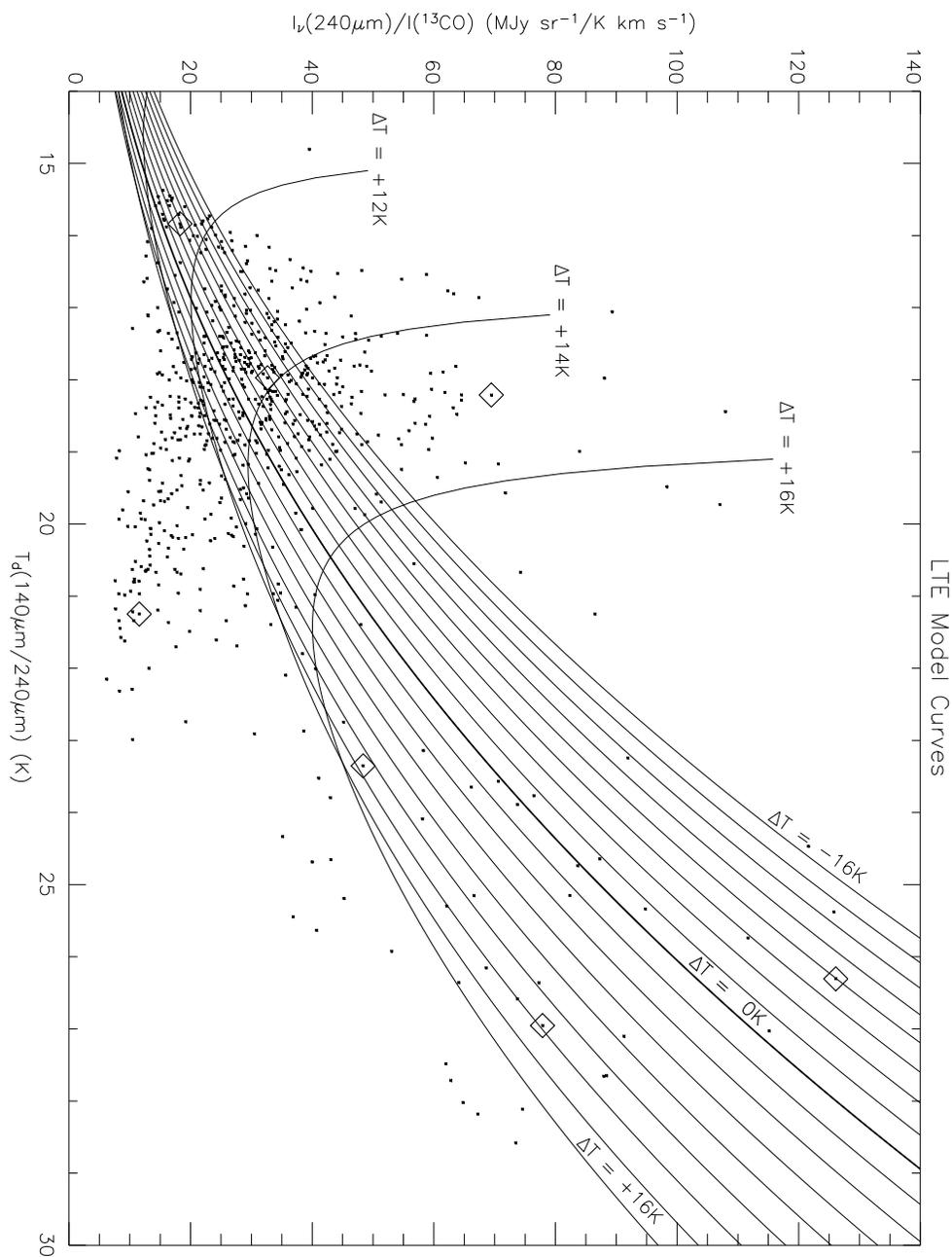}
\caption{The model curves of $\rd$ versus $\Td$ are shown for the LTE case.  
The plot includes the same data points as in Figure~\ref{fig6} with error 
bars excluded for clarity.  Each model curve is for a specific value of 
$\DT$ or $\Tk - \Td$.  The curves shown range from $\DT = -16\,K$ to 
$+16\unit K$ in steps of 2$\,$K.  The diamonds in this figure are the 
fiducial points discussed in Section~\ref{ssec35} and listed in 
Table~\ref{tbl-2x}. 
\label{fig7}}
\end{figure}

In this section the $\cOone$ line is assumed to be in LTE and optically thin.  
In practice, LTE means that the gas density is high enough that the line 
strength is not {\it explicitly\/} dependent on that density.  The critical 
density of this line, at which the downward collisional transition rate is 
equal to the spontaneous radiative transition rate, is $3\times 10^3\unit 
cm^{-3}$.  Hence, this line is ``close" to LTE for densities exceeding that 
critical density, or $10^4$--$10^5\unit cm^{-3}$ or higher.  Assuming LTE has 
the advantage that it is only necessary to fit one parameter, $\DT\equiv\Td - 
\Tk$. The model curve to be fit is  
\begin{equation}
\rd = 69.2 {Q(\Td-\DT)\over C_{_{BG}}}
\ {exp\biggl({5.28\over\Td-\DT}\biggr)
\over exp\biggl({59.9\over\Td}\biggr) - 1}\qquad ,
\label{mr3}
\end{equation}
where Q is the rotational partition function for $\cO$ and $\rm C_{_{BG}}$ 
is the correction for the cosmic background radiation.  The reader is 
referred to Appendices~\ref{appa} and \ref{appb} for details.  Figure~\ref{fig7} 
shows the model curves of equation~(\ref{mr3}) for $\DT = -16$ to $+16\unit K$.

\begin{figure}
\epsscale{0.77}
\plotone{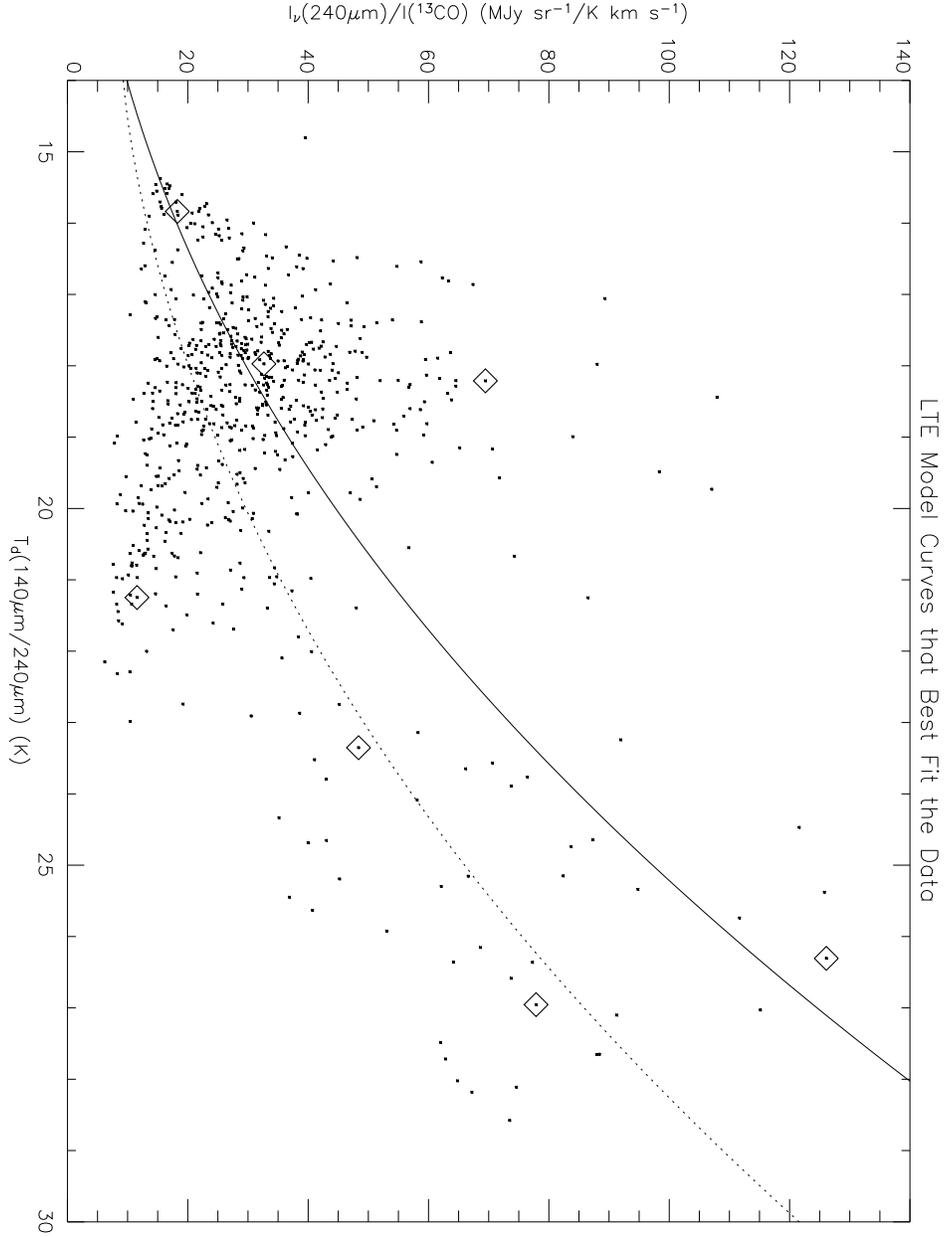}
\caption{The model curves of $\rd$ versus $\Td$ that best fit the data are 
shown for the LTE case. The plot includes the same data points as in 
Figure~\ref{fig6} with error bars excluded for clarity.   The solid curve 
represents $\DT = -4\,K$ and is the best fit to all the data points.  The
dotted curve represents $\DT = +9\,K$ and is the best fit to only the data 
points with $\Td>20\,K$.  The diamonds in this figure are the 
fiducial points discussed in Section~\ref{ssec35} and listed in 
Table~\ref{tbl-2x}.
\label{fig8}}
\end{figure}

   The optimum value found for $\DT$ for the LTE models, i.e. $\DT=-4\pm 1\,K$, 
is far from satisfactory: as is seen in Figure~\ref{fig8}, even though this 
curve goes through the center of the triangular cluster of points, it is 
systematically too high for the points with $\Td > 20\,K$.  Indeed, the 
chi-square per degree of freedom is $\chi_\nu^2=16.5$.  Fitting only to the 
points with $\Td > 20\,K$ gives the result $\DT=+9\pm 1\,K$ with $\chi_\nu^2=9.0$, 
but this curve is too low for the center of the triangular cluster of points.  In 
addition, neither the $\DT=-4\,K$ curve nor the $\DT=+9\,K$ curve match the 
high-$\rd$ or low-$\rd$ points in the triangular cluster. 
 
   The uncertainties specified for the above $\DT$ value are the formal 
uncertainties due to the random errors.  Normally, this uncertainty is determined 
by finding the values of $\DT$ for which the $\chi^2$ value is increased by 1.  
However, this is appropriate only if the chi-square per degree of freedom, 
$\chi_\nu^2$, is near 1 \citep[see][]{Press92}.  This is not the case here: the 
$\chi_\nu^2$ is significantly greater than 1.  Consequently, the uncertainty in 
$\DT$ was determined by finding the values of $\DT$ for which the $\chi^2$ was 
increased by the value $\chi_\nu^2$. This is a more conservative estimate of the 
error in the fitted quantity.  

   Larger than the random uncertainties are the systematic uncertainties;
these uncertainties include the calibration errors in the data and the 
uncertainties in the assumptions used in comparing data with theory. These
uncertainties are listed below:
\begin{enumerate}
\item The calibration of the observed $\rd$ or $\Ib/\Ic$ ratio.  The calibration 
uncertainty of $\Ib$ is 12\% \citep{Hauser98} and that of $\Ic$ is less
than 20\% \citep{Nagahama98}.  Combining the two uncertainties quadratically
would imply a calibration uncertainty of 24\% for $\rd$.  For simplicity, we
adopt a calibration uncertainty of 20\%. 
\item The uncertainty in the abundance of $\cO$, X($\cO$).  This affects the
relationship between the CO column density and the H$_2$ column density.  Again,
20\% is adopted as the contribution to the uncertainty in the model curve.
\item The uncertainty in the dust optical depth to total gas column density,
$\rm\tau_\nu/N(H)$.  This affects the relationship of $\Ib$ to N(H) (see
equation \ref{apb6}).  It is assumed that this contributes an uncertainty of 20\% 
to the model curve.
\end{enumerate}
For simplicity, all of the above have adopted uncertainties of 20\%, which is a
number that is roughly appropriate for the above cases.  Also, these uncertainties 
are independent of one another, so that they should be added quadratically.  
This gives an overall uncertainty of 30-40\% between the model curves and the data.

   Note that the systematic and random uncertainties are treated separately.  The
systematic uncertainties represent a simple scaling up or down of the axes in
plots of one quantity against another, while the overall distribution of points
in a plot does not change.  Hence, parameters determined from fits to the
data, like line slopes, are simply scaled up or down accordingly.
The random uncertainties, however, represent random reshapings of the distribution
of points in a plot and must be included in the model fitting procedure.  

\begin{figure}
\epsscale{0.7}
\plotone{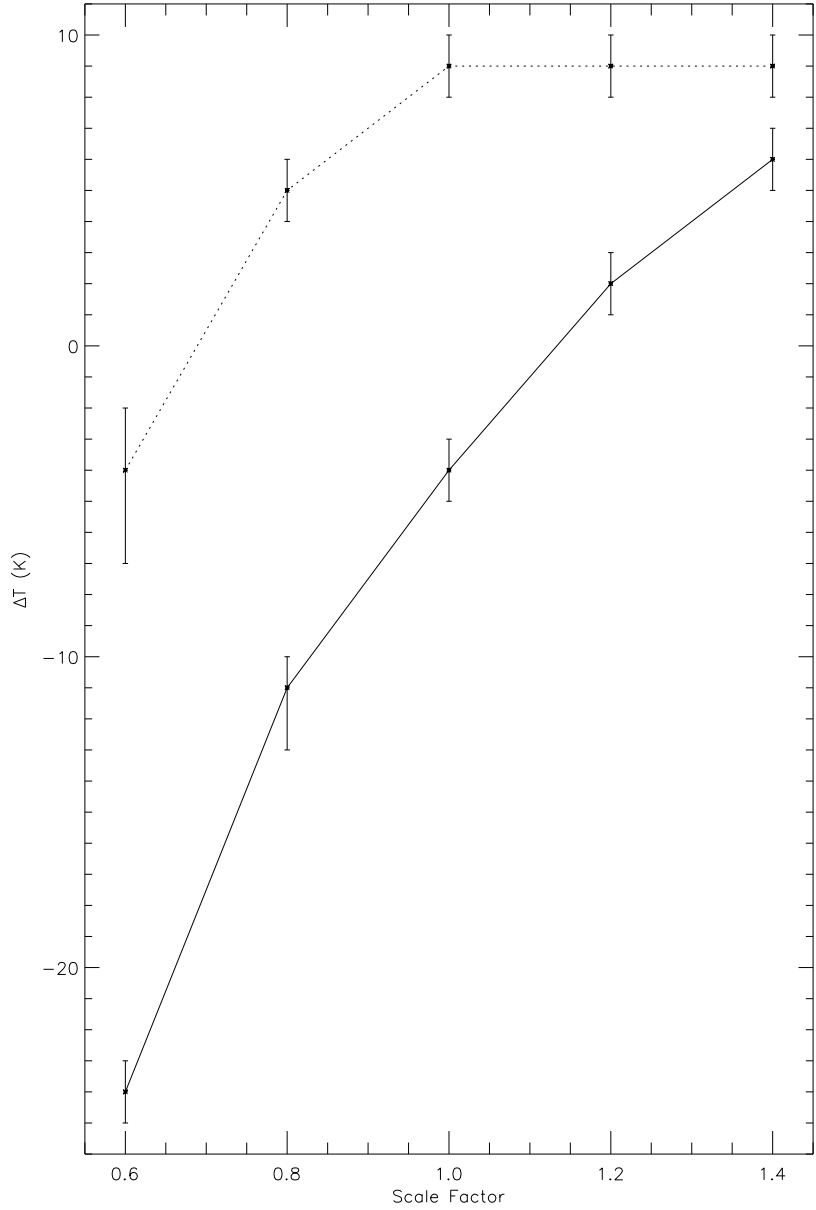}
\caption{The effect of the systematic uncertainties on the resultant $\DT$
from the fits of the LTE model curves is shown.  The effect of these 
uncertainties was tested by applying the scale factors 0.6, 0.8, 1.0, 1.2,
and 1.4 to the model curves and fitting the $\DT$ for each one.  The solid
line represents the resultant $\DT$ values for the fits to all the data
(i.e., all the data points shown in Figure~\ref{fig6}).  The dotted line
represents the resultant $\DT$ values for the fits to the data with $\Td > 
20\,K$.  The error bars represent the formal error bars for each model fit.
\label{fig9}}
\end{figure}

   To test the effect of the systematic uncertainties on the result, the model
fitting is repeated after applying scale factors to the model curves.  The
scale factors used in the new model fits deviate by as much as 40\% from unity 
and are in steps of 20\%: 0.6, 0.8, 1.0, 1.2, and 1.4.  As Figure~\ref{fig9}
clearly shows, the total variation in the resultant $\DT$ values due to the
systematic uncertainties is much larger than the error bars of the random or
formal uncertainties.  For the fits to all the high signal-to-noise positions,
the $\DT$ varies from $-24\unit K$ to $+6\unit K$.  For the fits to the high
signal-to-noise positions with $\Td > 20\,K$, the $\DT$ varies from $-4\unit K$ 
to $+9\unit K$.  Such a strong variation of $\DT$, a range of 30$\,$K for the 
fits to all the high signal-to-noise data, means that $\DT$ cannot be determined 
in this simple LTE case.  This is not surprising given that the fits 
are not very good.  Also, since only one parameter varies anyway, scaling the 
model curves varies that one parameter --- $\DT$ in this case. 

   Another test of the model fit is to compare the gas column density as
derived from the continuum observations, $\NHd$, to that derived from the 
$\cOone$ line, $\Ngth$.  The $\NHd$ values are determined from a trivial
rearrangement of equation (\ref{apb6}) in Appendix~\ref{appb} and the $\Ngth$
values come from 
\begin{equation}
\Ngth = N(H\,I) + 2\,\NHth \qquad ,
\label{mr3a}
\end{equation}
in analogy to equation~(\ref{apb3}).  The $\NHth$ comes from 
equation~(\ref{apa27r}).
The value of $\Tk$ used in (\ref{apa27r}) for each position is given by
the value of $\Td$ for that position minus the $\DT$ of the model fit.
As stated above, $\DT = -4\,K$, but $\DT = +9\,K$ was used for those 
positions with $\Td > 20\,K$.  A plot of $\Ngth$ versus $\NHd$ is shown
in Figure~\ref{fig10}.  The error bars represent the random uncertainties;
the details of which are discussed in Appendices~\ref{appc} and \ref{appd}.  
The systematic uncertainty in $\NHd$ is 30\%.  This includes the calibration 
uncertainty of 12\% for $\Ib$, the calibration uncertainty of 6\% for $\Ia/\Ib$, 
and the uncertainty due to the background subtraction of 10\%.  The systematic
uncertainty in $\Ngth$ is 20\% due to the calibration uncertainty in $\Ic$. 
The quadrature sum of these two uncertainties means that the systematic
uncertainty in the slope in the $\Ngth$ versus $\NHd$ plot is about 40\%. 
The slope of the plotted straight line and the y-intercept, as stated above, 
were determined from a routine that considers the uncertainties in both $x$ 
and $y$, similar to the orthogonal regression method discussed previously. 
The uncertainties in the slope and y-intercept were determined by finding
the changes in each that would raise the total $\chi^2$ by one, provided
that $\chi_\nu^2\simeq 1$.  Since $\chi_\nu^2$ was greater than 1, the
uncertainties in the slope and y-intercept were determined after scaling
up the errors in $x$ and $y$ by the same amount until $\chi_\nu^2\simeq 1$.
The slope of the plot, 1.06$\pm$0.03, suggests that the column densities 
derived from the $\cO$ line agree favorably with those determined from
the continuum.  However, the scatter is large: $\chi_\nu^2 = 16.6$.  This
is clearly a reflection of the bad fit of the single-component,
LTE model to the data.

    As a comparison, Figure~\ref{fig10} also includes a plot of the gas 
column density as derived from applying the usual conversion  factor to the 
integrated intensity of $\COone$, I($\CO$).  As stated previously, the 
conversion factor employed was $\rm N(H_2)/\int\Tr dv = 2.6\times 10^{20}
\cKkms$.  The slope in the fitted line is 1.40$\pm$0.04.  This
suggests that a more appropriate value for the conversion factor is
$1.9\times 10^{20}\cKkms$; this will discussed further in Section~\ref{ssec36}. 
This value for the X-factor is only a crude average for the Orion clouds, because 
there is large scatter about the straight-line fit: $\chi_\nu^2 = 28.3$.  
Despite this large scatter, the $\CO$-derived column densities are within a 
factor of 2 of those derived from the continuum observations for about 80\% of 
the points.  So even on sub-cloud scales, the X-factor seems to give column
densities that are correct to within a factor of 2 for most positions.  

    Comparison of the scatter in the upper and lower panels of Figure~\ref{fig10} 
may provide insights into the reliability of $\CO$-derived column densities.  
While the $\chi_\nu^2$ for the lower-panel straight-line fit 
with the $\CO$-derived column densities is nearly double that for the upper 
panel, the samples of points in the two 
panels are not the same.  The sample with the $\cO$-derived column densities 
includes only those positions with greater than 5-$\sigma$ in $\Ia$, $\Ib$, and 
$\Ic$.  This is 674 points.  The sample with the $\CO$-derived column densities 
includes those positions with greater than 5-$\sigma$ in $\Ia$, $\Ib$, and I($\CO$).  
This is 1053 points.  If the same 674-point sample is also used for the $\CO$-derived 
column densities, then $\chi_\nu^2 = 17.5$ for the straight-line fit in the bottom 
panel.  This is comparable to the $\chi_\nu^2 = 16.6$ found for the straight-line fit 
to the upper panel with the $\cO$-derived column densities.  At face value, the 
comparable $\chi_\nu^2$ values would suggest that the $\CO$-derived column densities 
are almost as reliable as those derived from $\cO$.  However, considering that the 
single-component LTE models used here do {\it not\/} properly characterize the data, 
this conclusion would be premature.  

    Another important point to consider is that of the area filling factors
of the gas.  The model fitting has the potential of determining the gas kinetic 
temperature.  Since the $\COone$ line is optically thick, this line could also
give the gas kinetic temperature, if we assume LTE.  The peak radiation 
temperature of this line, $\Tr(\CO)$, would give $\Tk$.  However, $\Tr(\CO)$-derived
$\Tk$ values differ from the model $\Tk$ values.  This discrepancy is
easily resolved by considering that the gas does not fill the beam within
each narrow velocity interval.  The radiation temperature would then be
given by
\begin{equation}
\rm\Tr(\nu) = \filf\,\,{\cal J}_\nu(\Tk)\,\,\left[1-exp(-\tau_\nu)\right]\qquad ,
\label{mr4}
\end{equation}
where
\begin{equation}
\rm{\cal J}_\nu(\Tk)\equiv {\strut h\nu\over
\strut k}\Biggl\{\,\Biggl[exp\Biggl({\strut h\nu\over
\strut k\Tk}\Biggr) - 1\Biggr]^{-1} - 
\ \Biggl[exp\Biggl({\strut h\nu\over\strut 
k\Tbg}\Biggr) - 1\Biggr]^{-1}\Biggr\}\quad .
\label{mr5}
\end{equation}
Expression (\ref{mr4}) assumes that, at each velocity within the line profile, 
the telescope beam has an area devoid of gas emission and an area with gas
emission.  The $\tau_\nu$ is the optical depth of the gas averaged in the 
appropriate way over the gas-emitting area at frequency $\nu$ within the line 
profile.  The $\filf$ is the area filling factor and is the fraction of the 
beam area that has gas emission at the frequency $\nu$ within the line profile.
Given that the $\COone$ line is optically thick, equation (\ref{mr4}) can be
written as
\begin{equation}
\rm\filf = {\Tr(\nu)\over {\cal J}_\nu(\Tk)}\qquad .
\label{mr6}
\end{equation}
Figure~\ref{fig11} shows the $\filf$ for the peak of the $\COone$ line
profile determined from (\ref{mr6}) plotted
against the continuum-derived column densities.  The gas kinetic temperatures
were computed in the same way as for the column density calculations.  
There are two important things to notice in this plot.
One is that, on average, positions with higher column densities have
higher filling factors.  This is expected: as the gas column
density increases, there is an increasing number of cloud substructures
or clumps along the line of sight.  On average, more and more of the 
beam area is filled with clumps.  Another thing to notice is that all
the area filling factors are less than unity, as expected for a geometric
filling factor.  

   In short, the single-component, LTE models are not very successful
in explaining the $\rd$ versus $\Td$ plot.  At best, these models
only crudely reproduce the overall trend in the data and allow only
a rough agreement between $\cO$-derived column densities and 
dust-derived column densities. 

\subsection{Non-LTE Models and Results\label{ssec32}}

In this section, non-LTE effects on the $\cOone$ line emission
are considered.  Also, the effects of non-negligible optical depths
are also considered.  To model the non-LTE line
emission, the Large Velocity Gradient (LVG) method was used 
\citep{Goldreich74}.  The advantage of this method is its simplicity.
Photodissociation region (PDR) models \citep[e.g., see][]{Mochizuki00,
Stoerzer00, Tielens85} are considerably more complicated and not  
necessary for the crude modeling done here. 
The LVG method assumes that every point in the gas is identical to every other 
point, with the exception of the velocities.  In the spherical geometry 
case, every point in the gas is surrounded by points moving away from it 
in a spherically symmetric pattern with more distant points moving away 
even more rapidly --- similar to a cosmological Hubble flow.  Hence, every 
point in the gas is radiatively coupled to, and at the center of, a 
spherical region.  Outside this region, the line emission of the gas is 
redshifted beyond that of the gas at the center of the sphere and, 
consequently, radiatively decoupled from it.  This spherical geometry can 
then estimate the line emission from 
a spherical clump.  This is only an estimate because the LVG method 
cannot accurately estimate the emission from the edges of a clump.  Also, 
the velocity pattern within the clump would not be spherically symmetric.  
Nevertheless, the geometry of the region that is radiatively coupled to each 
point within the clump may be roughly spherical.  Each clump is assumed 
to be radiatively decoupled from the other clumps; the other clumps 
that have radial velocities within the clump velocity width of a given clump 
are assumed to fill a small solid angle with respect to 4$\pi$ steradians.

    Once some basic parameters are chosen --- the molecule, cosmic
background temperature, geometry, and the maximum number of levels to be
included in the computation --- three parameters are specified: the kinetic 
temperature, gas density, and a third parameter.
This third parameter can be the abundance per unit velocity gradient,
$\Xdvdr$ for $\cO$, or the column density per velocity interval, $\NDv$. 
It is common to choose $\Xdvdr$ as the third parameter, but this has a
couple of problems.  One problem is that, by itself, $\Xdvdr$ has little
physical meaning.  It says, in effect, that a region of the gas with a
certain velocity width has a certain size (assuming fixed abundance);
this says very little about the excitation by itself; the density must
also be known.  On the other hand, $\NDv$ states that a region of gas
with a particular velocity width has a certain column density; this
provides strong clues to the level of radiative trapping.  For example,
in the range of parameter space described in the next paragraph, given the 
optical depth of the $\Jone$ line of CO for given $\Tk$ and $\NDv$ values, 
one can estimate the optical depth to within a factor of ten for 
other values of $\Tk$ and $\NDv$ about 80\% of the time {\it even without 
knowing the gas density\/} (provided that that unknown density is held 
constant).  Another problem with using $\Xdvdr$ is that the effects of 
collisions and radiative trapping are mixed together, making interpretation 
of observational data more difficult.  Specifically, the gas density, $\nH$, 
affects not only the collisions but also the column density.  Because 
$\NDv =\nH\,\Xdvdr$, raising $\nH$ for fixed $\Xdvdr$ also raises $\NDv$, 
thereby raising the line optical depth while also raising the collision 
rate.  This results in misleading impressions of the density sensitivity 
of line emission.  For example, it has been mistakenly found that the 
$\cOone/\COone$ intensity ratio is sensitive to $\nH$ \citep[e.g.,][]
{Sakamoto97,Tosaki02}.  This intensity ratio when plotted as a function 
of $\nH$ for fixed $\Xdvdr$ does show a strong variation with $\nH$, but 
this is because the rising $\nH$ also results in rising $\tau(\cOone)$ and, 
of course, the $\cOone/\COone$ intensity ratio {\it is\/} sensitive to 
$\tau(\cOone)$.  (In fact, $\tau(\cOone)$ is equivalent to this intensity 
ratio in the LTE, optically thin limit.)  Being unaware of this coupling 
between optical depth and density has led to errors in interpretation of the 
data.  In one case, the authors went so far as to conclude that the molecular 
gas in the galaxy M$\,$51 must have temperatures of several hundred kelvins 
on scales of hundreds of parsecs, based on a comparison of the $\Jone$ lines 
of HCN, $\cO$, and $\CO$ \citep{Tosaki02}; this despite the much simpler and 
much more likely interpretation of low-temperature gas with optically thin 
$\cOone$ emission.  Therefore we choose $\NDv$ to be the third parameter
used in the LVG models. 

The specific range of parameter space used in the 
modeling is as follows:
\begin{itemize}
\item $\Tk = 2.8$ to 40.7$\,$K in steps of 0.1$\,$K.
\item $log_{10}[\nH(cm^{-3})]$ = 1 to 5 in steps of 0.25.
\item $log_{10}\left[\rm{N(\cO)\over\Delta v}(\ckms)\right]$ = 15.5 to 18.3 
in steps of 0.2.
\end{itemize}
The reason for the chosen lower limit on $\NDv$ has to do with the observed
column densities of the Orion clouds.  The majority of positions, 93\%,
have $\rm N(H\,I + 2H_2) < 1\times 10^{22}\, cm^{-2}$.  Negligible
N(H$\,$I) and cloud line widths of 2$\kms$ mean
that this limit corresponds to $\rm N(H_2)/\Delta v = 2.5\times 10^{21}
\ckms$.  Using the abundances given in Appendix~\ref{appa}, this is equivalent
to $\NDv = 3.2\times 10^{15}\ckms$.  This column density per velocity
interval represents a large-scale cloud property, for some positions
in the Orion clouds, and must {\it not\/} be confused with the column
density per velocity interval within the individual substructures or
clumps.  The $\NDv$ values within the clumps can be higher than the
large-scale values for a cloud.  However, the values for $\NDv$ cannot be 
much smaller than the large-scale value without also having many clumps 
within a given velocity interval that are radiatively coupled.  Radiatively
coupled clumps means that the effective $\NDv$ is roughly
the sum of the $\NDv$ values for the individual clumps. 
Consequently, values much less than the large-scale cloud $\NDv$
value are not meaningful.  Therefore, a lower limit of $\NDv = 
3.2\times 10^{15}\ckms$ is adopted, which is not too low for 93\% of the
high signal-to-noise positions (i.e. those positions represented in 
Fig~\ref{fig6}) in the Orion clouds. 

   To develop the expression for $\rd$ that uses LVG model results, we
start with
\begin{equation}
\rm I_\nu = 2\, B_\nu(\Td)\,\kappa_\nu x_{_d}\mu_{_H} m_{_H} N(H_2) \qquad ,
\label{mr12}
\end{equation}
where $\rm I_\nu$ is the continuum intensity.  The above expression is
a combination of equations~(\ref{apb1}), (\ref{apb2}), and (\ref{apb3})
with $\rm N(H\,I)=0$.  The integrated strength of the $\cOone$ line can
be written as 
\begin{equation}
\Ic = \Tr(\cO)\,\Delta v \qquad ,
\label{mr13}
\end{equation}
which is exact if $\Delta v$ is defined appropriately.  The quotient of
(\ref{mr12}) and (\ref{mr13}) can be expressed as
\begin{equation}
\rm {I_\nu\over\Ic} = {2\, B_\nu(\Td)\,\kappa_\nu x_{_d}\mu_{_H} m_{_H}
\over X(\cO)}\ \left[{N(\cO)\over\Delta v}\right]_{\hbox{$_{mod}$}} 
[\Tr(\cO)]_{\hbox{$_{mod}$}}^{-1}\qquad .
\label{mr14}
\end{equation}
The subscript $mod$ means that the quantity is either an LVG model input, 
$[\NDv]_{\hbox{$_{mod}$}}$ or output, $[\Tr(\cO)]_{\hbox{$_{mod}$}}$. 
Using the values of the physical parameters from Appendices~\ref{appa} 
and \ref{appb}, equation~(\ref{mr14}) becomes
\begin{equation}
\rm {I_\nu(\MJsr)\over\Ic} = 0.440\,\lambda^{-5}(\mu m)
\biggl[exp\biggl({1.44\times 10^4\over\lambda(\mu m)\Td}\biggr) - 1\biggr]^{-1}
\left[{N(\cO)\over\Delta v}\right]_{\hbox{$_{mod}$}} 
\back\back\back [\Tr(\cO)]_{\hbox{$_{mod}$}}^{-1}\ .
\label{mr15}
\end{equation}
Setting $\lambda$ to 240$\um$ gives
\begin{equation}
\rm \rd = 5.53\times 10^{-13}\,\biggl[exp\biggl({59.9\over\Td}\biggr)
 - 1\biggr]^{-1}\left[{N(\cO)\over\Delta v}\right]_{\hbox{$_{mod}$}} 
[\Tr(\cO)]_{\hbox{$_{mod}$}}^{-1}\ .
\label{mr16}
\end{equation}
Even though $\DT$ and $\nH$ do not {\it explicitly\/} appear in (\ref{mr16}), 
the $[\Tr(\cO)]_{\hbox{$_{mod}$}}$ is an LVG model result, so 
it depends on $\Tk$($=\Td-\DT$), $\NDv$, and $\nH$.  
The effects of varying the parameters $\DT$, $\NDv$, and $\nH$ on the
model curves are shown in Figures~\ref{fig12}, \ref{fig13}, and \ref{fig14}.

   The best fitting curve is shown in Figure~\ref{fig15}.  The
solid curve is fit to all data at more than 5-$\sigma$ in
$\Ia$, $\Ib$, and $\Ic$.  The best fit parameters are $\DT = -1\pm 1\, K$,
$\NDv = (3.2\pm {1.8\atop ?})\times 10^{15}\ckms$, and $\nH = (1.0\pm 
{?\atop 0.4})\times 10^5\, cm^{-3}$.  Note that it is not possible to put a 
firm lower limit on $\NDv$ nor a firm upper limit on $\nH$.  In the former
case, $\NDv$ is low enough that $\cOone$ is close to the optically thin 
limit, and $\rd$ is insensitive to $\NDv$ in this limit.  In the latter 
case, $\nH$ high enough that $\cOone$ is close to LTE and $\rd$ is 
insensitive to $\nH$ in that limit.  The chi-square per degree of freedom,
$\chi_\nu^2$ is 16.9, similar to that for the LTE model fit.  The dashed
curve is fit to the same 5-$\sigma$ data, but only for $\Td
\geq 20\,K$.  The parameters for this curve are $\DT = -3\pm 4\, K$,
$\NDv = (3.2\pm {1.8\atop ?})\times 10^{15}\ckms$, and $\nH = (5.6\pm 
{4.4\atop 2.5})\times 10^3\, cm^{-3}$.  The goodness of fit is $\chi_\nu^2
=9.98$.  Again, as for the LTE case, the LVG one-component
model clearly does not well characterize the data. 

   The poorness of the fit is also illustrated in Figure~\ref{fig16}, which
shows plots of the parameter variations with the scale factors.  
The total range in $\DT$ is 13 or 
14$\,$K.  $\NDv$ varies by a factor of 7 for the solid curve and does not
vary for the dashed curve.  $\nH$ varies by nearly two orders of magnitude.
If we ignore the cases with limits instead of error bars, then
we see again that the systematic variations are larger than the random
errors (except for the dashed curve showing the $\NDv$ variation).  The
particularly large uncertainty in $\nH$ is understandable because the $\cOone$ 
line loses its density sensitivity for $\nH\gsim$$few\times 10^3\unit cm^{-3}$.
In short, as in the LTE case, it is difficult to determine the values of the
parameters with certainty. 

   Again the continuum-derived gas column density, 
$\NHd$, is compared with that derived from the $\cOone$ line, $\Ngth$.  The 
$\NHd$ values are determined as in the LTE case (see Section~\ref{ssec31}). The 
$\Ngth$ values come from equation~(\ref{mr3a}), where $\NHth$ is derived from
\begin{equation}
\NHth = {\left[\rm N(\cO)/\Delta v\right]_{\hbox{$_{mod}$}}
\over\rm X(\cO)\ \left[\Tr(\cOone)\right]_{\hbox{$_{mod}$}}}\ \Ic \qquad ,
\label{mr17}
\end{equation}
where the subscript $mod$ refers to LVG model quantities (i.e. model input or 
output parameters), $\rm X(\cO)$ is the value 
adopted in Appendix~\ref{appa}, and $\Ic$ is the observed intensity.  Each 
point on the model curve has particular values of 
$\left[\rm N(\cO)/\Delta v\right]_{\hbox{$_{mod}$}}$ and 
$\left[\Tr(\cOone)\right]_{\hbox{$_{mod}$}}$ associated with it.  Each data point was 
then associated with a point on the model curve.
This was done in the same way that the $\chi^2$ values were computed: for
each data point, the minimum of the distances to all the points on the model 
curve was the minimum of all values given by 
equation~(\ref{mr2}). (Also see discussion of this in Appendix~\ref{appe}.)  Note
that the model curve used is the solid curve in Figure~\ref{fig15} for 
points with $\Td<20\,K$ and is the dashed curve in Figure~\ref{fig15} for
points with $\Td\geq 20\,K$.  The random errors associated with the $\NHth$
column densities were computed as described in Appendix~\ref{appe}.  The plot of
$\Ngth$ versus $\NHd$ is shown in Figure~\ref{fig17}.  The straight-line   
fit to the points was done in same way as for Figure~\ref{fig10} in the
LTE case.  The slope of the line, $1.08\pm 0.02$, suggests that there
is good overall agreement between the continuum-derived and spectral 
line-derived column densities, except that the scatter is large: 
$\chi_\nu^2 = 15.5$.  This is only marginally smaller than the 
$\chi_\nu^2 = 16.6$ obtained for the LTE case.  Hence, the improvement
from LTE, one-component models to LVG, one-component models is
minor. 

   The filling factors as a function of spectral line derived column
densities in the LVG, one-component case are plotted in Figure~\ref{fig18}.
The filling factors were derived using equation~(\ref{mr6}).  The $\Tk$
used in that equation was determined from the observed $\Td$ at each
position minus the appropriate $\DT$ from the model fits.  (Even though
these are non-LTE models, the $\COone$ line is still essentially thermalized 
at these densities --- i.e., $\nH = 
1\times 10^5$ and $5.6\times 10^3\unit cm^{-3}$ --- due to its high opacity.
Consequently, the excitation temperature {\it of this transition\/} is
close to $\Tk$ and simply using $\Tk$ in equation~\ref{mr6} is appropriate.)
As in the LTE case, the filling factors, on average, 
increase with increasing column density.  Again there is a saturation effect: 
the slope of the plot decreases with increasing column density. 
   
   In short, the single-component, LVG models do not successfully
explain the $\rd$ versus $\Td$ plot and are, at best, only marginally
better than the single-component, LTE models.

\subsection{Two-Component, Non-LTE Models and Results\label{ssec33}}

As discussed in the previous sections, single-component models only 
poorly characterize the data.   In this section we assume that there
are two emission components.  ``Component~0" is present on all lines
of sight and has {\it constant\/} properties on all those lines of sight.
Specifically, the dust temperature, gas kinetic temperature, column density
per velocity interval, and gas density are all constant.  ``Component 1" 
is similar except that the dust and gas temperatures can vary from line
of sight to line of sight while maintaining a constant dust$-$gas temperature
difference, $\DT$, as stated in the basic assumption.  The physical motivation
for these components is straightforward.  Component~0 represents 
molecular gas at the surface of the Orion clouds and is heated only by the 
general interstellar radiation field (ISRF), giving it a constant temperature.
Component~1 is the molecular gas deeper into the Orion clouds and is largely 
shielded from the general ISRF by component~0.  Therefore, those positions 
in component~1 without stars will have colder dust temperatures than in
component~0.  In contrast, those positions in component~1 with stars 
will have warmer dust than in component~0.  In this simple picture, component~0
has a constant temperature and component~1 has a varying temperature from
position to position, varying above and below the dust temperature of 
component~0. 

We now need expressions relating the observed emission to the physical 
parameters of the gas and dust in order to construct the two-component models.  
The expressions derived here will be necessary for the simulations described in 
Paper~II.  Construction of the models requires certain simplifications.  For 
example, it is assumed that component~0 has the same $\DT$ as component~1.   Also, 
as discussed in Section~\ref{ssec32}, the clumps are assumed to be radiatively 
de-coupled from one another.  Specifically, for these two-component models, it is 
assumed that the clumps of component~0 are de-coupled from those of component~1.  
Therefore the emission of the two components can simply be added together.  
Consequently, the specific intensity of the continuum emission averaged over the 
beam, $\rm I_\nu$, is given by
\begin{equation}
\rm I_\nu = g_1 N_{c1} f_\nu(\Tdo)\ +\ g_0 N_{c0} f_\nu(\Tdz)\qquad ,
\label{mr18}
\end{equation}
where $\rm f_\nu(\Td)$ is defined as
\begin{equation}
\rm f_\nu(\Td)\equiv \kappa_{\nu 0} x_{_d}\mu_{_H} m_{_H}
\biggl({\nu\over\nu_0}\biggr)^\beta B_\nu(\Td)\qquad ,
\label{mr19}
\end{equation}
combining equations (\ref{apb1}), (\ref{apb2}), (\ref{apb5}) and removing
the column density factor. $\rm N_{c1}$ is the column density of hydrogen
nuclei (i.e. $\rm 2N(H_2)$ for purely molecular hydrogen) of each clump of 
component~1.  $\rm g_1$ is the number of clumps of component~1 on each
line of sight averaged over the beam area.  $\rm N_{c0}$ and $\rm g_0$ are
the corresponding quantities for component~0.  The total number of clumps
per line of sight averaged over the beam from both components, g, is given
by 
\begin{equation}
\rm g = {N\over a_1 N_{c1}\ +\ a_0 N_{c0}}\qquad ,
\label{mr20}
\end{equation}
where N is the observed column density (i.e. the column density averaged
over the beam) and $\rm a_1$, $\rm a_0$ are the fractions of clumps belonging
to components 1 and 0, respectively.  Obviously, then 
\begin{equation}
\rm a_0 + a_1 = 1\qquad .
\label{mr21}
\end{equation}
It follows then that
\begin{equation}
\rm g_1 = a_1 g = {N\over N_{c1}\ +\ c_0 N_{c0}}\qquad
\label{mr22}
\end{equation}
and
\begin{equation}
\rm g_0 = a_0 g = {c_0 N\over N_{c1}\ +\ c_0 N_{c0}}\qquad ,
\label{mr23}
\end{equation}
where 
\begin{equation}
\rm c_0 \equiv {a_0\over a_1}\qquad .
\label{mr24}
\end{equation}
Notice that $\rm g = g_1 + g_0$ follows naturally from (\ref{mr20}), 
(\ref{mr21}), (\ref{mr22}), and (\ref{mr23}). Equation~(\ref{mr18})
can now be written as
\begin{equation}
\rm I_\nu =  {N\over N_{c1}\ +\ c_0 N_{c0}}\, N_{c1} f_\nu(\Tdo)\ +
\ {N\over N_{c1}\ +\ c_0 N_{c0}}\, c_0 N_{c0} f_\nu(\Tdz)\qquad .
\label{mr25}
\end{equation}
To compare the continuum emission strength with that of a spectral line,
the cloud's velocity structure must be considered.  Even
though the dust grain velocities have no appreciable effect on
the continuum emission, we can imagine examining the emission
of only those dust grains associated with the $\cO$ molecules emitting
in a certain part of the $\cOone$ line.  As such, the right side of
(\ref{mr25}) is multiplied by ${\Dvc\over\Dv}{\Dv\over\Dvc}$, where
$\Dv$ is the observed line width and $\Dvc$ is the clump line width.
For simplicity it is assumed that {\it all\/} clumps have the same
velocity width.  Consequently, (\ref{mr25}) can be written as
\begin{equation}
\rm I_\nu =  c_1\left[\nvco f_\nu(\Tdo)\ +\ c_0 
\nvcz f_\nu(\Tdz)\right]\Dv\qquad ,
\label{mr26}
\end{equation}
where
\begin{equation}
\rm c_1\equiv  {(N/\Dv)\over \nvco\ +\ c_0 \nvcz}
\qquad .
\label{mr27}
\end{equation}
Analogous to $\rm g_1$, $\rm c_1$ is the number
of clumps of component~1 per line of sight averaged over the beam
{\it within a velocity interval equal to a clump velocity width.\/}
The parameter $\rm c_0$ (which is {\it not\/} analogous to $\rm g_0$)
is the ratio of the number of clumps of component~0 to those of 
component~1.  (Note that clump velocity widths of the two components
are assumed to be identical.  Generalizing to the case of different
clump velocity widths of the two components is trivial.)  Analogous to 
the continuum emission equation (i.e. equation~\ref{mr26}), the equation 
for the integrated radiation temperature of the $\cOone$ line is
\begin{equation}
\rm \Ic =  c_1\left[\Trone(\Tkone,\nvco,n_{c1})\ +\ c_0 
\Trz(\Tkz,\nvcz,n_{c0})\right]\Dv\qquad ,
\label{mr28}
\end{equation}
where $\Trone$ and $\Trz$ are the component-1 and component-0 radiation 
temperatures, respectively, $\Tkone$ and $\Tkz$ are the gas kinetic temperatures
in components 1 and 0, respectively, and $\rm n_{c1}$ and $\rm 
n_{c0}$ are the molecular gas densities in the component-1 and component-0 clumps, 
respectively.  Note that $\nvco$
is $\nvhco$ and $\nvcz$ is $\nvhcz$.  The LVG results, $\Trone$ and
$\Trz$, depend directly on $\nvtco$ and $\nvtcz$, respectively.  Consequently,
$\Trone$ and $\Trz$ as written in equation (\ref{mr28}) also implicitly depend
on the $\cO$ to $\rm H_2$ abundance ratio, $X(\cO)$. 
Dividing (\ref{mr26}) by (\ref{mr28}) gives the
desired $\rd$:
\begin{equation}
\rm\rd =  {\nvco f_\nu(\Tdo)\ +\ c_0\nvcz f_\nu(\Tdz)\over
\Trone(\Tdo-\DT,\nvco,n_{c1})\ +\ c_0\Trz(\Tdz-\DT,\nvcz,n_{c0})}\qquad ,
\label{mr29}
\end{equation}
where the following equations were used:
\begin{eqnarray}
\Tkone &=&\Tdo - \DT
\label{mr30}\\
\noalign{\noindent and}
\Tkz &=&\Tdz - \DT\qquad .
\label{mr31}
\end{eqnarray}
Again, the same $\DT$ is assumed for both components.  

Equation (\ref{mr29}) is used to determine the $\rd$ ratio for the two-component,
non-LTE case.  The $\Trone$ and $\Trz$ values are determined from the LVG method.  
These models have seven input parameters: $\DT$, $\rm c_0$, $\Tdz$, $\nvcz$,
$\rm n_{c0}$, $\nvco$, and $\rm n_{c1}$.  The component-1 dust temperature,
$\Tdo$, varies, resulting in model curves of $\rd$ versus dust temperature.
The dust temperature used in the plots of the model curves is the 140$\um$/240$\um$
color temperature, $\Tdc$.  To solve for $\Tdc$, the expression (\ref{mr26}) 
is evaluated at $\nu_9$ corresponding to $\lambda_9=140\um$ and at $\nu_{10}$
corresponding to $\lambda_{10}=240\um$ and these are divided to yield:
\begin{equation}
\rm {\cal R} =  {\nvco\, f_{\nu9}(\Tdo) K_{\nu9}(\Tdo,\beta)\ +\ c_0\,\nvcz\, f_{\nu9}(\Tdz)
K_{\nu9}(\Tdz,\beta)\over\nvco\, f_{\nu10}(\Tdo) K_{\nu10}(\Tdo,\beta)\ +\ c_0\,\nvcz 
\, f_{\nu10}(\Tdz) K_{\nu10}(\Tdz,\beta)}\qquad ,
\label{mr32}
\end{equation}
which uses the same notation as expression (\ref{dp04}) in Section~\ref{ssec21}.  The
color corrections, $\rm K_\nu(\Td,\beta)$, are used in order to correctly assess
the relative contributions of the two components to {\it DIRBE's\/} bands 9 and 10. 
Equation (\ref{dp04}), with $\Tdc$ in place of $\Td$, is then solved for $\Tdc$.
This allows a valid comparison of the model $\Tdc$ values with the {\it DIRBE\/} observed
$\Tdc$ values.

The master grid of $\Tk$ and $\nH$ values is the same as that for the one-component, LVG 
models.  The master grid values of $log_{10}\left[\rm{N(\cO)\over\Delta v}(\ckms)\right]$
is extended down from 15.5 to 14.3.  This does not necessarily violate the constraint
that the minimum column density per velocity interval be about that of the cloud on the
scale of a beam, because we are now dealing with two components.  Having the combined
column density per velocity interval of the two components above this minimum is still
required. 

   The best-fit curve is shown in Figure~\ref{fig26} and
is fitted to all the data at more than 5-$\sigma$ in $\Ia$, $\Ib$, 
and $\Ic$, and the parameters are listed in Table~\ref{tbl-1}.
The curve is characterized by
the triangular loop that exists between $\Tdc\simeq 15$ and $\simeq19\,$K and 
$\rd\simeq 7$ and $\simeq 40\MJkk$.  This loop is a consequence of the two
components emitting in the Wien limit.    The formal 
uncertainties are not listed because they are extremely small and not really 
meaningful when compared with the systematic uncertainties.  The formal uncertainties 
for $\DT$ and $\Tdz$ are less than $3\times 10^{-4}\unit K$ and for the other 
parameters the formal {\it relative\/} uncertainties are less than $3\times 
10^{-4}$.  Unlike the one-component model curve, the two-component
model curve is a reasonably good fit to the data, with $\chi_\nu^2 = 5.7$.  The 
F-test states that the two-component model is better than the
one-component models at the 99.9\% confidence level.  (Note that the number
of degrees of freedom adopted for estimating this confidence level was
about 1/9 the number in Table~\ref{tbl-1}, because the points in
Figure~\ref{fig26}, which represent individual map pixels, are 
{\it not\/} independent.  Given that the number of pixels per effective beam is 
about 9, the effective number of degrees of freedom is reduced accordingly.  
This reduction results in a {\it lower} confidence level than  
would have occurred without this reduction.)
Figure~\ref{fig26} explains the overall
trend in the data and {\it even explains a large part of the triangular cluster of
points\/} (see Section~\ref{sec3}).  Nevertheless,
this fit still has certain problems: the curve does not follow the curvature
in the data for $\Tdc\gsim 20\,K$ and it does not account for the low-$\rd$
points in the triangular cluster where $\Tdc > 18\,K$.  While the 
best-fit curve does not explain these low-$\rd$ points, 
some combination of parameter
values could account for these points:  if the component-0 parameters were 
allowed to vary from position to position (i.e., ignoring the basic assumption),
for example.  Also,
allowing $\Tdz$ to vary between about 18$\,$K and 20$\,$K could fit these points.
Nevertheless, given the simplicity of these two-component models, the current fit
is reasonable.  In addition, as mentioned in Section~\ref{ssec31}, these points could also
be explained by uncertainties in the observations.  

   Even though the curve in Figure~\ref{fig26} is an adequate fit, the parameters 
in Table~\ref{tbl-1} still have appreciable uncertainty. 
Figure~\ref{fig27} shows the systematic effects on the resultant parameter values
when the scale factor is changed.  Parameters like $\DT$ and $\Tdz$ are known
extremely well, $\pm 1\unit K$ or better, while parameters like $\nvtcz$ and 
$n_{c0}$ are uncertain by 2 or 3 orders of magnitude.  Still others like $c_0$,
$\nvtco$, and $n_{c1}$ are apparently known to within about an order of magnitude 
or better.  In the case of the one-component models, ascertaining
{\it any\/} of the parameter values was difficult.  In the case of the two-component
models, again {\it some\/} parameter values are very uncertain, but some are
apparently very well known indeed.

   The systematic uncertainties in the resultant parameter values were also
tested by varying the starting search grid.  There are enough parameters in
the two-component models that a search grid with {\it both\/} sufficient range 
in all parameters {\it and\/} sufficient resolution (i.e. grid spacing) in all 
parameters was not possible.  Therefore, the starting
search grid had a large range in all parameters and poor resolution (i.e. large
grid spacing).  After the initial search, a new grid with better resolution 
and smaller range moved closer to the final solution.  This was
repeated a few times to home in on the final solution.   In total there were 3 
to 5 search grids (including the starting grid) for each final solution.  Of 
course, one wonders about the final solution's stability: 
does the starting search grid affect the final answer?  In 
the ideal case the answer is no.  Nevertheless, this was tested by adjusting 
the starting search grid's position to see the effect on the final answer.
In total, four different starting search grids were tested and the results
were examined.  The results did indeed change depending on
the starting grid.  Nevertheless, the ranges in the results' variations had nearly
identical sizes (in a linear sense for $\DT$ and $\Tdz$ and in a
logarithmic sense for the other parameters) to those found from varying the 
scale factors. 

   As in the previous sections, the gas column density as derived from the 
continuum observations, $\NHd$, is compared with that derived from the $\cOone$ 
line, $\Ngth$.  The $\NHth$ values are derived from a modified version of
equation~(\ref{mr17}) that was used for the one-component, LVG model:  
\begin{equation}
\NHth = {\left[\nvtco\right]_{\hbox{$_{mod}$}} + c_0 
\left[\nvtcz\right]_{\hbox{$_{mod}$}}\over\rm X(\cO)\ 
\left\{\left[\Trone(\cOone)\right]_{\hbox{$_{mod}$}} + c_0 
\left[\Trz(\cOone)\right]_{\hbox{$_{mod}$}}\right\}}\ \Ic \ .
\label{mr33}
\end{equation}
Using this equation is similar to using 
equation~(\ref{mr17}) for the one-component, LVG models.  As for those
one-component models, the random errors associated with the $\NHth$
column densities were computed as described in Appendix~\ref{appe}.  The $\NHd$
values must now be computed and require corrections from the one-component
cases of the previous sections.  This correction is
\begin{equation}
{\NHd\over\NHdo} = \rm {N_{2M}\over N_{1M}}\qquad ,
\label{mr34}
\end{equation}
where $\NHd$ is the ``true" dust-derived column density inferred 
from the observations in the two-component case, $\NHdo$ is the dust-derived
column density that {\it would be\/} inferred from the observations assuming
one component only, $\rm N_{2M}$ is the model column density for a 
two-component model, and $\rm N_{1M}$ is the model column density that would
result if a single component were assumed.  Specifically, this single component
would have the color temperature, $\Tdc$, and the specific intensity
of the two-component model, as given by (\ref{mr26}).  So the right side of 
(\ref{mr34}) is the correction factor that would be applied to the model
column density to obtain the ``true" model column density.  
Equation~(\ref{mr34}) then applies this correction factor to the actual
observations.  If all the column densities in (\ref{mr34}) are beam-averaged
column densities, then
\begin{equation}
\rm N_{2M} = g_1(N_{c1} + c_0 N_{c0})
\label{mr35}
\end{equation}
and
\begin{equation}
\rm N_{1M} = g_1 N_c\qquad .
\label{mr36}
\end{equation}
Equation~(\ref{mr35}) is combination of equations (\ref{mr20}), (\ref{mr22}),
(\ref{mr23}), and (\ref{mr24}).  The $\rm N_c$ in equation~(\ref{mr36}) is
the effective column density of a single clump for the one-component 
``version" of the two-component model.  It is given by the expression,
\begin{equation}
\rm I_{\nu_{10}} =  c_1{N_c\over\Dvc} f_{\nu_{10}}(\Tdc)\Dv\qquad ,
\label{mr37}
\end{equation}
which is the one-component analog of expression~(\ref{mr26}) that uses
$\nu_{10}$ in place of $\nu$.  Equating (\ref{mr37}) with (\ref{mr26})
and rearranging yields
\begin{equation}
\rm {N_c\over\Dvc} = {\nvco f_{\nu_{10}}(\Tdo) + c_0\nvcz f_{\nu_{10}}(\Tdz)
\over f_{\nu_{10}}(\Tdc)}\qquad .
\label{mr38}
\end{equation}
Expression~(\ref{mr38}) is substituted into (\ref{mr36}), and (\ref{mr36})
and (\ref{mr35}) are substituted into (\ref{mr34}).  After this, a factor
of X($\cO$) is removed from both numerator and denominator to yield
\begin{equation}
\NHd = {\left\{\left[\nvtco\right]_{\hbox{$_{mod}$}} + 
c_0 \left[\nvtcz\right]_{\hbox{$_{mod}$}}\right\}f_{\nu_{10}}(\Tdc)\over
\left[\nvtco\right]_{\hbox{$_{mod}$}}\back f_{\nu_{10}}(\Tdo) + 
c_0\left[\nvtcz\right]_{\hbox{$_{mod}$}}\back f_{\nu_{10}}(\Tdz)}\ \NHdo
\label{mr39}
\end{equation}
As done in the previous section for the one-component, LVG models,
each data point was associated with the point on the model curve closest to it.  
The parameter values associated with that point on 
the model curve were then used in (\ref{mr33}) and (\ref{mr39}) to give
the column densities $\NHth$, and therefore $\Ngth$, and $\NHd$ corresponding
to that data point.  The uncertainties in $\NHd$ are discussed in Appendix~\ref{appf}.

Figure~\ref{fig28} is the plot of the $\cO$-derived column densities
versus the dust-derived gas column densities.  Clearly the agreement is
much better than for the one-component models: $\chi_\nu^2=7.4$.  The
slope of the straight-line fit, $1.08\pm 0.01$, is within the error of 
the corresponding slope for the one-component, LVG models.  {\it It is clear
that the two-component models are significantly better than the one-component
models.\/} 

   The filling factors as a function of spectral line derived column
densities in the LVG, two-component case are plotted in Figure~\ref{fig29}.
The filling factors were derived using a two-component analog of 
equation~(\ref{mr6}).  Specifically, an effective excitation temperature,
${\cal J}_\nu(eff)$, is determined from 
\begin{equation}
{\cal J}_\nu(eff) = {{\cal J}_\nu({\rm T_{x1}})\ +\ c_0 
{\cal J}_\nu({\rm T_{x0}})\over 1\ +\ c_0}\qquad ,
\label{mr40}
\end{equation}
where $\rm T_{x1}$ and $\rm T_{x0}$ are the excitation temperatures of the
$\COone$ line in components one and zero, respectively.  This effective 
excitation temperature is the brightness temperature that would be observed 
at frequency $\nu$ if the gas filled the beam at the velocity in the line 
profile corresponding to frequency $\nu$.  (Note that this assumes that the 
line is optically thick.  Note also that ${\cal J}_\nu(eff)$ includes the 
Rayleigh-Jeans correction and the correction for the background.)  The
excitation temperatures, $\rm T_{x1}$ and $\rm T_{x0}$, were computed 
with the following method.  Each observed position has an associated point on 
the model curve.  The point on the model curve has associated $\Tdo$, $\Tdz$,
and $\DT$ values.  These allow computation of the $\Tkone$ and $\Tkz$ values. 
The point on the model curve is also associated with the other model parameters,
which permit determination of the $\rm T_{x1}$ and $\rm T_{x0}$ values from 
LVG models of the $\CO$ line strengths.  In particular, the $\Jone$ line of
$\CO$ is optically thick and, as such, radiative trapping
keeps the line closer to LTE.  Therefore, the conversion from kinetic temperature
to excitation temperature is close to unity.  Specifically, the following 
approximations are accurate to within a few percent:
\begin{mathletters}
\begin{eqnarray}
\rm T_{x0} &\simeq& \rm\Tkz
\label{mr41a}\\
\noalign{\noindent and}
\rm T_{x1} &\simeq& \rm\Tkone\left[1-0.08\left({\Tkone\over 30}\right)^2\right]
\qquad .
\label{mr41b}
\end{eqnarray}
\end{mathletters}
Over the observed temperature range, the conversion of $\Tkone$ to $\rm T_{x1}$
deviates from unity by no more than about an 8\%.  The filling factor is then 
given by
\begin{equation}
\rm\filf = {\Tr(\nu)\over {\cal J}_\nu(eff)}\qquad .
\label{mr42}
\end{equation}
The value of $\filf$ at the line peak is plotted in
Figure~\ref{fig29}.  As in the one-component cases, the filling 
factors, on average, increase with increasing column density.  Again there is 
saturation with increasing column density.  However, for these models,
the area-filling factors can have non-physical values (i.e. $>1$).  In fact, some
observed positions {\it seem\/} to have $\filf$ up to about 6.  Since these
are area-filling factors, they cannot be greater than unity.  Even if we were to
reinterpret $\filf$ as the number of clumps per line of sight within a clump
velocity width, there would still be a problem:  the effective column density per 
velocity interval would be higher than, and therefore inconsistent with, the values 
of $\nvtco$ and $\nvtcz$ determined from the model fits.  Only for $\filf\leq 1$ is 
there no such problem, because some lines of sight within the beam, and within a clump 
velocity interval, can be free of clumps while others can have a single clump (with the 
properties given by the model fit).  When $\filf>1$, lines of sight that
have clumps will, on average, have more than one clump in a clump velocity
interval, thereby forcing the effective $\NDv$ to be higher than the fitted 
$\nvtco$ and $\nvtcz$ values.  In addition, given that this
is a filling factor within a clump velocity interval and that $\COone$ is optically
thick, detecting more than one clump per line of sight within a narrow velocity 
interval is not possible, because the clump in front would block the emission of the 
ones behind.  

   Nevertheless, the problem of the unphysical $\filf$ values has a solution.  
Both the numerator and
denominator of (\ref{mr29}) depend, explicitly or implicitly, on the product 
$c_0\nvcz$, and {\it not\/} on
the individual $c_0$ and $\nvcz$ values alone, over a large
range parameter space.  Accordingly, over that range of parameter space,
$\rd$ depends on that product and determining the parameters $c_0$ and $\nvcz$ 
separately is very difficult.  Indeed, Figure~\ref{fig27} shows
that $c_0$ and $\nvcz$ are roughly anticorrelated.  The
plot of their product shows that it is constant to within a
factor of roughly 3 about some geometric mean.  Consequently, restricting $c_0$ to 
some desired range can still yield physically reasonable
fits.  This will be examined in the next section.  
    
   In summary, the two-component, LVG models are reasonably good in explaining 
the $\rd$ versus $\Tdc$ plot, which is in stark contrast to the one-component models.
Examination of the systematic uncertainties suggests that $\DT$ and $\Tdz$
are known to within 1$\,$K or better and $n_{c1}$ is known to within a
factor of 2 about some geometric mean. Other parameters like $c_0$ and $\nvco$ 
are known to within factors of 4 or 5 about a geometric mean, while the 
remaining component-0 parameters, $\nvcz$ and $n_{c0}$ can vary by factors
of $\sim 30$ about the geometric mean.  While a different class of two-component 
models could possibly explain the entire triangular cluster of points in the $\rd$ 
versus $\Tdc$ plot, the two-component models described here only explain the triangular 
cluster for $\Tdc\lsim 18\,$K.  The points in the triangular cluster with $\Tdc\gsim 
19\,$K are not accounted for with the current models.  Another problem with the 
current models is that they predict area-filling factors that are unphysical for 
many of the observed positions.  And this problem will be addressed in the following 
section.

\subsection{Two-Component, Two-Subsample, Non-LTE Models and Results\label{ssec34}}

As mentioned in the previous section, the two-component models suffer from a 
non-trivial physical drawback: the area-filling factors for some positions are 
greater than unity.  Another problem is that the two-component models do not
fit the $\Tdc\geq 20\,$K points well.  As discussed in that section, keeping $c_0$ 
large enough will ensure that the filling factors have more reasonable values.  
Therefore, in this section we will consider models where $c_0$ is restricted to be 
$\geq 1$.  To better fit $\Tdc\geq 20\,$K points, we divide the points 
into two subsamples: one subsample with $\Tdc<20\,$K and the other with $\Tdc\geq 
20\,$K.  We fit the two-component, LVG models with $c_0\geq 1$ {\it only\/} to the 
$\Tdc<20\,$K subsample.  The $\Tdc\geq 20\,$K subsample has a relatively simply 
trend and we need only fit the one-component models that were used in 
Section~\ref{ssec32}. Notice that we are relaxing the basic assumption somewhat: 
we are allowing parameters, in addition to $\Td$ and $\Tk$, to change from one 
subsample to the other subsample.  However, within each subsample we still 
impose the basic assumption.  The parameter values resulting from the fits to 
the two subsamples are listed in Table~\ref{tbl-2}.  The resultant model curves are 
shown in Figure~\ref{fig30}.   

Casual inspection of the curves in Figure~\ref{fig30} suggests that
these models better characterize the data than the simple two-component models 
of the previous section.  However, the appropriate combination of the $\chi_\nu^2$
values in Table~\ref{tbl-2} into a reduced chi-square that characterizes {\it 
both\/} curves together yields $\chi_\nu^2=5.73$.  Table~\ref{tbl-1} shows 
that the two-component models of the previous section yielded a best fit of
$\chi_\nu^2=5.69$.  Hence, the model curves for the two subsamples fit the data
with essentially the same goodness of fit as for the simple two-component models.
To be consistent with the two-component model fit to the $\Tdc<20\,$K subsample,
we could try to fit a two-component model to the $\Tdc\geq 20\,$K subsample. 
This two-component fit to the $\Tdc<20\,$K subsample, when combined with that for 
the $\Tdc\geq20\,$K subsample, gives an overall $\chi_\nu^2=5.34$.  Again, this 
is {\it not\/} significantly better than the simple two-component models for the 
entire sample of points.  Also, this model requires fixing $\Tdz$ at 18$\,$K to get 
a reasonable fit.  Therefore, only the one-component model fit to the $\Tdc\geq20\,$K 
subsample is listed in  Table~\ref{tbl-2} and is used in the column density and
filling factor determinations. 

If we consider only that the goodness of fit is no better and that extra model 
parameters are required, then the models discussed in this section should be 
ignored in favor of the two-component models for the entire sample of points.
However, the two-component, two-subsample model described here has three
important advantages.  One is that the model fits the $\Tdc\geq 20\,$K points
better.  The second advantage is that the agreement between the dust-derived
column densities, $\NHd$, and the $\cO$-derived column densities, $\Ngth$,
is noticeably improved.  Figure~\ref{fig31} shows the plot of $\Ngth$ versus
$\NHd$.  The $\chi_\nu^2$ of the straight line fit to the points in this
plot is 5.3, compared with 7.4 in the corresponding plot for the simple 
two-component models.  This an improvement with a 90\% confidence level, 
according to the F-test. 

Another advantage to the two-component, two-subsample models is seen in 
Figure~\ref{fig32}, which shows the area-filling factors, $\filf$, versus the 
$\cO$-derived molecular gas column densities, $\NHth$.  The filling
factors are now all (or almost all) less than or equal to unity, as desired.
As was seen for the one-component models, the filling factors here again show
saturation with increasing column density. 

The uncertainties are estimated from the work of the previous sections.
For the $\Tdc<20\,$K subsample, the uncertainties of the 
two-component model parameters are largely systematic.
As such, the uncertainties should be similar to those for the 
two-component models fitted to the entire sample of points (see previous
section).  However, our estimates of those uncertainties for the component-1 
density, $n_{c1}$ must be modified.  The previous section
suggests that $n_{c1}$ is known to within a factor of 2 about a geometric
mean of possible densities.  However, the $n_{c1}$ for the current
section is 2 orders of magnitude larger than for the previous section.  Hence,
we can only place a lower limit of
$\sim few\times 10^2\unit cm^{-3}$ on $n_{c1}$.  We obtain
a higher density value in the current section, possibly because the densities 
in the $\Tdc<20\,$K subsample are 2 orders of magnitude higher than some kind of 
average density for the whole sample.  Indeed, we find that the 
$\Tdc\geq 20\,$K subsample has a density that is lower by more than an order of
magnitude.  However, the $\Tdc<20\,$K subsample has always dominated the 
results of the fits; the results obtained for this subsample are nearly
identical to those for the whole sample.  Therefore, this method likely places only
a rough lower limit on the density.
This is expected because the $\cOone$ transition is insensitive to densities
higher than about $10^3\unit cm^{-3}$.  Thus the uncertainties in the parameter
values for the $\Tdc<20\,$K subsample are the same as those for the entire
sample points (see the previous section) with the exception that the component-1
density only has a lower limit of $\sim few\times 10^2\unit cm^{-3}$. 

The parameter uncertainties for the $\Tdc\geq 20\,$K subsample
are again dominated by the systematics.  The $\DT$ versus scale factor for the 
one-component models, i.e. Figure~\ref{fig16}, shows that 
the fit to the $\Tdc\geq 20\,$K points varies from $-10$ to $+4\,$K. 
Hence we can say that $\DT = -3\pm 7\,$K.  This uncertainty is only slightly
larger (i.e. less than a factor of 2) than the formal uncertainty given in
Table~\ref{tbl-2}.  The plot of $\NDv$ versus scale factor suggests that only
the formal uncertainties in $\NDv$ are important.  However, the two-component
model fitted to the $\Tdc\geq 20\,$K sample suggests larger systematic
uncertainties.  The $\nvtco$ value found (which corresponds to the $\NDv$ value
in the one-component model) is less than a factor of 2 greater than the $\NDv$
value found for the one-component model.  The systematic uncertainties (see
the plot of $\nvtco$ versus scale factor in Figure~\ref{fig27}) suggest
that $\NDv$ varies between about $10^{15}$ and $10^{16}\ckms$.  (Recall that
it cannot be lower than about $10^{15}\ckms$ because the large-scale observed
$\NDv$ represents a lower limit.)  As for the systematic uncertainty of the 
density, Figure~\ref{fig16} suggests that, again, only a lower limit is applicable.  
This lower limit is $\sim 10^3\unit cm^{-3}$.  It is interesting that the two-component 
model results for the $\Tdc\geq 20\,$K subsample suggest a component-1 density
identical to that of the one-component model fit.

One noticeable problem with the current models is the abrupt
transition in the solid curve when crossing the $\Tdc=20\,$K boundary.  A more 
natural-looking curve would have a smooth transition across this boundary.
Examination of the different parameter values for both subsamples suggests that
gradually decreasing the $\nvco$ value by factors of 4 to 6 when going from the 
$\Tdc<20\,$K subsample to the $\Tdc\geq 20\,$K subsample (thereby becoming the 
$\NDv$ value for the one-component model for the latter subsample) could possibly
produce the desired smooth curve.  Note that other solutions, such as varying
$n_{c1}$ would {\it not\/} produce a satisfactory solution, because
it would not merge well with the low-$\Tdc$ curve.  Also, the
component-0 parameters $c_0$, $\nvcz$, and $n_{c0}$ have little affect on the
overall curve shape and are irrelevant for the $\Tdc\geq 20\,$K subsample. 
Therefore, the $\Tdc\geq 20\,$K subsample clumps are likely to have $\NDv$ values 
that are factors of a few lower than those in the $\Tdc<20\,$K subsample. 

The dashed curve on the $\Tdc>20\,$K side of Figure~\ref{fig30} suggests another 
possibility.  The current models
do not account for the vertical spread of the points (neither in the 
$\Tdc>20\,$K side nor in the triangular cluster mostly on the 
$\Tdc<20\,$K side) in the $\rd$ versus $\Tdc$ plot.  Also, they do not account 
for the horizontal spread of the triangular cluster,
although this spread is probably due to the large uncertainties in $\Tdc$ for
some of these points.  The dashed curve on the $\Tdc>20\,$K side 
passes through the high-$\rd$.  Accordingly, 
two or more subsamples could indeed account for these spreads {\it if\/} 
the curves for {\it all\/} the subsamples were applied to the whole range of 
observed color temperatures.   However, what is the 
optimum way of separating these two (or more) subsamples?  Separating them 
according to dust temperature is simplistic: the $\Tdc\geq 20\,$K solution, for 
example, does not pass through points on the $\Tdc < 20\,$K side (i.e. the 
dashed line on that side) that are far from the 
$\Tdc < 20\,$K solution (the solid curve on that side).  The optimum
separation of these subsamples would involve two (or more) curves for the whole 
color temperature range of that would correctly characterize all the points in the 
plot.  In other words, the basic assumption in its simplest
form is not correct.  Nonetheless, the basic assumption imposed
within each subsample would simplify the modeling.  For now, we keep
the current two-component, two-subsample model parameters because they represent
an adequate characterization of the data.  
 
In summary, the two-component, two-subsample models produce a similarly good
fit to the $\rd$ versus $\Tdc$ plot as the simple two-component models, but
fix the problems of the unphysical filling factors and the poor fit to the high-$\Td$
points.  Also, the dust-derived and $\cO$-derived column densities agree better
for these two-subsample models than for the models fitted to the whole
sample.  Also, the current two-component, two-subsample models hint that
the optimum choice of subsamples has yet to be found.   This will be left to
future work. 

\subsection{The Fiducial Points\label{ssec35}}

As a rough illustration of the limitations of the basic assumption and of the
modeling, a few reference or fiducial points were chosen in the
$\rd$ versus $\Tdc$ plots.  These points roughly follow the
overall shape of the sample in these plots.  Specifically, four of the fiducial
points approximately represent the corners and centroid of the 
triangular cluster of points and the other three fiducial points represent the 
overall trend in the high-$\Tdc$ branch.  The fiducial points are represented
as diamonds in the $\rd$ versus $\Tdc$ plots (i.e., Figures~\ref{fig6}, 
\ref{fig7}, \ref{fig8}, \ref{fig12}, \ref{fig13}, \ref{fig14}, \ref{fig15}, 
\ref{fig26}, and \ref{fig30}).  Table~\ref{tbl-2x} lists the
fiducial points (in order of increasing $\Tdc$) and the physical properties
that pertain to each.  Because of the basic assumption, the derived physical
properties would be the same for all the points (except $\Tk$ and $\Td$). 
However, when a high-$\Tdc$ solution was found, as for all but the LVG 
two-component case, the physical conditions derived for the four fiducial
points with $\Tdc\geq 20\,K$ were those of the high-$\Tdc$ solution.  The
table lists these physical conditions for the four cases of the previous 
sections:
\bupskip
\begin{itemize}
\bupskip
\item[]Case 1: LTE, one-component models.
\bupskip
\medskip
\item[]Case 2: LVG, one-component models.
\bupskip
\medskip
\item[]Case 3: LVG, two-component models.
\bupskip
\medskip
\item[]Case 4: LVG, two-component, two-subsample models.
\bupskip
\end{itemize}
\bupskip

Despite representing overall trends, the fiducial points do not necessarily 
represent the sample as a whole: e.g., the second, third, and fifth
points as listed in Table~\ref{tbl-2x}, which have distances of 3 or more sigmas
from the best-fitting model curves.  The deviations of these three points from each
model curve are not just statistical uncertainties; the models used
here simply cannot account for the total vertical spread of the points in the
sample.  The basic assumption does not allow models that would ``fill'' the space
occupied by the sample of points in the $\rd$ versus $\Tdc$ plot.  One alternative
is to have two or three of these two-compt model curves fit to all the 
sample simultaneously.  In practice, this is time-consuming
and is left to a later date.

\subsection{Masses\label{ssec36}}

The derived column densities now allow us to estimate the
masses of the Orion clouds.  Maps of the column densities are presented in
Figures~\ref{fig33} to \ref{fig35}.  Figure~\ref{fig33} shows the column density
maps derived from the observations using the LVG, one-component model results, 
Figure~\ref{fig34} the analogous maps for the two-component, two-subsample case,
and Figure~\ref{fig35} the maps of column density for just component~1 on those
lines of sight where this component has a temperature of 10$\,$K or less (i.e.,
$\Tdo=\Tkone\leq 10\,$K).  The maps of column densities for the LTE case are
not shown, because they closely resemble the maps of the LVG, one-component case. 
The upper panel in each of these figures shows the continuum-derived (dust-derived) 
gas column densities and each lower panel shows the $\cO$-derived gas column densities.
Summing these column densities appropriately and using the adopted distance of
450$\,$pc gives the masses ({\it not\/} including Helium) listed in Tables~\ref{tbl-3}, 
\ref{tbl-4}, \ref{tbl-5}, and \ref{tbl-6}.  The random uncertainties in these masses are 
small: relative uncertainties of the order of 0.5\% for the continuum-derived masses, 
0.25\% for the $\CO$-derived masses, and 6.5\% for the $\cO$-derived masses (although 
the smaller Orion$\,$Nebula and NGC$\,$2024 Fields have relative uncertainties of 20\%
for the $\cO$-derived masses).  The systematic uncertainties (i.e., the uncertainties
due to calibration and background-subtraction uncertainties) are more appreciable:
30\% for the continuum-derived masses and 20\% for the $\cO$-derived masses.  
The model results in each of the four cases were combined with the $\cOone$ and
H$\,$I observations to yield the column densities, as described previously, and these 
were summed to yield the masses in the tables.

All the $\cO$-derived column density maps show a plateau with steep edges. This
plateau is roughly 20$^\circ$ wide in longitude and about 6$^\circ$ in latitude.
For example, one of the edges of this plateau can be seen running from galactic 
longitude 216$^\circ$ to 208$^\circ$ at latitude $-10.25^\circ$.  These are
approximately the mapped area in $\cO$.  Specifically, these 
edges occur just inside the mapped area's edges, where the observed $\cOone$ 
line strength falls to zero. These plateau edges are also partly visible 
in the upper panel of Figure~\ref{fig34}, even though this panel does not have the 
$\cO$-derived column densities.  This is because these ``observed" column densities 
depend on the two-component, two-subsample model results, which, in 
turn, require knowledge of the $\cO$ line strengths.  This model dependence is 
especially true of the cold gas column density maps in Figure~\ref{fig34}, resulting 
in the clear visibility of the plateau edges in the northward and eastward (i.e. 
towards the top and left) directions.  The southward and westward edges
are the arbitrary boundaries imposed by selecting only those lines of 
sight with $\Tdo\leq 10\,$K; column densities of component~1 on lines of sight 
with $\Tdo > 10\,$K are set to zero. 

\begin{figure}
\epsscale{0.60}
\plotone{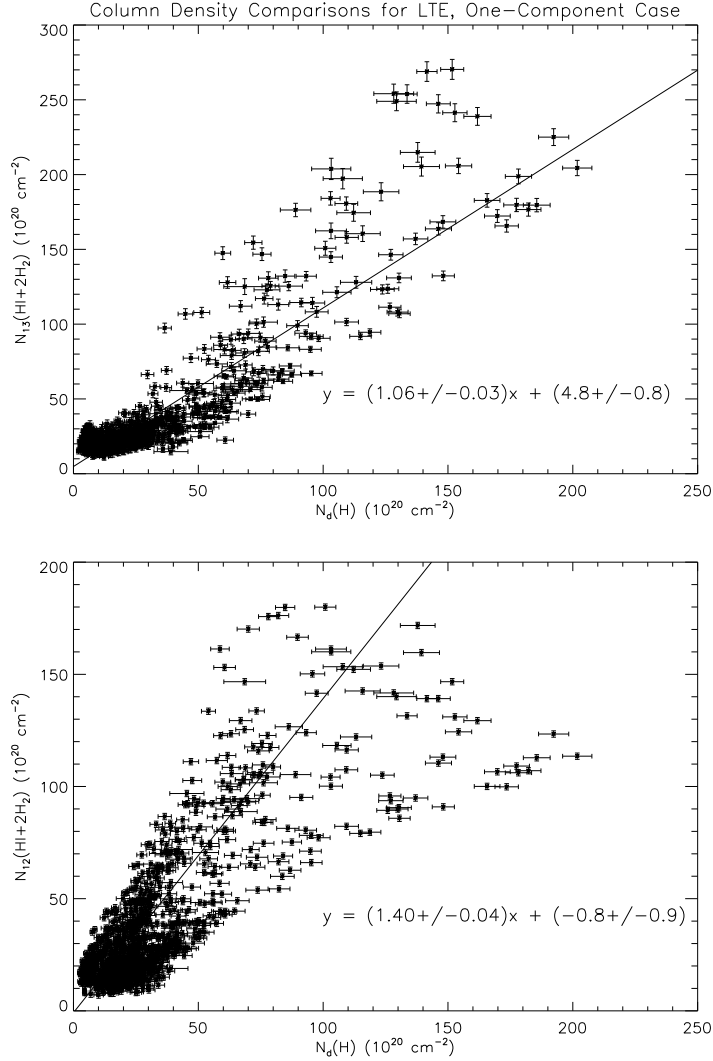}
\caption{The plots above compare the gas column densities as derived
from gas tracers and those derived from dust continuum emission for
the case of LTE gas emission and a single component.  The upper panel 
shows the gas column density as derived from the $\cOone$ in the LTE,
optically thin case versus the gas column density as derived from the
140$\um$ and 240$\um$ continuum emission.  The points in the upper 
panel plot represent the positions where $\Ia$, $\Ib$, and $\Ic$ are 
at more than 5-$\sigma$, as in Figure~\ref{fig6}.  The lower panel 
shows the gas column density as derived from the $\COone$ line assuming 
the conversion factor discussed in the text.  The points in the lower
panel plot represent the positions where $\Ia$, $\Ib$, and $\rm I(\CO)$ 
are at more than 5-$\sigma$.  Consequently, this sample has more points
than in the upper panel: specifically the upper panel has 674 points
and the lower panel has 1053 points. 
\label{fig10}}
\end{figure}

\begin{figure}
\epsscale{0.8}
\plotone{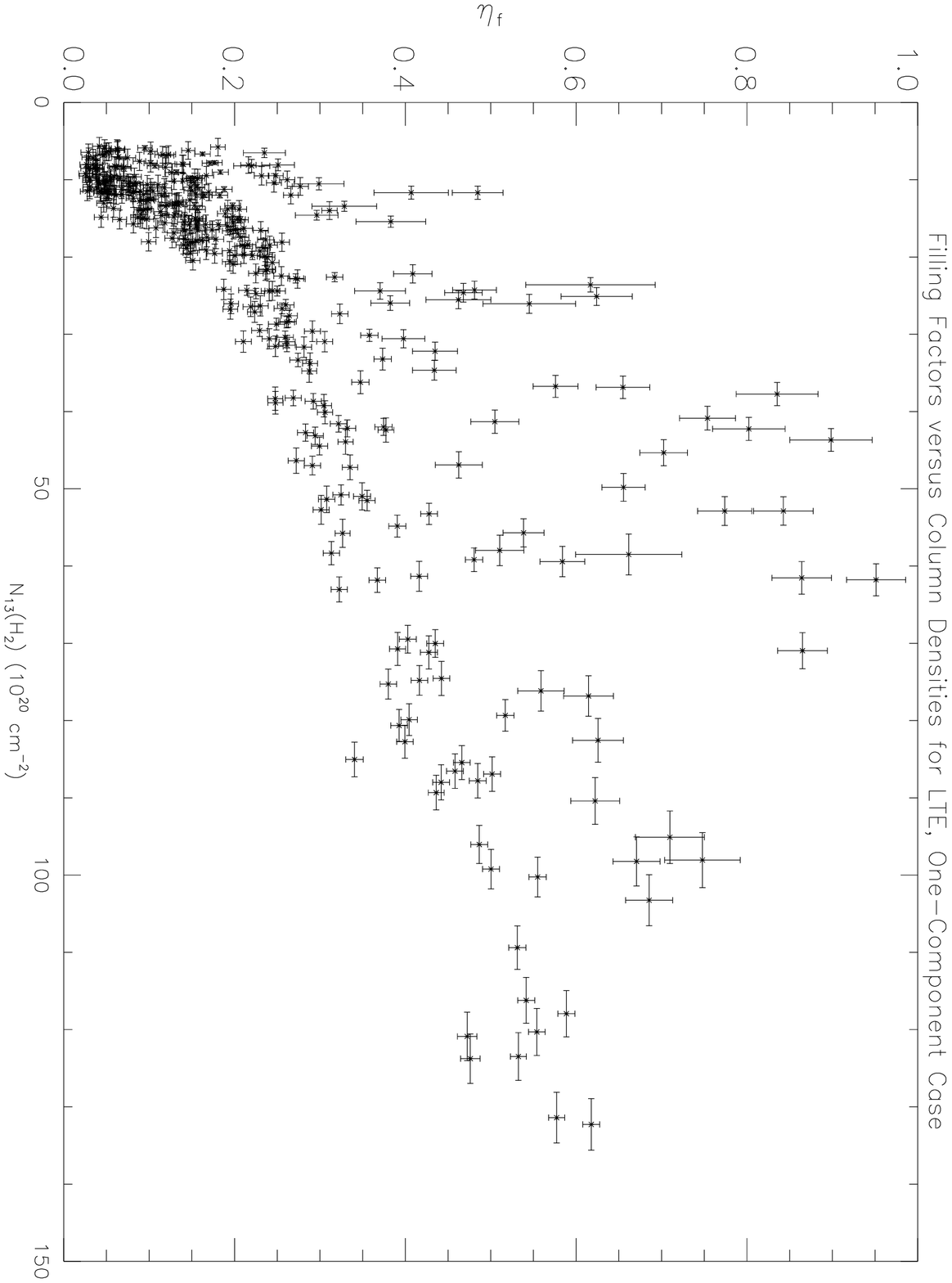}
\caption{The area filling factors of substructures in the Orion
clouds are plotted against the molecular gas column densities in
the LTE, one-component case.  The points in this plot represent the 
positions where $\Ia$, $\Ib$, and $\Ic$ are at more than 5-$\sigma$ 
and where the peak radiation temperature of $\COone$ is at more than 
3-$\sigma$.  This is a total of 372 points. 
\label{fig11}}
\end{figure}

\begin{figure}
\epsscale{0.68}
\plotone{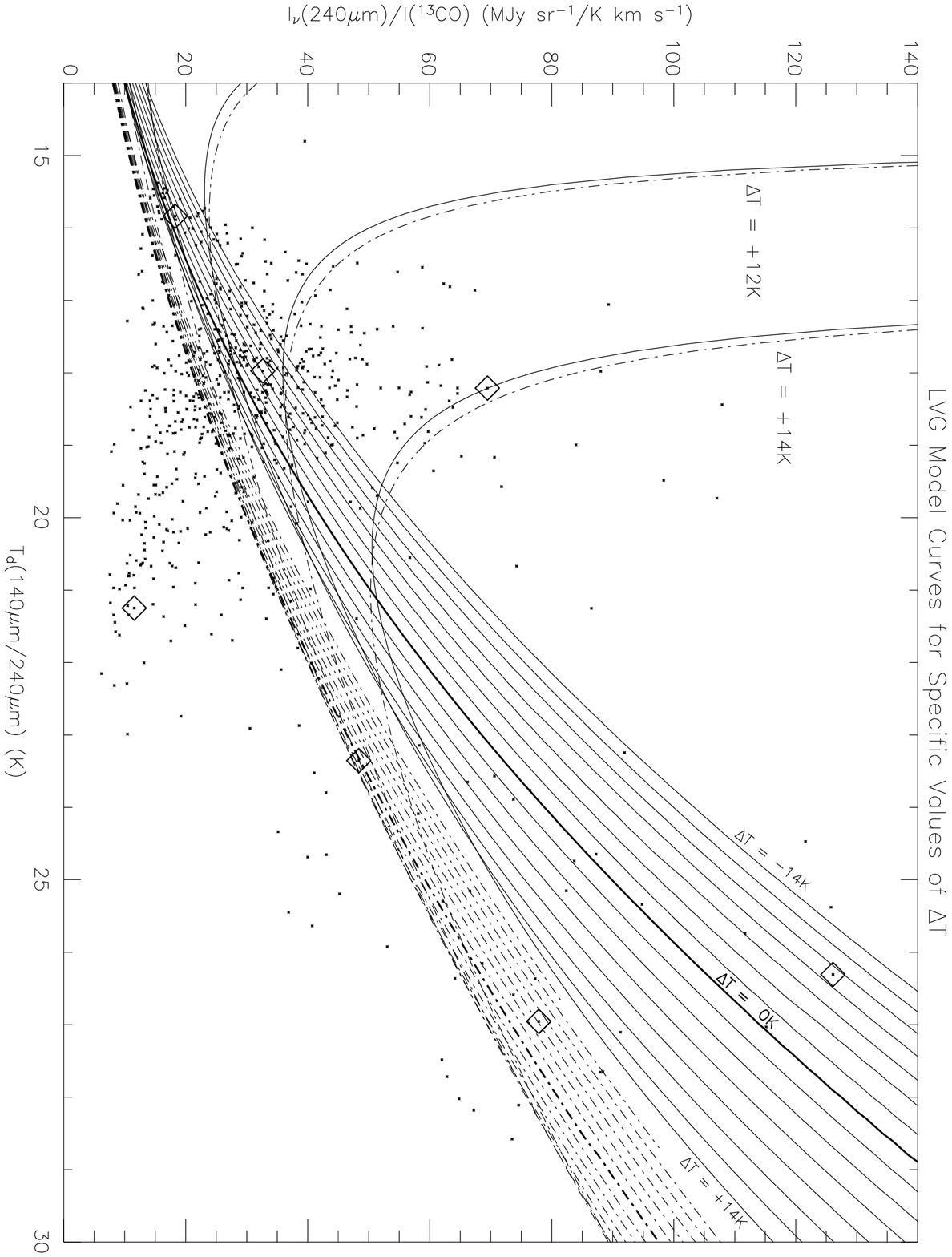}
\caption{The model curves $\rd$ versus $\Td$ for different $\DT$ values
are shown for the non-LTE, one-component case.  For the solid curves, the 
$\NDv$ value is fixed at $3.2\times 10^{15}\ckms$ and the $\nH$ value is 
fixed at $1\times 10^5\unit cm^{-3}$ --- the best fit $\NDv$ and $\nH$ 
values for model fits to all the points in the plot.  For the dashed-dotted 
curves, the $\NDv$ value is fixed at $3.2\times 10^{15}\ckms$ and the $\nH$ 
value is fixed at $5.6\times 10^3\unit cm^{-3}$ --- the best fit $\NDv$ and 
$\nH$ values for model fits to only the points with $\Td>20\,K$.  The curves 
(solid and dashed-dotted) shown range from $\DT = -14\,K$ to $+14\unit K$ in 
steps of 2$\,$K.   The plot includes the same data points as in Figure~\ref{fig6} 
with error bars excluded for clarity.   The diamonds in this figure are the 
fiducial points discussed in Section~\ref{ssec35} and listed in
Table~\ref{tbl-2x}. 
\label{fig12}}
\end{figure}

\begin{figure}
\epsscale{0.68}
\plotone{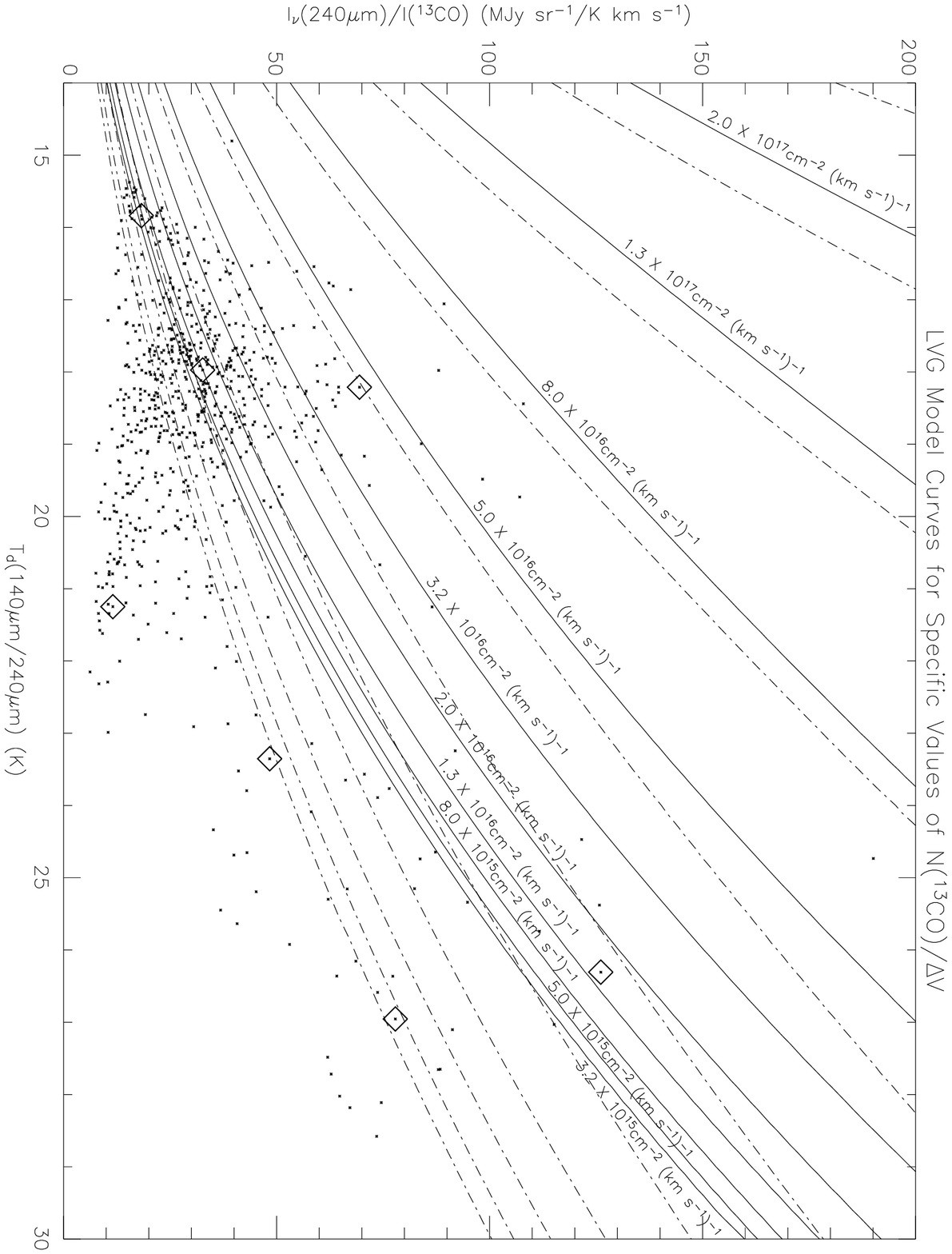}
\caption{The model curves $\rd$ versus $\Td$ for different $\NDv$ values
are shown for the non-LTE, one-component case.  For the solid curves, the 
$\DT$ value is fixed at $-1\unit K$ and the $\nH$ value is fixed at $1\times 
10^5\unit cm^{-3}$ --- the best fit $\DT$ and $\nH$ values for model fits 
to all the points in the plot.  For the dashed-dotted curves, the $\DT$ 
value is fixed at $-3\unit K$ and the $\nH$ value is fixed at 
$5.6\times 10^3\unit cm^{-3}$ --- the best fit $\DT$ and $\nH$ values for 
model fits to only the points with $\Td>20\,K$.  The solid curves are 
labelled with their corresponding values of $\NDv$.  The $\NDv$ values for
the dashed-dotted curves follow the same order as for the solid curves,
i.e., from lower right to upper left.  The plot includes the same data 
points as in Figure~\ref{fig6} with error bars excluded for clarity. 
The diamonds in this figure are the fiducial points discussed in 
Section~\ref{ssec35} and listed in Table~\ref{tbl-2x}.
\label{fig13}}
\end{figure}

\begin{figure}
\epsscale{0.70}
\plotone{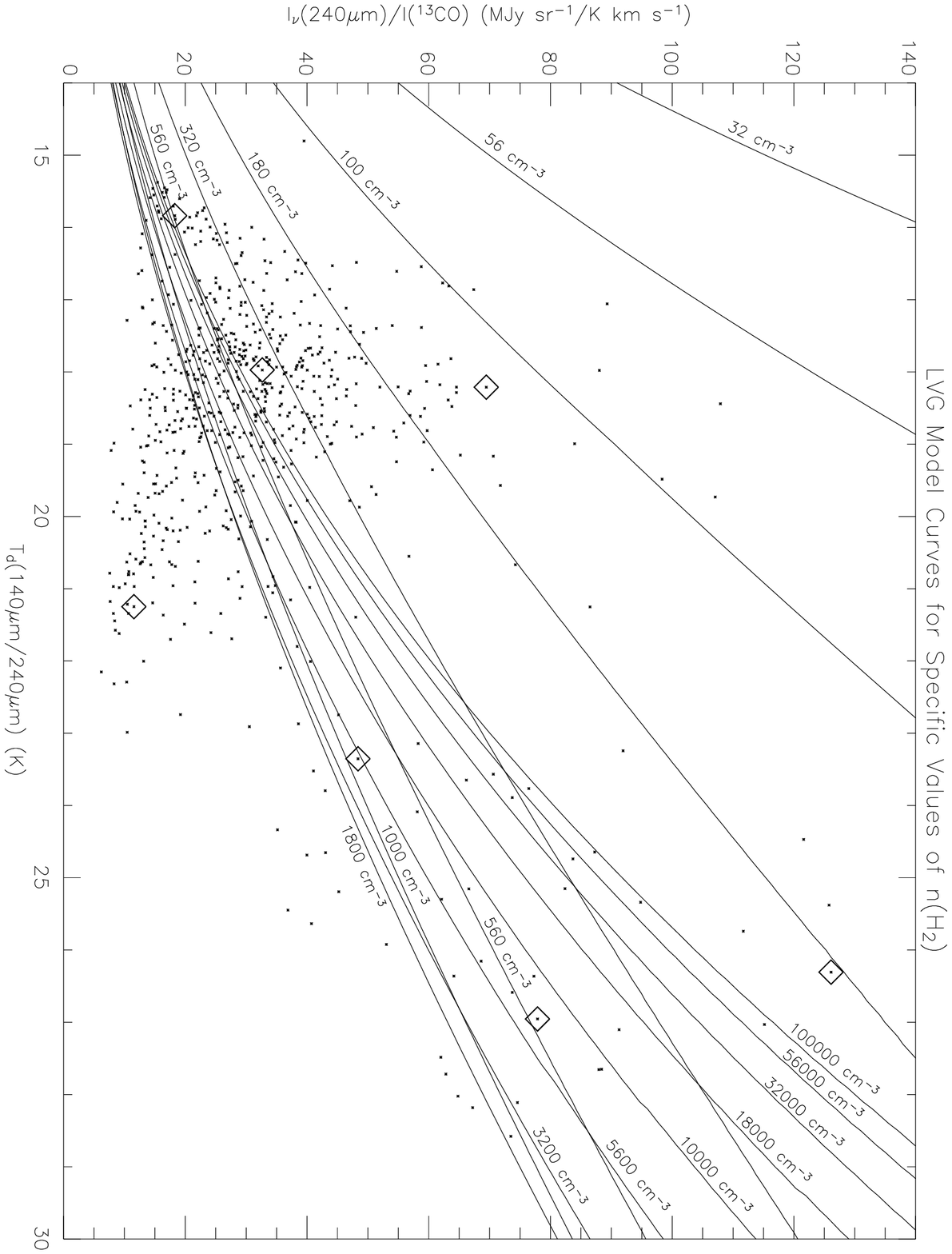}
\caption{The model curves $\rd$ versus $\Td$ for different $\nH$ values
are shown for the non-LTE, one-component case.  For the curves shown, the 
$\DT$ value is fixed at $-1\unit K$ and the $\NDv$ value is fixed at 
$3.2\times 10^{15}\ckms$ --- the best fit $\DT$ and $\NDv$ values for model 
fits to all the points in the plot.  The best fit $\DT$ and $\NDv$ values 
for model fits to the points with $\Td>20\,K$ are similar or identical to 
those for all the points; consequently, those curves would be virtually 
identical to the solid curves above and are not shown.  The plot includes 
the same data points as in Figure~\ref{fig6} with error bars excluded for 
clarity.  The diamonds in this figure are the fiducial points 
discussed in Section~\ref{ssec35} and listed in Table~\ref{tbl-2x}.
\label{fig14}}
\end{figure}

\begin{figure}
\epsscale{0.73}
\plotone{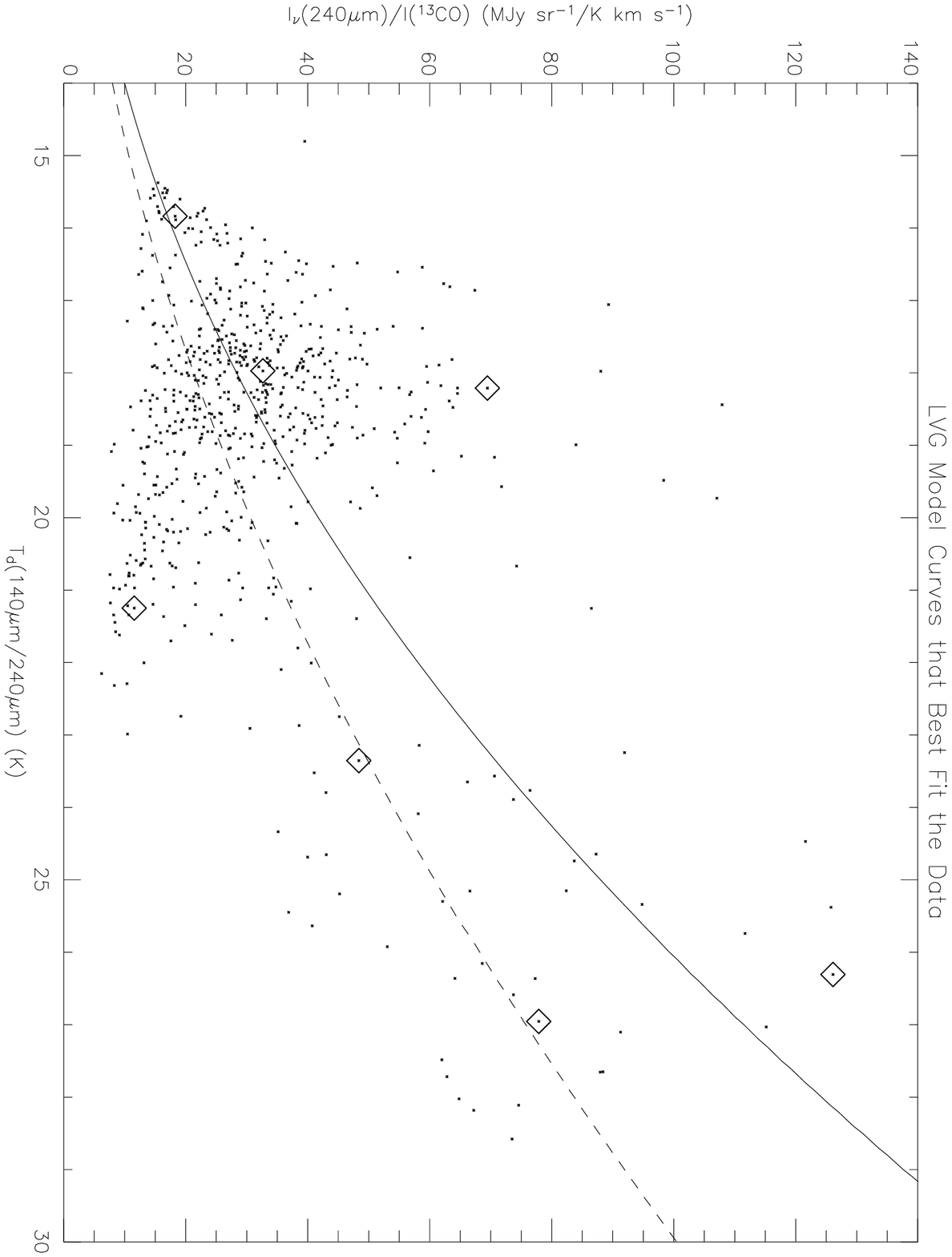}
\caption{The model curves of $\rd$ versus $\Td$ that best fit the data are 
shown for the non-LTE, one-component case. The plot includes the same data 
points as in Figure~\ref{fig6} with error bars excluded for clarity.   The 
solid curve represents $\DT = -1\,K$, $\NDv = 3.2\times 10^{15}\ckms$, and 
$\nH = 1\times 10^5\unit cm^{-3}$ and is the best fit to all the data points.  
The dotted curve represents $\DT = -3\,K$, $\NDv = 3.2\times 10^{15}\ckms$, 
and $\nH = 5.6\times 10^3\unit cm^{-3}$ and is the best fit to only the data 
points with $\Td>20\,K$.  The diamonds in this figure are the fiducial points 
discussed in Section~\ref{ssec35} and listed in Table~\ref{tbl-2x}.
\label{fig15}}
\end{figure}

\begin{figure}
\epsscale{0.62}
\plotone{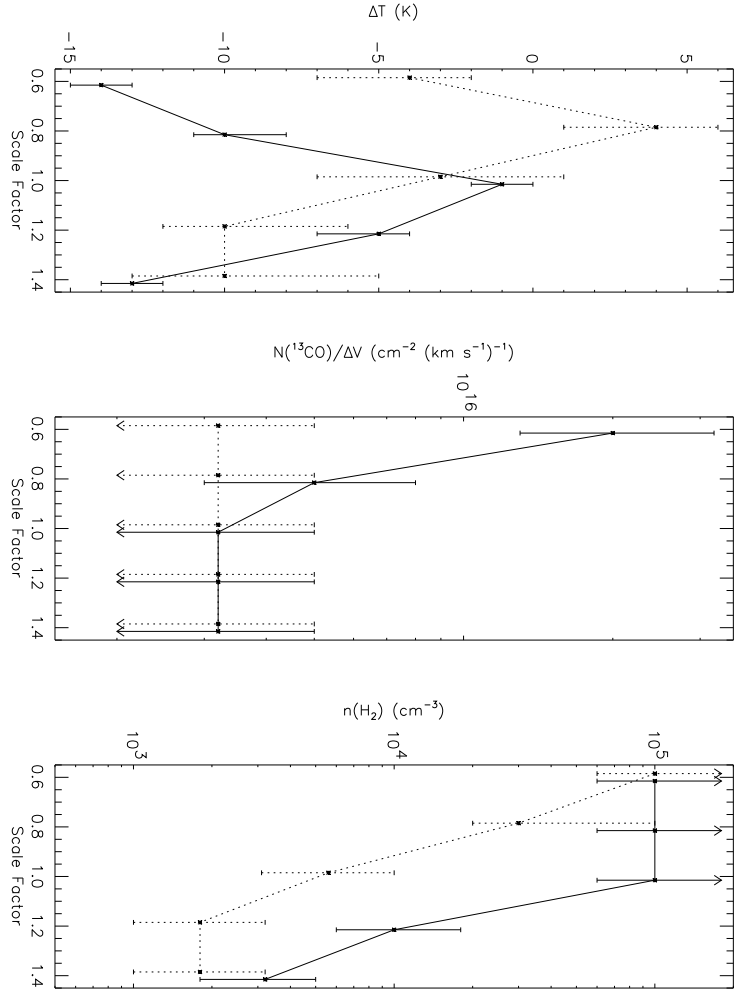}
\caption{The effect of the systematic uncertainties on the resultant
parameters from the fits of the LVG model curves is shown.  The effect of 
these uncertainties was tested by applying the scale factors 0.6, 0.8, 1.0, 
1.2, and 1.4 to the model curves and fitting the parameters for each scale
factor.  The left panel shows the resultant $\DT$ values, the center panel
shows the resultant $\NDv$ values, and the right panel shows the $\nH$ values.
The solid line in each panel represents the resultant parameter values for the 
fits to all the data (i.e., all the data points shown in Figure~\ref{fig6}).  
The dotted line represents the resultant parameter values for the fits to the 
data with $\Td\geq 20\,K$.  Notice that the plotted points have been slightly 
displaced horizontally from their true scale factor values for clarity. The 
error bars represent the formal error bars for each model fit and are the 
minimum grid spacing, for the grid of LVG models used, necessary to increase 
$\chi^2$ by a {\it minimum\/} of $\chi_\nu^2$.  These formal errors are 
therefore very conservative estimates of the true formal errors.   
\label{fig16}}
\end{figure}

\begin{figure}
\epsscale{0.7}
\plotone{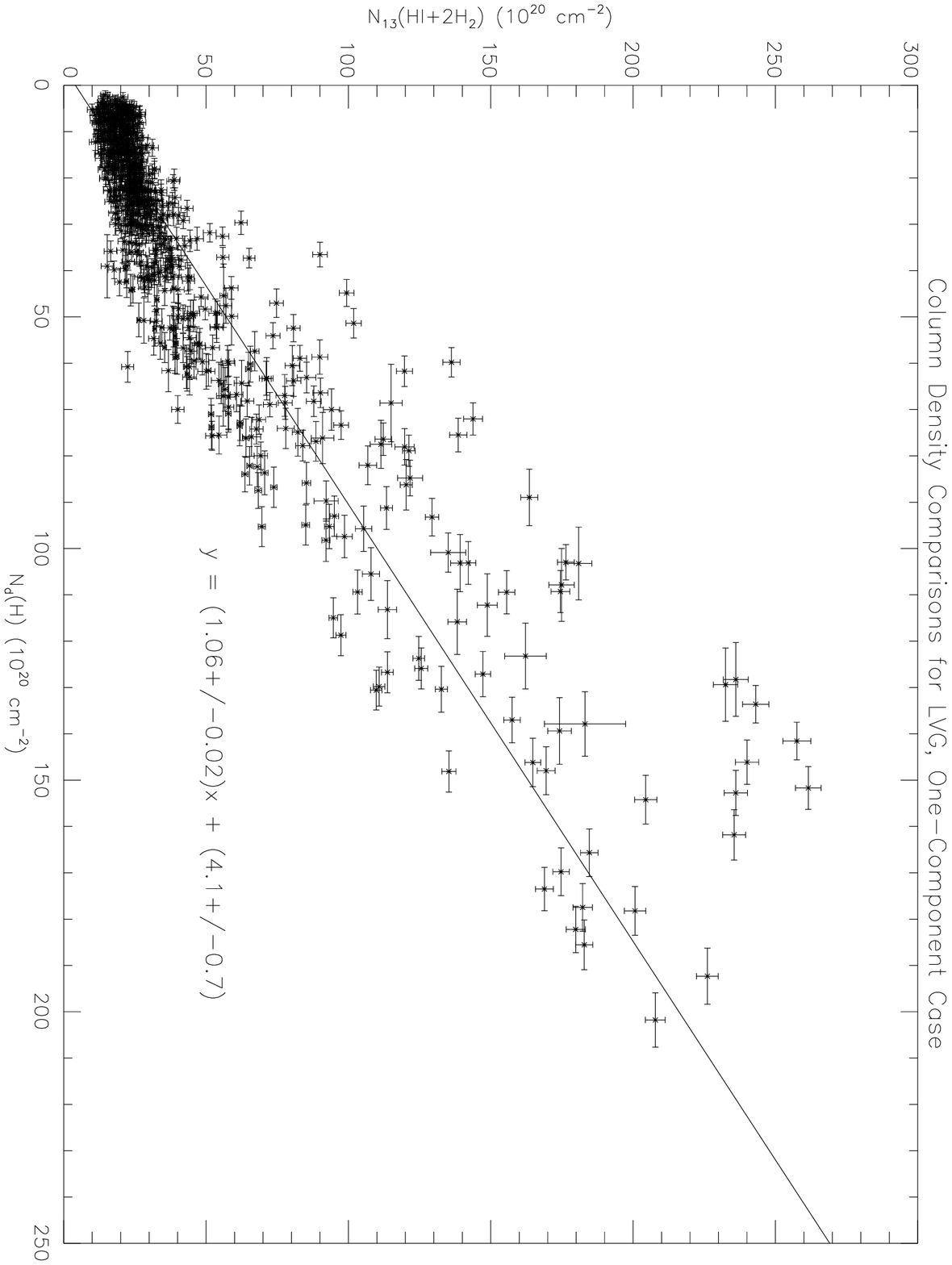}
\caption{The plot above compares the gas column densities as derived
from gas tracers and those derived from dust continuum emission for
the case of non-LTE gas emission and a single component.  The plot 
shows the gas column density as derived from $\cOone$ using the 
results from the LVG mehtod versus the gas column density as derived 
from the 140$\um$ and 240$\um$ continuum emission.  The points in 
this plot represent the positions where $\Ia$, $\Ib$, and $\Ic$ are 
at more than 5-$\sigma$, as in Figure~\ref{fig6}.  
\label{fig17}}
\end{figure}

\begin{figure}
\epsscale{0.8}
\plotone{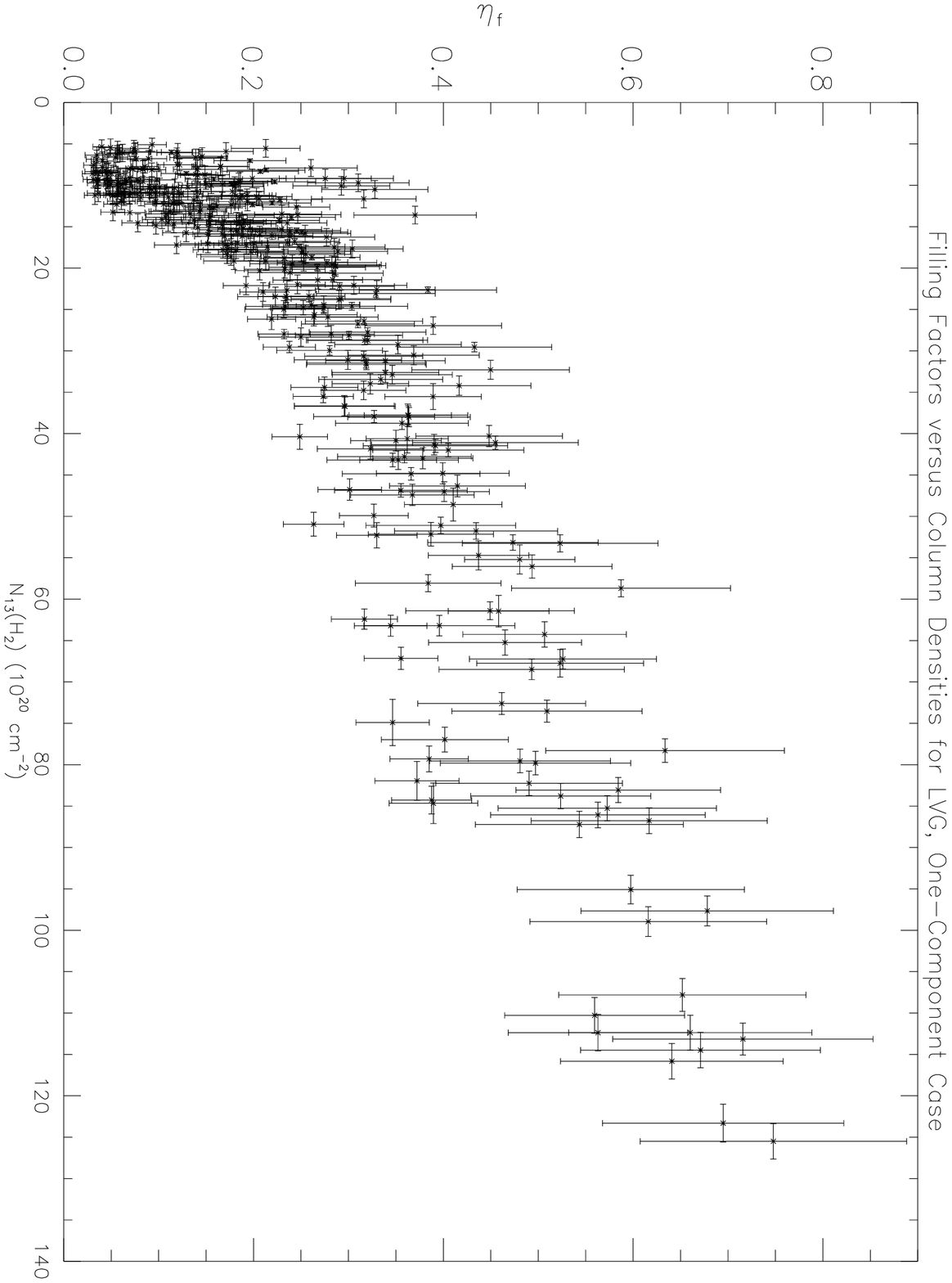}
\caption{The area filling factors of substructures in the Orion
clouds are plotted against the molecular gas column densities in the
LVG, one-component case.  The points in this plot represent the positions 
where $\Ia$, $\Ib$, and $\Ic$ are at more than 5-$\sigma$ and where the 
peak radiation temperature of $\COone$ is at more than 3-$\sigma$.  This 
is a total of 372 points. 
\label{fig18}}
\end{figure}

\begin{figure}
\epsscale{0.75}
\plotone{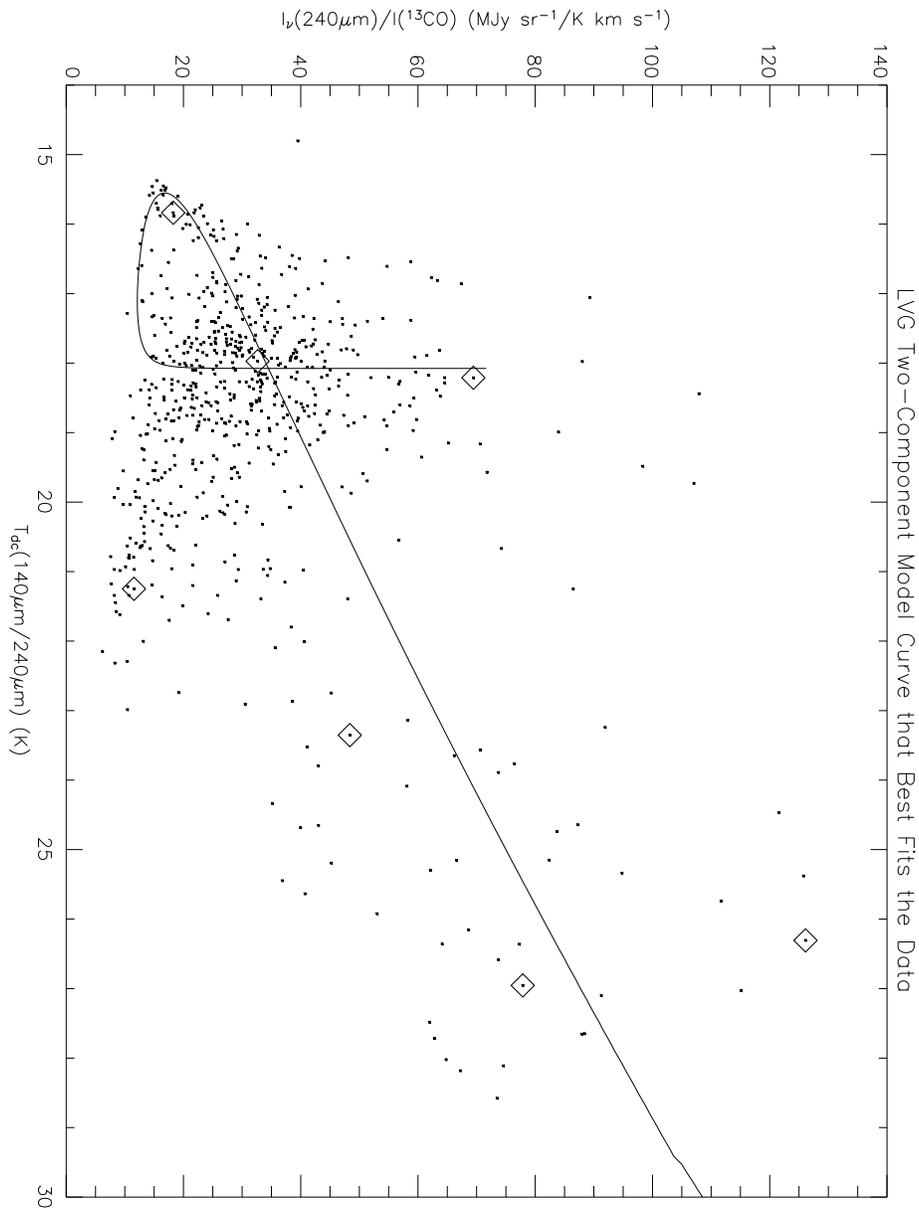}
\caption{The model curve of $\rd$ versus $\Tdc$ that best fits the data is 
shown for the non-LTE, two-component case. The plot includes the same data 
points as in Figure~\ref{fig6} with error bars excluded for clarity.  The
parameters and their values are listed in Table~\ref{tbl-1}.  The diamonds 
in this figure are the fiducial points discussed in 
Section~\ref{ssec35} and listed in Table~\ref{tbl-2x}.
\label{fig26}}
\end{figure}

\begin{figure}
\epsscale{0.75}
\plotone{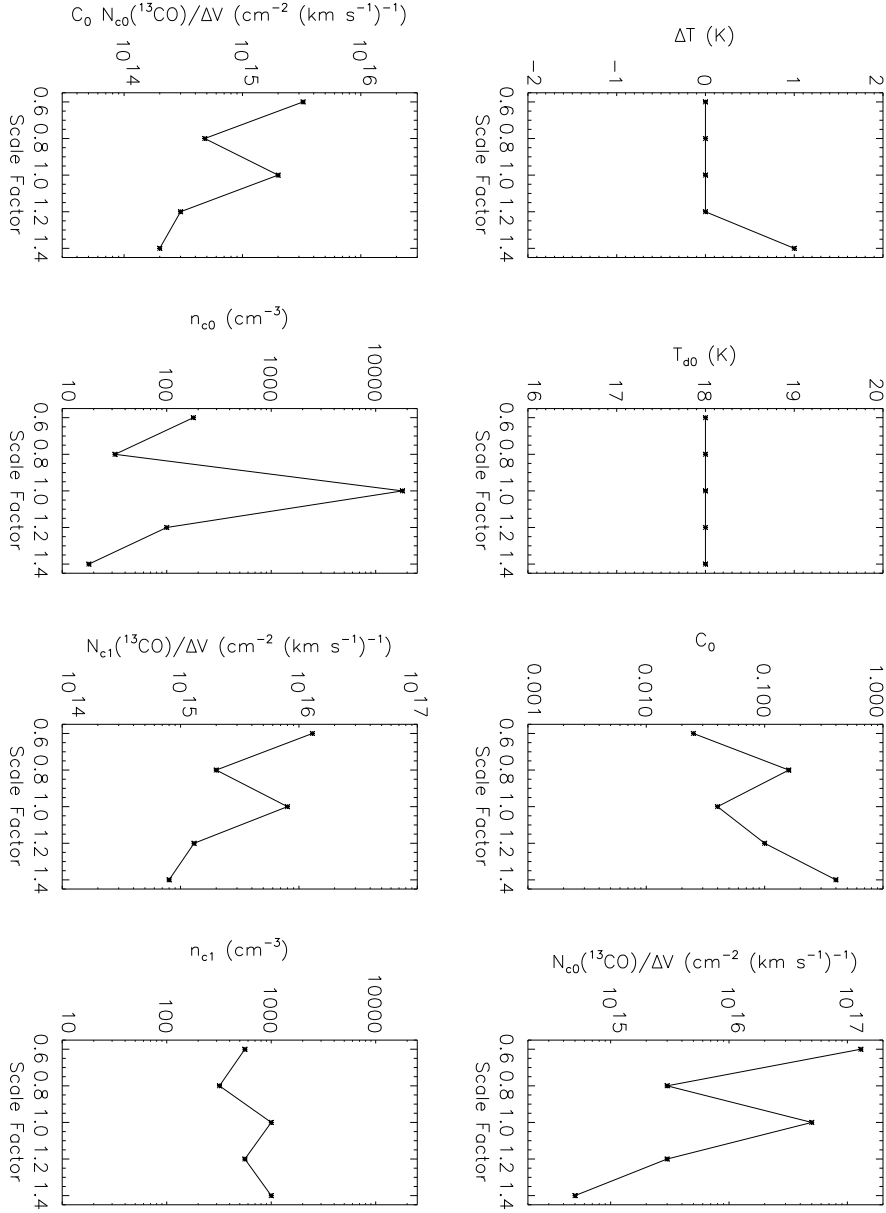}
\caption{The effect of the systematic uncertainties on the resultant
parameters from the fits of the two-component, LVG model curves is shown.  
The effect of these uncertainties was tested by applying the scale factors 
0.6, 0.8, 1.0, 1.2, and 1.4 to the model curves and fitting the parameters 
for each scale factor.  Except for the plots for $\DT$ and $\Tdz$, all
plots are semi-logarithmic where the vertical axes cover about the same 
logarithmic difference in range (about 3 orders of magnitude).  This allows 
easy visual determination of which parameters have the smallest systematic 
uncertainties.  
\label{fig27}}
\end{figure}

There are apparent discrepancies between the derived masses (i.e. Tables~\ref{tbl-3}
and \ref{tbl-4}) and the column density versus column density plots (i.e., 
Figures~\ref{fig10}, \ref{fig17}, \ref{fig28}, and \ref{fig31}).  For example, the 
ratio of the $\CO$-derived masses in Table~\ref{tbl-3} compared to the 
continuum-derived masses of Cases~1 \& 2 in Table~\ref{tbl-4} is 1.08, whereas the 
linear fit in the lower panel of Figure~\ref{fig10} suggests a ratio of 1.40. 
There are two reasons for the discrepancy that are related to the 
signal-to-noise ratio of the observations.  One reason is that the plot of 
column densities in Figure~\ref{fig10} only considers positions with high 
signal-to-noise data (i.e. more than 5-$\sigma$ in $\COone$ and in the continuum 
bands).  If all the positions in the Orion fields were included in the plot, the
slope would change to about 1.3.  The rest of the discrepancy is because of the 
difference between the slope of a line from linear regression and
the ratio of the averages of the ordinate values and abscissa values.
The former represents a sum of values that are weighted by $1/\sigma^2$ and 
divided by another sum of similarly weighted values.  Determining the total 
masses by simply summing the column densities in the fields and dividing by 
another similarly determined sum represents the latter.  In other words, the
masses in the tables are not determined from weighted sums, whereas the slopes
in the column density plots are indeed determined from such sums.  Therefore,
the slopes in the column density plots are biased toward the points with higher 
signal-to-noise ratios {\it even\/} when these plots include the low 
signal-to-noise ratio points.  This discrepancy in the ratio between the tables 
and the plots is simply a consequence of the scatter of the points in
these plots.  If there were less scatter, then the tables and the plots would
give similar ratios for the $\CO$-derived over continuum-derived masses (assuming
similar coverage in both types of maps).  This affects the accuracy 
of our X-factor estimates.  The slope from the lower panel of 
Figure~\ref{fig10} suggests an X-factor of $1.9\times 10^{20}\cKkms$, whereas  
comparing the total masses in the tables for the one-component cases suggests an 
X-factor of $2.4\times 10^{20}\cKkms$. Combining these two numbers gives an estimate 
of the X-factor of $(2.1\pm 0.3)\times 10^{20}\cKkms$.  This is an uncertainty of 
about $\pm$15\%.  Thus specifying the X-factor to greater accuracy (i.e., a smaller 
uncertainty) is meaningless.  

An interesting, but crude, comparison can be made with the Orion X-factor estimated from 
Figure~8 of \citet{Dame01}.  In that figure, the predicted I(CO) 
profile that cuts along the Orion region is {\it on average\/} about 20-30\% higher than 
that of the observed profile.  Accordingly, the predicted X-factor of $1.8\times 
10^{20}\cKkms$ for the entire Galaxy must be increased by factors of about 1.2 to 
1.3 to give the {\it average\/} X-factor for Orion.  This gives an average X-factor for 
Orion of about 2.2-2.3$\times 10^{20}\cKkms$.  This is consistent with the value derived
above. 

There is another apparent discrepancy between the plots and the tables; the total masses
derived from $\cO$ (i.e., in Table~\ref{tbl-3}) disagree with those derived from 
the continuum (i.e., in Table~\ref{tbl-4}).  In this case, the ratios derived from
the tables are more extreme (i.e., further from 1) than from the column density
plots.  The problem here is that the $\cO$ map does not cover the full area covered
by the continuum maps.  This results in two problems: a problem with the 
continuum-derived gas masses and a problem with the $\cO$-derived gas masses.  The
former problem is that the two-component-model results cannot give
the continuum-derived gas column densities outside the area mapped in $\cO$.  
Specifically, an observed data point in the $\rd$ versus $\Tdc$ plot can{\it not\/} 
be associated with a point on the theoretical curve, because the observed data point 
can{\it not\/} even be placed in this plot without knowing the $\cOone$ line strength.  
Consequently, the correction factor in equation~(\ref{mr39}) is assumed to be unity 
(i.e., $\NHd=\NHdo$ is assumed) for those positions outside the area mapped in $\cOone$.   
The latter problem is that the $\cO$ observations alone do {\it not\/} yield accurate 
{\it total\/} masses for these clouds; the mass estimates from $\cOone$ must be 
corrected for the missing coverage towards the edges of the clouds.  These corrections 
were estimated by determining the total of the gas masses of the three big fields from 
the continuum observations for {\it only those positions inside the $\Ic=0$ contour\/} 
for each of the four cases (where Cases 1 and 2 are identical for the continuum-derived 
masses). These were then compared with the continuum-derived masses for all the gas in 
those fields (i.e., the numbers listed as ``Total" in Table~\ref{tbl-4}).  

Before discussing this comparison, assessing the accuracy of this method is possible
by comparing the continuum-derived masses inside the $\Ic=0$ contour with the 
$\cO$-derived masses in Table~\ref{tbl-3}.  These latter masses represent, in 
practice, the molecular gas mass inside the $\Ic=0$ contour plus the H$\,$I mass
throughout the entire Orion fields.   That is, the masses listed in the ``Case~1",  
``Case~2", ``Case~3", and ``Case~4" columns of Table~\ref{tbl-3} are the $\cO$-derived 
molecular gas masses within the $\Ic=0$ contour plus the H$\,$I mass both inside and 
outside this contour.   Therefore, to assess the accuracy of this correction for the lack 
of $\cO$ coverage, the continuum-derived masses inside this contour must be corrected
upwards for the H$\,$I mass that lies outside this contour.  This H$\,$I mass is
34000$\unit M_\odot$.  After adding this mass to the continuum-derived masses inside
the $\Ic=0$ contour, the continuum-derived masses agree with the $\cO$-derived masses of 
Table~\ref{tbl-3} to within 5 to 10\%, depending on the case.   

Now to correct the total masses in Table~\ref{tbl-3} for the gas mass that is missing 
outside the $\Ic=0$ contour, we compare the continuum-derived masses within this contour 
with the masses in Table~\ref{tbl-4}.  This gives mass differences of 88000$\unit 
M_\odot$ in the one-component cases and 87000$\unit M_\odot$ in the two-component cases.  
Allowing again for the 34000$\unit M_\odot$ of H$\,$I outside the $\Ic=0$ 
contour that is already implicit in Table~\ref{tbl-3} requires the masses in 
Table~\ref{tbl-3} to be corrected upwards by 54000$\unit M_\odot$ in the one-component 
cases and by 53000$\unit M_\odot$ in the two-component cases.  The resultant best estimates
for the Orion gas masses in the one-component case (for Case~2) and in the two-component
case (for Case~4) are in Table~\ref{tbl-5}, where the
continuum-derived and $\cO$-derived masses agree to within about 5\%. 

Table~\ref{tbl-6} shows that roughly 40 to 50\% of the two-component gas mass is in  
the form of cold ($\Tk < 10\,$K) gas (compare with Table~\ref{tbl-4}).  Consequently, 
the mass of gas or dust being detected at wavelengths much more sensitive to warm,
rather than cold, gas or dust (e.g., in the Wien limit) must be 
corrected upward by a factor of at least 1.6. 
Figure~\ref{fig35} shows that the cold component is on one edge of each of the small 
2$^\circ$ HII-region fields (i.e. Orion$\,$Nebula and NGC$\,$2024 Fields) and is
spread out along the northern edge of the Orion$\,$B Field.  There is also some cold
dust towards the north-eastern edge of the $\lambda\,$Orionis Field.  A natural 
question at this point is about a realistic lower limit to the temperature: are gas 
and dust temperatures as low as 3$\,$K
really necessary for explaining the data?  To test this, two-component models 
were applied, where $\Tkone$ was restricted to be $\geq 10\,$K.   This gave $\DT=-7\,$K 
and $\chi_\nu^2=10.5$.  In other words, the minimum $\Tdo$
was {\it still\/} found to be 3$\,$K.  Also, the $\chi^2$ is a factor of 1.9 higher
than the best fit models that had no such restriction on $\Tkone$.  This is a worse 
fit at the 99\% confidence level.  Restricting {\it both\/} $\Tdo$ {\it and\/}
$\Tkone$ would have given an even worse fit.  A less extreme restriction on $\Tkone$
of $\Tkone\geq 5\,$K gives $\DT=-2\,$K and $\chi_\nu^2=7.7$.  This fit is worse at
almost the 90\% confidence level and {\it again\/} the minimum $\Tdo$ is 3$\,$K. 
In short, avoiding having at least some dust or gas with temperatures near 3$\,$K is very 
difficult, provided that the basic assumption is valid.  Further discussion of this is 
deferred to Paper~III \citep{W05a}.

Since the failings of the basic assumption could be relevant here, longer-wavelength 
observations confirming the existence of such cold gas or dust would be helpful.  The longer 
wavelength {\it FIRAS\/} data were used to test this, but the results were inconclusive.  
The {\it FIRAS\/} data extend to a wavelength of about 4$\,$mm, making them much 
less sensitive to the warm dust and, as such, much less {\it in}sensitive
to the cold dust.  However, the {\it FIRAS\/} data also have a spatial resolution of 
7$^\circ$; the {\it FIRAS\/} beam solid angle is a 
{\it factor of 100 larger\/} than that of the {\it DIRBE\/} beam.  Consequently, a direct 
comparison with the {\it DIRBE\/} data and the Orion fields as defined here is difficult.  
Nevertheless, such a comparison was attempted as follows:
\begin{enumerate}
\item The spectra for 183 positions towards and around the Orion fields were
obtained.  Each position had two spectra: a low-frequency spectrum and a 
high-frequency spectrum.  These were joined to form a single spectrum.  Data
in the channels that overlapped between the two spectra were averaged with
$1/\sigma^2$ weights.  This yields spectra that cover frequencies of from 68 to 
2911$\,$GHz (i.e. wavelengths from 4410 to 103$\um$) with channels 13.6$\,$GHz
wide. 
\item The 60 highest frequency channels were then binned into 5 channels to
improve the signal-to-noise ratio.  The 12 channels that went into each of
the 5 new channels were averaged with $1/\sigma^2$ weights.
\item A background with the form $I_{BG}= a_0 + a_1\, csc(b_{eff})$ was fitted to each 
channel from those positions outside the Orion fields and farther than half a {\it FIRAS\/} 
beam from the Galactic plane.  The $a_0$ represents an isotropic background component 
and is the cosmic background radiation.  The $a_1$ represents a scaling 
factor for the Galactic plane emission.  The $b_{eff}$ is the effective Galactic 
latitude for emission that follows a cosecant($b$) law averaged within the {\it FIRAS\/}
beam.  Specifically, $b_{eff}\equiv arccsc(\langle csc(b)\rangle)$ where $\langle 
csc(b)\rangle$ is the average of $csc(b)$ within the 7$^\circ$ tophat beam of {\it FIRAS\/}.  
A good empirical approximation to $b_{eff}$ is $b_{eff} = |b| - 5|b|^{-1.15}$ for 
$|b|$ in the range of 4$^\circ$ to 27$^\circ$.  After determining the best fit $a_0$ 
and $a_1$ values for all the channels, the resultant background spectrum was 
subtracted from the {\it FIRAS\/} spectra centered on the Orion field centers.
\item Two-component models of modified blackbodies were fitted to the\hfil\break
background-subtracted spectra.  These were of the form $I_\nu = \tau_a 
(\nu/\nu_0)^\beta B_\nu(T_{da}) + \tau_b (\nu/\nu_0)^\beta B_\nu(T_{db})$, where
$\tau_a$ and $\tau_b$ are the optical depths of components $a$ and $b$, respectively,
at reference frequency $\nu_0$. 
\end{enumerate} 
The best fitting curves had a warm component with temperatures between 17 and 20$\,$K.
The cold component was very hard to define.  Assuming that this component even exists,
the temperature is around 5$\,$K, but
could not be pinned down.  Because of this temperature uncertainty, placing limits on 
the cold component mass was difficult.  The cold-component mass was anywhere between
0 and 10 times the warm-component mass.  In any event, confirming the existence of this 
cold dust must await long-wavelength spectra with greater spatial resolution and 
sufficient spatial and wavelength coverage.

\section{Summary and Discussion\label{sec4}}

We used the 240$\um$ continuum to the $\cOone$ intensity ratio plotted against
the 140$\um$/240$\um$ dust color as a diagnostic of the dust and gas physical 
conditions in the Orion$\,$A and B molecular clouds.  Model curves were fitted
to the data with orthogonal regression.  Two sets of one-component models were
applied to the data: Local Thermodynamic Equilibrium (LTE) and Large Velcocity
Gradient (LVG).  Two sets of two-component models were also applied to the data: 
two-component, LVG models and two-component, two-subsample, LVG models.  In each
case, the model fit gave the conversion of the observed intensities into gas/dust
column densities and beam-area filling factors.  The results of the fitting are 
summarized below:
\begin{itemize}
\item {\it One-Component, LTE Models:\/} These models fit the data only poorly
with a reduced chi-square of $\chi_\nu^2=16.5$.  When considering systematic
uncertainties, the only parameter to be fitted, $\DT$, was very poorly determined:
$\DT$ lies anywhere between $-24\,$K and $+6\,$K.  Comparison of the 
continuum-derived gas column densities with those derived from the 
$\cOone$ spectral line showed considerable scatter with $\chi_\nu^2=16.6$.
The filling factors exhibit an understandable increase with increasing column
density and are all less than unity.  At best, the one-component, LTE models
only crudely reproduce the overall trend in the data.
\item {\it One-Component, LVG Models:\/} These models also fit the data only 
poorly with a reduced chi-square of $\chi_\nu^2=16.9$.  When considering systematic
uncertainties, the $\DT$ was again very poorly determined: $\DT$ lies anywhere 
between $-14\,$K and $-1\,$K.  Comparison of the continuum-derived versus 
spectral line-derived gas column densities showed comparable scatter to the
LTE model case, with $\chi_\nu^2=15.5$.  The filling factors behaved similarl
to those in the case of the LTE modeling.  The LVG model results overall are,
at best, only marginally better than the single-component, LTE models.
\item {\it Two-Component, LVG Models:\/} In contrast to the one-component models, 
these models explain the overall trend in the data with $\chi_\nu^2=5.7$; this is 
better than the one-component models at a confidence level of better than 99.9\% 
according to the F-test.  Also in contrast to the one-component model results, 
$\DT$ is known to be within 1$\,$K of 0$\,$K even when the systematic uncertainties 
are considered.  The component-0 dust temperature {\it seems\/} to be accurately 
known and is consistent with the large-scale interstellar radiation field, i.e. 
18$\,$K.  Other parameters, such as gas densities and column densities per velocity 
interval, are known from factors of a few to more than an order of magnitude about 
some geometric mean.  The continuum-derived versus spectral line-derived gas
column densities showed much better agreement for these models than for the 
one-component component models, the former having $\chi_\nu^2=7.4$.  Despite their
advantages, these models suffer from two problems.  The first is that many of the
filling factors have unphysical values --- i.e., greater than unity.  The second 
problem is that the model curves do not fit well to the data with 140$\um$/240$\um$ 
color temperatures greater than 20$\,$K. 
\item {\it Two-Component, Two-Subsample, LVG Models:\/} Here the sample of data
points was split into two subsamples: one with dust color temperatures $<20\,$K
and the other with such temperatures $\ge 20\,$K.  These models have all the
advantages of the simpler two-component models described previously, and also
fix the problems of the unphysical filling factors and the poor fit to the data
points with 140$\um$/240$\um$ color temperatures $\ge 20\,$K.  The only 
disadvantage is that this approach is ad hoc.  Nevertheless, the derived
parameter values agree with those of the simpler two-component models to within 
the systematic uncertainties.
\item {\it Cloud Masses:\/} After correcting for systematic effects, such as the
lack of coverage of the $\cOone$ map, the total {\it hydrogen\/} gas mass derived in 
the Orion fields is $4.2\times 10^5\unit M_\odot$, as found in the more realistic
two-component case (and is $2.5\times 10^5\unit M_\odot$ for the less realistic 
one-component case).  Due to the systematics, the uncertainty in this mass is 
$\pm 40\%$.  The extra mass in the two-component case is due to the apparent presence 
of cold dust/gas with temperatures of 3-10$\,$K.  Attempting to confirm the existence 
of this cold dust/gas with the long-wavelength FIRAS data was inconclusive due to the 
insufficient spatial resolution. 
\end{itemize}

The above results have a number of astronomical implications, which are mentioned
here only briefly:
\begin{enumerate}
\item {\it Dust Temperatures Equal to Gas Temperatures:\/} If this surprising result 
applies to the gas and dust on large scales in our Galaxy, then models of gas-to-dust 
thermal coupling must be revised.  Typically, these models assume spherical grains 
with a single large grain size and a smooth grain surface \citep[e.g., see][]{Goldsmith01, 
Burke83}.  Clearly, relaxing these simple assumptions will increase the surface 
area-to-volume ratio of the grains and improve the dust-gas thermal coupling, reducing 
$\DT$ (see Papers~III and IIIa for more details). Having dust and gas temperatures equal 
on galactic scales implies that large-scale molecular gas temperatures are nearly double 
the temperatures previously believed (i.e. closer to 20$\,$K than to 10$\,$K).  Also, one 
dust grain alignment mechanism would be ruled out. 
\item {\it At least two temperatures are needed along each sightline to give reasonable
column densities:\/}  The two-component models show good agreement between spectral
line-derived and continuum-derived column densities.   The main difference between the
two components is that one component has unvarying temperature from sightline to sightline
and the other component has varying temperature.  Therefore, except for those few sightlines
where the two components have the same temperature, each sightline has two temperatures
(i.e. one temperature --- gas and dust --- for each component).   Requiring more than one 
temperature along each sightline is supported by the work of \citet{Schnee06}, who find that 
assuming a single temperature along each sightline does not give reliable column densities in 
the Perseus and Ophiuchus molecular clouds.   
\item {\it Cold Gas/Dust:\/} If similar two-component models are correct and are applicable 
to other large clouds in the Galaxy, then the total mass of molecular gas in the Galaxy 
must be revised upwards by about 60\%.  
\item {Millimeter Continuum to $\cOone$ Line Ratio as a Temperature Diagnostic:\/} For 
the millimeter continuum to $\cOone$ intensity ratio above some threshold, this ratio
can serve as a diagnostic of the gas/dust temperature, even in the high-temperature
limit.
\item {\it X-Factor:\/} An improved explanation for the X-factor is applicable.  This
will be discussed in detail in \citet{W05b}.
\end{enumerate}

Discussing the above results in depth would be premature at this point.  We must first
examine the modeling and various systematic effects more carefully.  Even though we 
have tested the effects of some systematics on the model curves (i.e., by applying scale 
factors or changing the starting grid when fitting the models), we need to find and 
understand any inherent biases or flaws that the model fitting may have.  For example,
the two-component models consistently find a $\DT$ equal to or near 0$\,$K, in spite of
the scale factor applied to the data.  Is this because the dust-gas temperature difference
really is zero?  Or is it simply because these models are biased towards $\DT=0\,$K despite
its true value?  And what of the reliability of the other parameter values?  The component-1
density, for example, is supposed to be known to within a factor of a few of 600$\,H_2$$\unit 
cm^{-3}$, according to the simple two-component models (see Figure~\ref{fig27}).  And yet, 
the two-subsample, two-component models suggest densities that are one to three orders
of magnitude higher than this.  Clearly, the methods used thus far to examine the systematics
are incomplete.  Consequently, Paper~II examines the results of running the models on
simulated data and will answer the concerns raised here. 

In addition to those concerns, there are other questions not yet addressed by the methods used
here.  Specifically, what is the effect of the background subtractions used?  How will dust 
associated with HI affect the results?  Does changing the spectral emissivity index $\beta$ 
appreciably affect the results?  Are there alternative kinds of models that would also 
explain the data?  The cloud positions modeled here only represent 26\% of the area of the
Orion clouds; how representative are the results of the clouds as a whole?  Paper~III
examines these questions. 

All these questions and concerns are addressed in Papers~II and III.  Paper~III also discusses
the scientific implications of the results in more detail.






\acknowledgments  This work was supported by CONACyT grants \#211290-5-0008PE and 
\#202-PY.44676 to W.~F.~W. at {\it INAOE.\/} I am very grateful to W.~T.~Reach
for his comments and support.  I owe a great debt of thanks to Y.~Fukui and 
T.~Nagahama of Nagoya University for supplying the $\cO$ data that made 
this work possible. The author is grateful to R.~Maddalena and T.~Dame, who 
supplied the map of the peak $\COone$ line strengths and provided important 
calibration information.  I thank P.~F.~Goldsmith, D.~H.~Hughes, R.~Padman, 
W.~T.~Reach, Y. Fukui, M.~Greenberg, T.~A.~.D.~Paglione, G.~MacLeod, E.~Vazquez 
Semadeni, and others for stimulating and valuable discussions.



\appendix

\section{Column Densities from Optically Thin Rotational Lines of CO\label{appa}}

    In this appendix, the relationship between the molecular gas column
density and integrated line strength of an optically thin rotational
line of CO, or one of its isotopologues, is derived.  The derivation
detailed below is more appropriately applied to the isotopologues of CO,
such as $\cO$ and $\Co$, than to CO itself (i.e. $\COex$), because
the low-J lines of CO (i.e. $\Jone$, 2$\too$1, 3$\too$2) are usually
optically thick, whereas the corresponding lines of its isotopologues are
usually optically thin for molecular gas on the scales of many
parsecs.  Such a derivation is helpful for a number of reasons.  One
reason is that this relationship is central to this paper and knowing
which approximations were {\it not\/} used, as well as which ones were 
used, is important.  Another reason is that the full derivation is
difficult to find in the literature.  Furthermore, aside from the usual
assumptions made in determining column densities (i.e., optically thin
line, non-varying gas physical conditions along the line of sight, and
sometimes LTE), the equations in the literature are sometimes
based on additional invalid or unnecessary assumptions. For example,
the Rayleigh-Jeans approximation is sometimes used
and is {\it both\/} invalid {\it and\/} unnecessary. (Ironically
enough, using the Rayleigh-Jeans approximation actually results in a
more complicated expression than {\it not\/} using this approximation.)

    The optical depth of the $\rm J\too J-1$ rotational transition,
$\rm\tau_\nu(J\too J-1)$, is related to the volume absorption
coefficient, $\rm\kappa_\nu(J\too J-1)$, by
\begin{equation}
\rm\tau_\nu(J\too J-1) = \int \kappa_\nu(J\too J-1)\ ds
\label{apa1}
\end{equation}
\noindent or
\begin{equation}
\rm\tau_\nu(J\too J-1) = {\strut 1\over\strut 4\pi}\int \Bigl(
B_{_{J-1,J}}\,n_{_{J-1}} - B_{_{J,J-1}}\,n_{_J}\Bigr) h\nu_{_{J,J-1}} 
\phi_\nu(J\too J-1)\ ds\quad .
\label{apa2}
\end{equation}
The integral is over the line of sight through the cloud.  The $\rm
n_{_{J-1}}$ and $\rm n_{_J}$ are the volume densities of those CO
molecules in the J$-$1 and J rotational levels, respectively.  The $\rm
B_{_{J-1,J}}$ and $\rm B_{_{J,J-1}}$ are the Einstein B-coefficients
defined in terms of the mean intensity of the radiation field, rather 
than in terms of the energy density.  The $\rm B_{J-1,J}$ is the
absorption rate coefficient for absorptions from level J$-$1 to level
J.  The $\rm B_{J,J-1}$ is the rate coefficient for stimulated emission
from level J to level J$-$1 and is the stimulated emission correction to 
the optical depth in equation (\ref{apa2}).  The $\rm
h\nu_{_{J,J-1}}$ is the energy difference between levels J$-$1 and J,
where h is Planck's constant and $\rm \nu_{_{J,J-1}}$ is the frequency
of the $\rm J\too J-1$ transition.  The quantity $\rm \phi_\nu(J\too
J-1)$ is the line profile function in terms of frequency, $\nu$.  It is
normalized such that
\begin{equation}
\int\limits_{\rlap{$\scriptscriptstyle spectral\ line$}}
\rm \phi_\nu(J\too J-1)\ d\nu = 1\quad .
\label{apa3}
\end{equation}
The integration in (\ref{apa2}) converts the volume densities,  $\rm n_{_{J-1}}$ 
and $\rm n_{_J}$, into the corresponding column densities, $\rm N_{_{J-1}}$ 
and $\rm N_{_J}$, yielding
\begin{equation}
\rm\tau_\nu(J\too J-1) = {\strut 1\over\strut 4\pi} 
\Bigl(B_{_{J-1,J}}N_{_{J-1}} - B_{_{J,J-1}}N_{_J}\Bigr) h\nu_{_{J,J-1}} 
\phi_\nu(J\too J-1)\quad .
\label{apa4}
\end{equation}
Note that the assumption of LTE has not been used.  Similarly,
assuming constant gas physical properties
along the line of sight has not been applied. (Strictly
speaking, the line profile function can change along the line of
sight.  Nevertheless, the $\rm\phi_\nu(J\too J-1)$ that appears in \ref{apa4}
is the appropriately weighted average along the
line of sight of the $\rm\phi_\nu(J\too J-1)$ that appears in \ref{apa2}.)  The
following relations eliminate the Einstein
B-coefficients in favor of the spontaneous emission rate coefficient,
$\rm A_{_{J,J-1}}$:
\begin{mathletters}
\begin{eqnarray}
\rm B_{_{J-1,J}} &=&\rm {\strut g_{_J}\over\strut g_{_{J-1}}}
B_{_{J,J-1}}\label{apa5a}\\
\noalign{\noindent and}
\rm B_{_{J,J-1}} &=&\rm {\strut c^2\over\strut 2h\nu_{_{J,J-1}}^3} A_{_{J,J-1}}\quad ,
\label{apa5b}
\end{eqnarray}
\end{mathletters}
where c is the speed of light and $\rm g_{_J}$ and $\rm g_{_{J-1}}$ are the
degeneracies of levels J and J$-$1, respectively.  Using (\ref{apa5a}) and
(\ref{apa5b}), equation (\ref{apa4}) becomes,
\begin{equation}
\rm\tau_\nu(J\too J-1) = {\strut c^2\over\strut 8\pi\nu_{_{J,J-1}}^2} 
\, g_{_J}A_{_{J,J-1}}\,\Biggl({\strut N_{_{J-1}}\over\strut g_{_{J-1}}} - 
{\strut N_{_J}\over\strut g_{_J}}\Biggr)\,\phi_\nu(J\too J-1)\quad .
\label{apa6}
\end{equation}
The frequency profile function, $\rm\phi_\nu(J\too J-1)$, is replaced
with the velocity profile function, $\rm\phi_v(J\too J-1)$.  Since
$\rm\phi_\nu(J\too J-1)\,d\nu=\phi_v(J\too J-1)\,dv$, 
$\rm\phi_\nu(J\too J-1)=\lambda_{_{J,J-1}}\phi_v(J\too J-1)$, where 
$\rm\lambda_{_{J,J-1}}$ is the wavelength of the $\rm J\too J-1$ 
transition.  Substituting this into (\ref{apa6}) yields
\begin{equation}
\rm\tau_v(J\too J-1) = {\strut g_{_J}A_{_{J,J-1}}\lambda_{_{J,J-1}}^3
\over\strut 8\pi}\,{\strut N_{_{J-1}}\over\strut g_{_{J-1}}}
\,\Biggl(1 - {\strut N_{_J}\over\strut N_{_{J-1}}}
{\strut g_{_{J-1}} \over\strut g_{_J}}\Biggr)\,\phi_v(J\too J-1)\quad .
\label{apa7}
\end{equation}
The excitation temperature between levels J and J$-$1, $\rm T_x(J\too J-1)$,
is defined by
\begin{equation}
\rm {\strut n_{_J}\over\strut n_{_{J-1}}} = {\strut g_{_J}\over\strut 
g_{_{J-1}}}\, exp\Biggl(-{\strut h\nu_{_{J,J-1}}\over\strut kT_x(J\too J-1)}
\Biggr)\quad ,
\label{apa8}
\end{equation}
where k is Boltzmann's constant.  Up to now, the assumption of constant 
gas physical properties along the line of sight and the assumption of LTE 
have not been used.

   The term ``excitation
temperature" is often used incorrectly in the literature.  Under the
assumptions of some optical depth limit and LTE, a ratio of two
molecular lines gives a temperature.  In the LTE limit, this
line-ratio temperature is the kinetic temperature of the gas.  However,
it is often called the ``excitation temperature of the gas", as though
excitation temperature is a physical property of the gas itself, like 
kinetic temperature, rather than a parameter determined {\it by}
the physical properties.  The excitation temperature properly represents the
population ratio of two {\it particular\/} levels of a {\it particular\/} 
molecule and {\it not\/} a physical property that characterizes all the
levels of all the molecules in a region of a molecular cloud.  A
line-ratio temperature represents three or four levels and is {\it not}
an excitation temperature, but depends on the {\it two\/} excitation
temperatures of the lines forming the ratio (as well as their optical
depths).  In general, temperatures from line ratios and the excitation
temperatures of the various transitions are numerically distinct
from one another, except in the limits of either low or high density.
In the former case, all excitation temperatures equalize to the
temperature of the ambient radiation field (i.e., the cosmic background
temperature in the absence of other sources) and the temperatures
derived from line ratios are equal to this temperature.  In the
latter case, all the excitation and line ratio temperatures are
the same as the kinetic temperature of the gas.  As such, many articles
actually mean ``line-ratio temperature" or ``kinetic temperature" when 
they say ``excitation temperature".  In particular, the term 
``kinetic temperature" is appropriate when such articles compute
column densities from a {\it single\/} optically thin rotational line of 
CO. Unless a non-LTE radiative transfer code (e.g., the Large Velocity
Gradient or LVG code) is used, along with some constraints on the gas
physical conditions, to explicitly consider the excitation temperature 
variation with position in the rotational ladder, the excitation 
temperatures are implicitly assumed to be {\it numerically\/} constant 
throughout the rotational ladder in computing the partition function.  
Therefore, this adopted and unvarying (with J) 
excitation temperature is actually the adopted kinetic temperature 
(or the temperature of the ambient radiation field in the case of very 
low density). 

   Now the gas physical conditions are assumed constant
along the line of sight, so that $\rm N_{_J}/N_{_{J-1}} = n_{_J}/n_{_{J-1}}$.
Substituting (\ref{apa8}) into (\ref{apa7}), integrating over the line velocity
profile, and rearranging gives
\begin{equation}
\rm N_{_{J-1}}
= {\strut 8\pi g_{_{J-1}}\over\strut g_{_J}A_{_{J,J-1}}\lambda_{_{J,J-1}}^3}
\,\Biggl[1 -  exp\Biggl(-{\strut h\nu_{_{J,J-1}}\over\strut kT_x(J\too J-1)}
\Biggr)\Biggr]^{-1}\int
\limits_{\rlap{$\scriptscriptstyle spectral\ line$}}
\tau_v(J\too J-1)\ dv\quad ,
\label{apa9}
\end{equation}
where we used the normalization condition (\ref{apa3}), but applied to
$\rm\phi_v$ instead of $\rm\phi_\nu$.  Following Kutner and Ulich
(1981), the radiation temperature, $\Tr$, which is the brightness
temperature defined in the Rayleigh-Jeans sense (as opposed to
the Planck sense), is given by
\begin{eqnarray}
\rm\Tr(J\too J-1) &=& \rm{\strut h\nu_{_{J,J-1}}\over
\strut k}\Biggl\{\,\Biggl[exp\Biggl({\strut h\nu_{_{J,J-1}}\over
\strut kT_x(J\too J-1)}\Biggr) - 1\Biggr]^{-1}\nonumber\\
\noalign{\medskip}
&-& \rm\,\Biggl[exp\Biggl({\strut h\nu_{_{J,J-1}}\over\strut 
k\Tbg}\Biggr) - 1\Biggr]^{-1}\Biggr\}\Bigl[1 - exp\Bigl(-\tau_v(J\too J-1)\Bigr)
\Bigr]\quad ,
\label{apa10}
\end{eqnarray}
where $\Tbg$ is the cosmic background temperature.  Note that, even
though this is a Rayleigh-Jean brightness temperature, the 
Rayleigh-Jeans approximation is not used.  For optically thin lines,
(\ref{apa10}) can be rewritten as
\begin{equation}
\rm {\strut k\Tr(J\too J-1)\over\strut h\nu_{_{J,J-1}}} = 
\Biggl[exp\Biggl({\strut h\nu_{_{J,J-1}}\over
\strut kT_x(J\too J-1)}\Biggr) - 1\Biggr]^{-1}\ \tau_v(J\too J-1)\ C_{_{BG}}
\quad ,
\label{apa11}
\end{equation}
where $\rm C_{_{BG}}$ is a correction for the cosmic background radiation
and is given by
\begin{equation}
\rm C_{_{BG}} = 1\ -\ {\strut exp\Biggl({\strut h\nu_{_{J,J-1}}
\over\strut kT_x(J\too J-1)}\Biggr)\ -\ 1\over\strut exp\Biggl({\strut
h\nu_{_{J,J-1}}\over\strut k\Tbg}\Biggr)\ -\ 1}\quad .
\label{apa12}
\end{equation}
Notice that this correction factor is zero when $\rm T_x(J\too J-1) =
\Tbg$, because a line with an excitation temperature
equal to that of the cosmic background is not seen against this
background.  

     Many expressions in the literature for CO column density implicitly 
assume that $\rm C_{_{BG}}$ is
unity.  This is a reasonable assumption when $\rm T_x(J\too
J-1)$ is appreciably greater than $\Tbg$.  For example, for the
$\cOone$ line ($\nu_{_{10}}=110.201\unit GHz$) the  $\rm T_x(1\too 0)$
is probably close to the kinetic temperature, $\Tk$, which is often
no lower than 10$\,$K in most cases.  For $\Tk=10\, K$, $\rm
C_{_{BG}} = 0.88$, thereby affecting the column density derivation by 
about 12\%.  For most purposes, this accuracy is
sufficient and is even better at higher temperatures:  for $\Tk=15\,
K$, $\rm C_{_{BG}} = 0.93$ and for $\Tk=20\, K$, $\rm C_{_{BG}} =
0.95$.  The accuracy is also better at higher frequencies:  for the
$\cOtwo$ line ($\nu_{_{10}}=220.399\unit GHz$), again adopting $\rm
T_x(2\too 1) = \Tk$ for simplicity, the $\rm C_{_{BG}} = 0.96$, 0.98,
0.99 for $\Tk = 10$, 15, and 20$\,$K, respectively.  Nevertheless, for 
the work described in this paper, the $\rm C_{_{BG}}$ correction was 
included in the computations.

     Rearranging (\ref{apa11}) gives
\begin{eqnarray}
\rm \Biggl[1 -  exp\Biggl(-{\strut h\nu_{_{J,J-1}}\over\strut 
kT_x(J\too J-1)}&\rm\Biggr)&\rm\Biggr]^{-1}\tau_v(J\too J-1)\nonumber\\
\noalign{\medskip}
&=&\rm {\strut k\Tr(J\too J-1)\over\strut h\nu_{_{J,J-1}}\, C_{_{BG}}}\, 
exp\Biggl({\strut h\nu_{_{J,J-1}}
\over\strut kT_x(J\too J-1)}\Biggr)\quad .
\label{apa13}
\end{eqnarray}
Substituting (\ref{apa13}) into (\ref{apa9}) results in
\begin{equation}
\rm N_{_{J-1}}
= {\strut 8\pi g_{_{J-1}}\over\strut g_{_J}A_{_{J,J-1}}\lambda_{_{J,J-1}}^3}
\,{\strut k\over\strut h\nu_{_{J,J-1}}\, C_{_{BG}}}\, exp\Biggl({\strut 
h\nu_{_{J,J-1}}\over\strut kT_x(J\too J-1)}\Biggr)\int
\Tr(J\too J-1)\ dv\quad .
\label{apa14}
\end{equation}
The excitation temperature, $\rm T_x(J\too J-1)$, is not known in general,
but is now eliminated by using (\ref{apa8}) and the 
expression $\rm N_{_J}/N_{_{J-1}} = n_{_J}/n_{_{J-1}}$:
\begin{equation}
\rm N_{_J}
= {\strut 8\pi\over\strut A_{_{J,J-1}}\lambda_{_{J,J-1}}^3}
\,{\strut k\over\strut h\nu_{_{J,J-1}}\, C_{_{BG}}}\int
\Tr(J\too J-1)\ dv\quad .
\label{apa15}
\end{equation}
Even though $\rm T_x(J\too J-1)$ is still implicitly present in 
$\rm C_{_{BG}}$, the dependence of $\rm C_{_{BG}}$
on $\rm T_x(J\too J-1)$ is very weak for $\rm T_x(J\too J-1) >>\Tbg$.

    Equation (\ref{apa15}) can be simplified further using 
\begin{equation}
\rm A_{_{J,J-1}} = {\strut 64\pi^4\nu_{_{J,J-1}}^3\over\strut 3hc^3}\,
\Biggl({\strut J\over\strut 2J+1}\Biggr)\,\mu^2\quad,
\label{apa16}
\end{equation}
where $\mu$ is the permanent dipole moment of the molecule.  This is 
0.11$\,$Debye for CO, where $1\unit Debye = 10^{-18} esu\cdot cm$ in
CGS units or $3.33564\times 10^{-30}\unit m\cdot C$ in SI units 
\citep[see][]{CRC}.  The energy above the ground 
state of level J, $\rm h\nu_{_{J,0}}$, is well represented by 
\begin{equation}
\rm  h\nu_{_{J,0}} = hB\,J(J+1) \quad ,
\label{apa17}
\end{equation}
where B is the rotational constant of the molecule (not to be confused
with the Einstein B-coefficients).  From (\ref{apa17}) it follows that
\begin{equation}
\rm \nu_{_{J,J-1}} = 2BJ \quad .
\label{apa18}
\end{equation}
Equations (\ref{apa17}) and (\ref{apa18}) assume that the molecule is a 
rigid rotor and neglect the effects of centrifugal stretching.  
Nevertheless, (\ref{apa18}) gives the correct observed frequencies of 
the rotational lines of CO up to the $\rm J=10\too 9$ line to within 
0.06\% or better.  (Note that this is still {\it very in}accurate for 
obtaining radial velocities during actual observations, but is 
more than sufficient accuracy for column density determinations.)  Using 
(\ref{apa18}), the rotational constant can be easily determined from 
$\rm B=\nu_{_{10}}/2$, where $\nu_{_{10}}$ is the observed frequency of 
the $\Jone$ line. For CO and its isotopologues, these values are 
$\rm B(CO) = 57.64\, GHz$, $\rm B(\cO) = 55.10\, GHz$, and 
$\rm B(\Co) = 54.89\, GHz$.  Equation(\ref{apa16}) is now written as
\begin{equation}
\rm A_{_{J,J-1}} = {\strut 512\pi^4 B^3\over\strut 3hc^3}\,
\Biggl({\strut J^4\over\strut 2J+1}\Biggr)\,\mu^2\quad .
\label{apa19}
\end{equation}

    Substituting equations (\ref{apa19}) and (\ref{apa18}) into 
(\ref{apa15}) yields the useful expression
\begin{equation}
\rm N_{_J} = {\strut 3k(2J+1)\over\strut 16\pi^3 B J^2 \mu^2\, 
C_{_{BG}}}\ I_{_{J,J-1}}\quad ,
\label{apa20}
\end{equation}
where $\rm I_{_{J,J-1}} = \int\Tr(J\too J-1)\ dv$ is often called
the integrated intensity of the line (although it is really the
velocity-integrated radiation temperature of the line).  Normally,
CGS units are used, but $\rm I_{_{J,J-1}}$ is in units of $\Kkms$;
consequently (\ref{apa20}) becomes
\addtocounter{equation}{-1}
\begin{mathletters}
\begin{equation}
\rm N_{_J} = {\strut 3k(2J+1)\over\strut 16\pi^3 B J^2 \mu^2\, 
C_{_{BG}}}\cdot 10^5\cdot I_{_{J,J-1}}(\Kkms)\quad ,
\label{apa20a}
\end{equation}
\end{mathletters}
\noindent where all quantities with units, except $\rm I_{_{J,J-1}}$, 
are in CGS units.  So far, the gas physical conditions have been assumed
constant along the line of sight.  LTE has not yet been assumed.
Therefore, for general non-LTE conditions, the total column density
N(CO) can be estimated by observing a number of optically thin
rotational lines of CO, using (\ref{apa20a}) to determine the column density 
in the upper levels of the observed transitions, and sum the $\rm N_{_J}$
values.  This sum must then be corrected for the column
densities in the upper levels of unobserved transitions and for the
column density in the ground state, $\rm N_{_0}$. (Notice that \ref{apa20a} is
not defined for $\rm N_{_0}$.)  As such, the fraction of the total column 
density in the ground state and in the upper levels of unobserved lines must 
be estimated by modeling the physical conditions from the observed line ratios.  
This fraction then corrects the sum of the observed $\rm N_{_J}$ 
values to an estimate of the true N(CO).

    When only one or two optically thin lines of CO are observed,
LTE is often assumed.  Notice from (\ref{apa19}) that the
spontaneous emission rates are lowest for the lowest-J transitions,
implying that these lines are closest to LTE.  So if the lines 
observed are low-J lines like $\Jone$ or $\Jtwo$, then the LTE assumption
might reasonably estimate the total CO column density
from the integrated intensity of a single observed line.  The total 
column density of CO is given by
\begin{equation}
\rm N(CO) = \sum_{J'=0}^\infty N_{_{J'}} = N_{_J}\,\Biggl(
{\strut N_{_0}\over\strut N_{_J}}\Biggr)\sum_{J'=0}^\infty 
{\strut N_{_{J'}}\over\strut N_{0}}\quad .
\label{apa21}
\end{equation}
The J in this expression is the upper-J of the observed transition. 
Replacing all occurrences of J$-$1 in (\ref{apa8}) by 0 and, because LTE
is now assumed, replacing $\rm T_x(J\too J-1)$ by the kinetic
temperature, $\Tk$, for all J, converts (\ref{apa21}) to
\begin{equation}
\rm N(CO) = {\strut N_{_{J}}\over 2J+1}\, Q(\Tk)\, exp\Biggl({\strut 
h\nu_{_{J,0}}\over\strut k\Tk}\Biggr)\quad .
\label{apa22}
\end{equation}
(\ref{apa22}) uses $\rm g_{_J} = 2J+1$ and $\rm g_{_0} = 1$.
The $\rm Q(\Tk)$ is the rotational partition function given by
\begin{equation}
\rm Q(\Tk) = \sum_{J'=0}^\infty (2J'+1)\, exp\Biggl(-{\strut 
hBJ'(J'+1)\over\strut k\Tk}\Biggr)\quad ,
\label{apa23}
\end{equation}
where (\ref{apa17}) was used.  When the kinetic temperature is high 
enough, the summation variable, $\rm J'$, can be treated as a
continuous variable (i.e. real instead of integral) and the summation
in (\ref{apa23}) is treated as an integral over $\rm J'$.  This gives the 
simplified, but approximate,
\begin{equation}
\rm Q(\Tk) \simeq {\strut k\Tk\over\strut hB}\quad .
\label{apa24}
\end{equation}
For CO, $\cO$, and $\Co$, (\ref{apa24}) is $\Tk/2.766$, $\Tk/2.644$, and
$\Tk/2.634$, respectively.  

    As is the case for the cosmic background correction factor,
$\rm C_{_{BG}}$, the approximate form of the partition function, (\ref{apa24}),
is often used in the literature.  Again, this approximate
form is appropriate most of the time, because usually $\Tk>>hB/k$.  For
$\Tk = 10$, 15, and 20$\,$K, the ratios of the approximate $\rm
Q(\Tk)$, given by (\ref{apa24}), to the exact $\rm Q(\Tk)$, 
given by (\ref{apa23}), are
0.91, 0.94, and 0.96, respectively.  Again, for the work described in
this paper, the approximate form was {\it not\/} used.  It is easy to
see why: the effects of neglecting the cosmic background correction and
using the approximate rotational partition function reinforce
each other and result in a theoretical I($\cO$) (for the $\Jone$ line)
that is too high by 14\% at $\Tk = 15\,K$.  This is non-negligible
for the work here.

    Substituting (\ref{apa17}) and (\ref{apa20}) into (\ref{apa22}) results in 
\begin{equation}
\rm N(CO) = {\strut 3k\over\strut 16\pi^3 B J^2 \mu^2\, C_{_{BG}}}\,
Q(\Tk)\, exp\Biggl({\strut hB\, J(J+1)\over\strut
k\Tk}\Biggr)\ I_{_{J,J-1}}\quad .
\label{apa25}
\end{equation}
Using the same units as in (\ref{apa20a}) gives
\addtocounter{equation}{-1}
\begin{mathletters}
\begin{equation}
\rm N(CO) = {\strut 3\times 10^5 k\over\strut 16\pi^3 B J^2 \mu^2\, 
C_{_{BG}}}\, Q(\Tk)\, exp\Biggl({\strut hB\, J(J+1)\over\strut
k\Tk}\Biggr)\ I_{_{J,J-1}}(\Kkms)\quad ,
\label{apa25a}
\end{equation}
\end{mathletters}
where N(CO) is in units of CO molecules per cm$^2$.  Note that the
$\rm C_{_{BG}}$ that appears in (\ref{apa25}) and (\ref{apa25a}) is now a function
of $\Tk$ instead of $\rm T_x(J\too J-1)$.  Substituting in the
fundamental physical constants gives
\begin{equation}
\rm N(CO) = 6.90\times 10^{24} {\strut Q(\Tk)\over\strut B J^2 C_{_{BG}}}\ 
exp\Biggl(4.80\times 10^{-11}\,{\strut B\, J(J+1)\over\strut\Tk}\Biggr)
\ I_{_{J,J-1}}(\Kkms)\quad .
\label{apa26r}
\end{equation}
Choosing a particular isotopologue of CO, $\cO$, then applying the value
for the rotational constant for $\cO$ results in
\begin{equation}
\rm N(\cO) = 1.25\times 10^{14}\ {\strut Q(\Tk)\over\strut J^2 C_{_{BG}}}\ 
exp\Biggl(2.64\,{\strut J(J+1)\over\strut\Tk}\Biggr)
\ I_{_{J,J-1}}(\cO)\quad ,
\label{apa27r}
\end{equation}
where $\rm I_{_{J,J-1}}(\cO)$ is still in units of $\Kkms$.  Equation (\ref{apa27r})
is used in the current paper for determining the molecular gas column densities 
in the LTE case.  When $\Tk>>2.644\,K$, (\ref{apa27r}) can be simplified using the
asymptotic form for Q($\Tk$), (\ref{apa24}), and $\rm C_{_{BG}}\simeq 1$ so that
\begin{equation}
\rm N(\cO) = 4.73\times 10^{13}\ {\strut \Tk\over\strut J^2}\ 
exp\Biggl(2.64\,{\strut J(J+1)\over\strut\Tk}\Biggr)
\ I_{_{J,J-1}}(\cO)\quad .
\label{apa28r}
\end{equation}
Note that (\ref{apa28r}) was not used in the current paper, but is still a useful
approximation: the upward corrections for this equation are 14\%, 6.5\%, and 1.8\%
for $\Tk = 15$, 30, and 100$\,$K, respectively. 

Using the appropriate abundances with (\ref{apa26r}) and (\ref{apa27r})
allows determination of N(H$_2$).  Reasonable values for these abundances are 
$\rm X(\CO) = 8\times 10^{-5}$, $\rm X(\CO)/X(\cO) = 25$ to 65, and $\rm 
X(\cO)/X(\Co) = 5$ to 6 \citep[e.g.,][]{Encrenaz75, Dickman75, Knapp76, Frerking82,
Langer90, Taylor89}.  Note that the abundances of the isotopologues near cloud 
surfaces can be different from the numbers given here, which are the abundances 
for cloud interiors \citep[see, e.g.,][]{Taylor89, Frerking82}.  Simply dividing 
the CO column density by the abundance can give a reasonable estimate of N(H$_2$), 
but \citet{Frerking82} find that a zero-point N(H$_2$) must also be added
to explain their observations of the Taurus and $\rho\,$Ophiuci molecular clouds.
Such a zero-point column density was not used in the current work because it
varies by a factor of a few from cloud to cloud \citep[see][]{Frerking82}
and because it would make little difference to the results and conclusions.
For the $\cO$ abundance, we adopt $\rm X(\CO)/X(\cO) 
= 63$.  This is an average of the value observed at the position (0.5, 2.5) observed 
by \citet{Langer90} and the interpolated value expected for the galactocentric radius 
of the Orion clouds (see their Figure 6).  This average is probably more 
representative of the Orion clouds as a whole and only differs by 6\% from the observed 
value at position (0.5, 2.5).  Note that the observed value at the (0,0) is probably
peculiar because it lies within a region with strong UV fields \citep[see][]{Langer90}.   
 
Briefly discussing other approaches for determining column densities with 
optically thin CO lines is worthwhile.  For example,
sometimes the Rayleigh-Jeans approximation is used.  In that case, the 
excitation temperature of the transition is assumed to be sufficiently
large that the approximation $\rm h\nu_{_{J,J-1}}/kT_x(J\too J-1) << 1$
is valid.  Equation (\ref{apa11}) would have the much simpler form,
\begin{equation}
\Tr(J\too J-1)\simeq T_x(J\too J-1) \tau_\nu(J\too J-1)
\label{apa26}
\end{equation}
However, adopting this simpler expression actually complicates the final formula 
(i.e., \ref{apa25}) for the column density because the $\rm [1 - exp(- h\nu_{_{J,J-1}}
/kT_x(J\too J-1))]^{-1}$ factor is not replaced with a simpler form.  Consequently, 
the LTE expression (\ref{apa25}) has an additional factor,  $\rm (h\nu_{_{J,J-1}}/k\Tk)
\,[exp(h\nu_{_{J,J-1}}/k\Tk) - 1]^{-1}$.  For some reason, this additional factor 
is retained, even though the Rayleigh-Jeans approximation used to validate the use 
of (\ref{apa26}) requires its replacement with {\it unity\/}. 
Because the high-$\Tk$ limit applies, $\rm C_{BG}$ is set to 1 and the approximation 
for the partition function, (\ref{apa24}), is used.  The latter allows cancellation of 
the extra $\Tk^{-1}$.  The resultant expression reasonably estimates the 
column density only in the high-$\Tk$, low-J limit.  The ratios of the column density
derived from the expression using the Rayleigh-Jeans approximation to that of
(\ref{apa25}) for the $\cOone$ line are 0.61, 0.79, 0.86, and 0.91 for $\Tk = $
10, 20, 30, and 50$\,$K, respectively.  For the $\cOtwo$ line the corresponding
ratios are 0.49, 0.72, 0.80, and 0.88.  For the $\cOthree$ line these ratios
are 0.37, 0.63, 0.74, and 0.83.  If column density estimates within a 
factor of 2 of those from (\ref{apa25}) are considered acceptable, then, even for
the rather common kinetic temperature of 10$\,$K, the estimates based on the 
Rayleigh-Jeans approximation are unacceptable for all transitions
higher than $\Jone$.  Even for $\Jone$, the column density estimate must
be corrected upwards by 64\%.  A similarly large upward correction, 59\%, is 
necessary even for a kinetic temperature of 20$\,$K for the $\Jthree$ line. 
Avoiding the Rayleigh-Jeans approximation
keeps the final expression more accurate {\it and\/} less complicated.  (Of
course, if the Rayleigh-Jeans approximation is applied correctly, then the
additional factor is unity anyway.  The resultant expression {\it would\/} be
correct in the high-$\Tk$ limit, where $\rm C_{_{BG}}$ is set to unity, which 
is appropriate for this limit.)

Another approach is to use the corresponding line of $\CO$, which is optically
thick, assume LTE and that the gas fills the beam in the velocity interval about
the $\CO$ line peak, and solve for the optical depth $\tau(\cO)$ or $\tau(\Co)$ 
from the $\cO/\CO$ or $\Co/\CO$ intensity ratios.  In addition, the $\CO$ line
peak intensity estimates the gas kinetic temperature. (
Note that this is often inappropriately referred to as the ``excitation
temperature".  It is inappropriate because LTE was assumed in computing 
$\tau$ and in the expression for the column density.)  
The optical depth, kinetic temperature, and line velocity-width are then used
in an LTE expression for the column density that assumes the high-$\Tk$ limit for
the rotational partition function.  This method has the advantages that assuming 
the optically thin limit is not necessary and that the additional information 
provided by the $\CO$ line is used.  However, it also has the disadvantages 
that the high-$\Tk$ limit for the partition function is used and that the gas fills 
beam in the interval about the $\CO$ line peak is assumed.  The former 
assumption does not produce very large errors: even down to $\Tk=2.8\,K$, the upward
correction to the column density found from using the full partition function 
is only 40\%.  The latter assumption is likely wrong for external galaxy 
observations, because the molecular gas is unlikely to fill the beam in any single 
velocity interval.  (As discussed in this paper, it is also likely wrong for 
observations of Galactic molecular clouds, but may still be roughly correct to within 
a factor of about 2.) In fact, this assumption can lead, in certain cases,
to column density estimates that are wrong by factors of a few or more.  If we 
consider a few cases with $\rm I(\Coone)$ a factor of 20 down from $\rm I(\COone)$
and if $\Tr(\COone)$ is weak, then the column density estimates 
are too high by factors of 2 to 5.  For example, if the filling factor in 
each velocity interval is 0.1 and if $\Tr(\COone) = 0.1\,K$, the column density
estimate is too high by a factor of 5.  Even in the less extreme case of
$\Tr(\COone) = 0.5\,K$ and a filling factor of 0.2, the estimate is too 
high by a factor of 2.  Even if $\Tr(\COone)$ is much stronger, 10$\,$K, the
estimate is still a factor of 2 too high for a filling factor of 0.1.  Despite
this shortcoming, the correction factor to the column density estimate is often
much less extreme than the filling factor itself.  In any event, this method
is easily improved by appropriate application of the filling factor to
the equations used to derive the optical depth, kinetic temperature, and the
column density itself.

One final point is that elaborate non-LTE radiative transfer models suggest
that LTE-derived CO column densities can be underestimates of the true column 
densities by factors of 1.3 to 7 \citep{Padoan00}.  As shown in the
current paper, the assumption of LTE can indeed underestimate
the true column density; similar factors are derived using much simpler 
two-component models that employ the LVG code.    

\bigskip
\bigskip

\section{Ratio of the Submillimeter or Millimeter Continuum to an Optically
Thin $\cObf$ Rotational Line in the LTE Case\label{appb}}

The optically thin continuum emission of the dust has a specific intensity, $\rm I_\nu$,
given by
\begin{equation}
\rm I_\nu = \tau_\nu B_\nu(\Td)\qquad .
\label{apb1}
\end{equation}
The dust optical depth $\tau_\nu$ is given by
\begin{equation}
\rm \tau_\nu = \kappa_\nu x_{_d}\mu_{_H} m_{_H} N(H)\qquad ,
\label{apb2}
\end{equation}
where $\kappa_\nu$ is the dust mass absorption coefficient at frequency $\nu$, $\mu_{_H}$ is 
the correction for the presence of helium, $\rm x_{_d}$ is the dust-to-gas mass ratio 
(where ``gas" means hydrogen and helium), $\rm m_{_H}$ is the hydrogen atom mass, and 
N(H) is the column density of hydrogen nuclei.  Specifically, 
\begin{equation}
\rm N(H) = N(H\,I) + 2N(H_2)\qquad ,
\label{apb3}
\end{equation}
where N(H$\,$I) is the column density of atomic hydrogen and N(H$_2$) is the column
density of molecular hydrogen.  Substituting (\ref{apb2}) into (\ref{apb1}), using
the explicit form for the Planck function, and converting from cgs units to $\MJsr$
yields
\begin{equation}
\rm I_\nu(\MJsr) = 2\times 10^{17}\ {h\nu^3\kappa_\nu x_{_d}\mu_{_H} m_{_H}\over c^2 
[exp({h\nu\over k\Td})-1]}\ N(H)\qquad .
\label{apb4}
\end{equation}
The mass absorption coefficient at frequency $\nu$ expressed in terms of that
coefficient at a reference frequency $\nu_0$ is
\begin{equation}
\kappa_\nu = \biggl({\nu\over\nu_0}\biggr)^\beta\kappa_{\nu 0}\qquad ,
\label{apb5}
\end{equation}
where $\beta$ is the spectral emissivity index of the dust.  Substituting (\ref{apb5})
into (\ref{apb4}) gives
\begin{equation}
\rm I_\nu(\MJsr) = 2\times 10^{17}\ {h\nu^{3+\beta}\kappa_{\nu 0} x_{_d}\mu_{_H} m_{_H}\over 
\nu_0^\beta c^2 [exp({h\nu\over k\Td})-1]}\ N(H)\qquad .
\label{apb6}
\end{equation}

    Now that the continuum intensity is expressed in terms of temperature and
column density, the same must be done for the $\cO$ rotational line.  In the
LTE, optically thin case, such an equation is simply the rearrangement of (\ref{apa25a}).
At this point the assumption N(HI) = 0 is adopted because the molecular gas column 
densities dominate for the majority of the positions studied here.  (Nonetheless, 
there are positions for which the atomic gas column density is appreciable, introducing
possible complications in the interpretion of the results.  These complications are
examined in detail in Paper~III.)   Applying the abundance of $\cO$ to change N($\cO$) to 
N(H$_2$), using (\ref{apb3}) with N(HI) = 0 to change N(H$_2$) to N(H), and rearranging 
(\ref{apa25a}) results in
\begin{equation}
\rm I_{J,J-1}(\cO) = {8\pi^3BJ^2\mu^2C_{_{BG}}X(\cO)\over 3\times 10^5\ kQ(\Tk)}  
\ exp\Biggl(-{hBJ(J+1)\over k\Tk}\Biggr)\ N(H)\qquad ,
\label{apb7}
\end{equation}
where $\rm I_{J,J-1}(\cO)$ is in units of $\Kkms$.  Dividing (\ref{apb6}) by (\ref{apb7})
gives
\begin{equation}
\rm r_{_{\nu j}}\equiv {I_\nu(\MJsr)\over I_{J,J-1}(\cO)} 
= 7.5\times 10^{21} {hk\nu^{3+\beta}\kappa_{\nu 0} x_{_d}\mu_{_H} m_{_H} Q(\Tk)\over
\pi^3BJ^2\mu^2\nu_0^\beta C_{_{BG}}X(\cO)c^2}\ {exp\biggl({hBJ(J+1)\over k\Tk}\biggr)
\over exp\biggl({h\nu\over k\Td}\biggr) - 1}\qquad ,
\label{apb8}
\end{equation}
where $\rm r_{_{\nu j}}$ is in units of $\MJkk$.  Substituting in the 
fundamental physical constants and for the rotational constant of $\cO$, 
\begin{equation}
\rm r_{_{\lambda j}} = 1.67\times 10^4 {\lambda_0^\beta(\mu m)\kappa_{\nu 0} x_{_d}\mu_{_H} 
Q(\Tk)\over\lambda^{3+\beta}(\mu m) J^2 X(\cO) C_{_{BG}}}
\ {exp\biggl({2.64\,J(J+1)\over\Tk}\biggr)
\over exp\biggl({1.44\times 10^4\over \lambda(\mu m)\Td}\biggr) - 1}\qquad ,
\label{apb9}
\end{equation}
where $\lambda_0(\rm\mu m)$ and $\lambda(\rm\mu m)$ are the wavelengths in $\mu m$ 
corresponding to frequencies $\nu_0$ and $\nu$, respectively.  Note that 
$r_{_{\lambda j}}$ is still in units of $\MJkk$.  Although not explicitly stated in 
every case in W96, that paper adopted the following values: $\kappa_{\nu 0}= 
40\unit cm^2/g$ for $\nu_0$ corresponding to $\lambda_0=100\um$, $\rm x_{_d}=0.0075$, 
$\mu_{_H}=1.4$, and $\beta=2$.   $\rm X(\CO)/X(\cO) = 63$ was adopted for the $\cO$ 
abundance (see Appendix~\ref{appa}), so that equation (\ref{apb9}) becomes
\begin{equation}
\rm r_{_{\lambda j}} = 5.51\times 10^{13} {Q(\Tk)\over\lambda^5(\mu m) J^2 C_{_{BG}}}
\ {exp\biggl({2.64\,J(J+1)\over\Tk}\biggr)
\over exp\biggl({1.44\times 10^4\over \lambda(\mu m)\Td}\biggr) - 1}\qquad .
\label{apb10}
\end{equation}
If $\lambda$ is set to 240$\um$ and J to 1, then
\begin{equation}
\rd = 69.2 {Q(\Tk)\over C_{_{BG}}}
\ {exp\biggl({5.28\over\Tk}\biggr)
\over exp\biggl({59.9\over\Td}\biggr) - 1}\qquad ,
\label{apb11}
\end{equation}
where the second subscript on the r (i.e. the J) was dropped for clarity.  
Equation~(\ref{apb11}) is used in the current paper for the LTE modeling 
of the continuum to line ratio as a function of $\Td$.

It is interesting to examine the behavior of $\rd$ in the low- and
high-temperature limits.  For simplicity we assume that $\Td$ and $\Tk$
are the same and equal to some T.  We also assume that $\rm T >> 2.644$ in both 
the low- and high-T limits, so that the simpler form of the partition
function, (\ref{apa24}), can be used and so that $\rm C_{_{BG}}\simeq 1$.  The 
low-T limit means that $\rm T<<59.9\,K$ and so
\begin{equation}
\rd = 26.2\ T \exp\biggl(-{54.6\over T}\biggr)\qquad .
\label{apb12}
\end{equation}
The upward correction to this formula is 16\% and 23\% at T=15$\,$K and 30$\,$K,
respectively.  The high-T limit is where $\rm T>>59.9\,K$ and we have 
\begin{equation}
\rd = 0.437\ T^2\qquad .
\label{apb13}
\end{equation}
The downward correction to this formula is 20\% at T=100$\,$K.  In short, the
continuum to line ratio has a strong Wien-like rise at low temperatures and
a {\it relatively\/} weaker $\rm T^2$ rise at high temperatures.  This 
high-temperature rise is still very strong compared to that of ratios of CO 
rotational lines from the same isotopologue and to that of ratios of continuum 
intensities at two different wavelengths.  

\bigskip
\bigskip

\section{Uncertainty Analysis of the LTE Derivation of the $\cObf$ Column Density\label{appc}}

To derive the uncertainty in the $\cO$ column density in the LTE case, we start
with equation (\ref{apa27r}) with J=1:
\begin{equation}
\rm N(\cO) = N_{_0}\,Q(\Tk)\,C_{_{BG}}^{-1}(\Tk)\,exp\left({T_{_{10}}\over\Tk}
\right)\,I(\cO)\qquad ,
\label{apc1}
\end{equation}
where $\rm N_{_0} = 1.25\times 10^{14}$ and $\rm T_{_{10}}=h\nu_{_{10}}/k = 
5.28\, K$.  The uncertainty in the derived column density, $\rm\sigma(N)$, is 
given by
\begin{equation}
\rm\sigma^2(N) = \left({\partial N\over\partial\Tk}\right)^2 \sigma^2(\Tk) +
\left({\partial N\over\partial I}\right)^2 \sigma^2(I)
\qquad ,
\label{apc2}
\end{equation}
Substituting equation (\ref{apc1}) into (\ref{apc2}) and then dividing by
(\ref{apc2}) yields
\begin{equation}
\rm {\sigma^2(N)\over N^2} = \left({\Tk\,Q'(\Tk)\over Q(\Tk)} - 
{\Tk\,C_{_{BG}}'(\Tk)\over C_{_{BG}}} - {T_{_{10}}\over\Tk}\right)^2
{\sigma^2(\Tk)\over\Tk^2} + {\sigma^2(I)\over I^2}\qquad .
\label{apc3}
\end{equation}
Differentiating (\ref{apa12}) gives $\rm C_{_{BG}}'(\Tk)$:
\begin{equation}
\rm C_{_{BG}}'(\Tk) = {{T_{_{10}}\over\Tk^2}\ exp\left({{T_{_{10}}\over\Tk}}
\right)\over exp\left({T_{_{10}}\over T_{_{BG}}}\right)-1}\qquad .
\label{apc4}
\end{equation}
Similarly, differentiating (\ref{apa23}) gives $\rm Q'(\Tk)$:
\begin{equation}
\rm Q'(\Tk) = {B_{_T}\over\Tk^2}\,\sum_{J=1}^\infty J(J+1)(2J+1)\, 
exp\Biggl(-{\strut B_{_T}J(J+1)\over\strut\Tk}\Biggr)\quad ,
\label{apc5}
\end{equation}
where $\rm B_{_T}=hB/k=2.64\,K$ and is the molecular rotational constant in units 
of Kelvins.  To simplify the above expression, only the first
two terms in the summation are necessary at low temperatures, i.e. $\Tk<2B_{_T}$,
to achieve 5\% accuracy or better.  At high temperatures, i.e. $\Tk\geq 2B_{_T}$,
the approximate expression (\ref{apa24}) can be used.  This results in the 
following approximations for $\rm Q'(\Tk)$:
\begin{mathletters}
\begin{eqnarray}
\rm Q'(\Tk) &=&\rm {B_{_T}\over\Tk^2}\,\sum_{J=1}^2 J(J+1)(2J+1)\, 
exp\Biggl(-{\strut B_{_T}J(J+1)\over\strut\Tk}\Biggr)\ ,\quad \Tk<2B_{_T}\\
\label{apc6a}
\rm Q'(\Tk) &=&\rm {1\over B_{_T}}\,\phantom{\rm\sum_{J=1}^\infty J(J+1)(2J+1)\, 
exp\Biggl(-{\strut B_{_T}J(J+1)\over\strut\Tk}\Biggr)}\ ,\quad\Tk\geq 2B_{_T}\quad .
\label{apc6b}
\end{eqnarray}
\end{mathletters}
The error in this approximation is no worse than 5\% for $\Tk$ just below
2B$\rm_{_T}$ or 5.28$\,$K and better than this at higher and lower temperatures.
The uncertainty in $\Tk$ itself, $\sigma(\Tk)$, is
\begin{equation}
\rm \sigma^2(\Tk) = \sigma^2(\Td) + \sigma^2(\DT)\quad .
\label{apc7}
\end{equation}
Therefore, the uncertainty in the derived $\cO$ column density is given by
equation (\ref{apc3}), with equations (\ref{apc4}), (\ref{apc5}), (\ref{apc6a}),
(\ref{apc6b}), and (\ref{apc7}) substituted in. 

\bigskip
\bigskip

\section{Uncertainty Analysis of the Column Density Derived from the Dust Continuum
Emission\label{appd}}

The Wien limit is used for estimating the uncertainties in the dust-derived
column densities.  Combining equations (\ref{apb1}) and (\ref{apb2}) and
using the Wien approximation gives
\begin{equation}
\rm N = I_\nu\, D\ exp\left({F\over\Td}\right)\quad ,
\label{apd1}
\end{equation}
where $\rm D = [(2h\nu^3/c^2) \kappa_\nu \mu_{_H} m_{_H} x_{_d}]^{-1}$ and
$\rm F = h\nu/k$.  The uncertainty in the dust-derived column density, 
$\sigma(N)$, is 
\begin{equation}
\rm\sigma^2(N) = \left({\partial N\over\partial I_\nu}\right)^2 \sigma^2(I_\nu) + 
\left({\partial N\over\partial\Td}\right)^2 \sigma^2(\Td)\qquad .
\label{apd2}
\end{equation}
Evaluating (\ref{apd2}) using (\ref{apd1}) and then dividing by (\ref{apd1})
results in
\begin{equation}
\rm{\sigma^2(N)\over N^2} = {\sigma^2(I_\nu)\over I_\nu^2} + 
{\sigma^2(\Td)\over\Td^2}{F^2\over\Td^2}\qquad .
\label{apd3}
\end{equation}
For $\nu$ corresponding to $\lambda = 240\um$, $\rm F=60.0\,K$.  Equation
(\ref{dp05}) gives $\sigma(\Td)$. 

\bigskip
\bigskip

\section{Uncertainty Analysis of the LVG Derivation of the $\cObf$ Column Density\label{appe}}

As stated in Section~\ref{ssec32}, the column density of molecular gas derived from the
$\cOone$ line, $\NHth$, using LVG model results can be written as
\begin{equation}
y = R(a_1, a_2, a_3; x_1, x_2, x_3)\ x_3 \qquad ,
\label{ape1}
\end{equation}
where
\begin{equation}
y\equiv\rm N(\cO)
\label{ape2}
\end{equation}
and 
\begin{equation}
R(a_1, a_2, a_3; x_1, x_2, x_3)\equiv {\left[\rm N(\cO)/\Delta v\right]_{mod}
\over\rm\left[T_{_R}(\cOone)\right]_{\hbox{$_{mod}$}}}
\label{ape3}
\end{equation}
The $\left[\rm N(\cO)/\Delta v\right]_{mod}$ and 
$\rm\left[T_{_R}(\cOone)\right]_{\hbox{$_{mod}$}}$
are the column density per velocity interval and radiation temperature of the
$\cOone$ line, respectively, from the model fit.  Just as there is a family of
model curves of $\rd$ versus $\Td$, the $R(a_1, a_2, a_3; x_1, x_2, x_3)$ can
be thought of as a family of curves where each $R$ is given by a particular
model $\rd$.  The $\{a_i\}$ are the curve parameters that specify a particular $R$ 
curve from the family of $R$ curves and the $\{x_i\}$ specify, {\it indirectly\/},
a particular point on that curve.  Specifically, the $\{a_i\}$ are
\begin{mathletters}
\begin{eqnarray}
a_1 &\equiv& \DT = \Td - \Tk\qquad ,
\label{ape4a}\\
a_2 &\equiv& \left[{N(\cO)\over\Delta v}\right]_{mod} \qquad ,
\label{ape4b}\\
\noalign{\noindent and}
a_3 &\equiv& \left[\nH\right]_{mod} \qquad .
\label{ape4c}
\end{eqnarray}
\end{mathletters}
The $\{x_i\}$ represent a particular observed data point and that point is 
associated with the point on the $R$ versus $\Td$ curve that is closest to it, where 
\begin{mathletters}
\begin{eqnarray}
x_1 &\equiv& \Td \qquad ,
\label{ape5a}\\
x_2 &\equiv& \Ib \qquad ,
\label{ape5b}\\
\noalign{\noindent and}
x_3 &\equiv& \Ic \qquad .
\label{ape5c}
\end{eqnarray}
\end{mathletters}
The uncertainty in the derived column density, $\sigma(y)$, is given by
\begin{equation}
\sigma^2(y) = \sum_{i=1}^3 \left({\partial y\over\partial a_i}\right)^2\sigma^2(a_i)
\ +\ \sum_{i=1}^3 \left({\partial y\over\partial x_i}\right)^2\sigma^2(x_i)
\label{ape6}
\end{equation}
The first summation in (\ref{ape6}) represents the contribution of the model-fit 
uncertainties to the computed column density uncertainty.  The second
summation in (\ref{ape6}) represents the contribution of the observational 
uncertainties to the computed column density uncertainty.  Specifically, the
observational uncertainties result in uncertainties in choosing the nearest point
on the model curve.  To evaluate equation (\ref{ape6}), we will use $r = x_2/x_3$, 
where the subscript ``240" on the $r$ was dropped for brevity.  This gives us
\begin{mathletters}
\begin{eqnarray}
{\partial y\over\partial a_i} &=& {\partial R\over\partial a_i}\ x_3
\label{ape7a}\\
{\partial y\over\partial x_1} &=& {\partial R\over\partial x_1}\ x_3
\label{ape7b}\\
\noalign{\medskip}
{\partial y\over\partial x_2} &=& {\partial R\over\partial x_2}\ x_3
\nonumber\\
&=& \left({\partial R\over\partial r}{\partial r\over\partial x_2}\right)\ x_3
\nonumber\\
&=& {\partial R\over\partial r}
\label{ape7c}\\
\noalign{\medskip}
{\partial y\over\partial x_3} &=& {\partial R\over\partial x_3}x_3 + R
\nonumber\\
&=& \left({\partial R\over\partial r}{\partial r\over\partial x_3}\right)x_3 + R
\nonumber\\
&=& -{\partial R\over\partial r}r + R \qquad .
\label{ape7d}
\end{eqnarray}
\end{mathletters}
Substituting equations (\ref{ape7a}) through (\ref{ape7d}) into (\ref{ape6})
yields
\begin{eqnarray}
\sigma^2(y) &=&\left[\sum_{i=1}^3 \left({\partial R\over\partial a_i}\right)^2
x_3^2\,\sigma^2(a_i)\right]
\ +\ \left({\partial R\over\partial x_1}\right)^2 x_3^2\,\sigma^2(x_1)
\ +\ \left({\partial R\over\partial r}\right)^2 \sigma^2(x_2)
\nonumber\\
&&\ +\ \left[\left({\partial R\over\partial r}\right)^2 r^2
\ -\ 2rR\left({\partial R\over\partial r}\right)\ +\ R^2\right]\sigma^2(x_3)
\label{ape8}
\end{eqnarray}
Employing
\begin{equation}
\sigma^2(r) = r^2\left({\sigma^2(x_2)\over x_2^2} + {\sigma^2(x_3)\over x_3^2}
\right)
\label{ape9}
\end{equation}
changes (\ref{ape8}) to 
\begin{eqnarray}
\sigma^2(y) &=&\left[\sum_{i=1}^3 \left({\partial R\over\partial a_i}\right)^2
\,\sigma^2(a_i)\right]x_3^2
\ +\ \left[\left({\partial R\over\partial x_1}\right)^2\,\sigma^2(x_1)
\ +\ \left({\partial R\over\partial r}\right)^2 \sigma^2(r)\right]x_3^2
\nonumber\\
&&\ +\ \left[R^2\ -\ 2rR\left({\partial R\over\partial r}\right)\right]
\sigma^2(x_3)
\label{ape10}
\end{eqnarray}
Equation (\ref{ape10}) has the advantage that there is no explicit dependence
on $\sigma(x_2)$, $\partial R/\partial x_2$, nor $\partial R/\partial x_3$.
Instead, there is an explicit dependence on $r$ and on $\sigma(r)$, both of
which have already been computed for the LTE modeling.  Equation~(\ref{ape10}) 
was used in determining the uncertainties in the derived column densities. 

   We must now compute the derivatives of $R$.  The derivatives
with respect to the $a_i$, $\partial R/\partial a_i$, are determined from
the difference in $R$, $\Delta R$, that results for a given difference in 
a specific $a_i$, $\Delta a_i$, from the family of model curves.  In other 
words, the following approximation is used:
\begin{equation}
{\partial R\over\partial a_i}\simeq {\Delta R\over\Delta a_i}
\label{ape11}
\end{equation}
The uncertainties in the $\sigma(a_i)$ 
were determined by finding the values of $a_i$ that resulted in the total
$\chi^2$ rising by a value of $\chi_\nu^2$, since this value was greater
than 1 (see discussion of this in Section~\ref{ssec31}).  Since the model 
curves were on a grid, changes in $a_i$ occur in discrete jumps that 
increase $\chi^2$ by amounts less than or greater than $\chi_\nu^2$.  For 
Figure~\ref{fig16}, the error bars represent the {\it minimum\/} change in 
$a_i$ that increased $\chi^2$ by {\it more\/} than $\chi_\nu^2$.  The
$\sigma(a_i)$ used for deriving $\sigma(y)$ was the change in $a_i$ 
interpolated linearly between the grid points that gave the appropriate
increase in $\chi^2$.  Where there was a limit instead of an upper (lower) 
error bar, an error bar the same size as the lower (upper) error bar was
assumed. 

    The derivatives $\partial R/\partial x_1$ and $\partial R/\partial r$
are with respect to the {\it observed\/} $x_1$ (i.e., $\Td$) and the {\it 
observed\/} $r$ (i.e., $\Ib/\Ic$).  The former derivative is the amount that 
$R$ changes when the observed $x_1$ changes (for fixed $r$).  In other words, 
when a point in the $r$ versus $x_1$ plot changes in $x_1$, the closest point 
on the model curve to this point will also change, 
thereby changing $R$.  The latter derivative is similar, but 
with $r$ in place of $x_1$.  Determining the closest point on the model curve 
to the observed data point requires defining distances in the $r$ versus $x_1$ 
plane.  The distance from any observed point can be defined by normalizing changes
in $x_1$ to its observational uncertainty, $\sigma(x_1)$, and changes in $r$ to
$\sigma(r)$.  Thus we define a coordinate system for a given data point as
\begin{mathletters}
\begin{eqnarray}
X &\equiv& {x_1\over\sigma(x_1)}
\label{ape12a}\\
Y &\equiv& {r\over\sigma(r)}
\label{ape12b} \qquad .
\end{eqnarray}
\end{mathletters}
The distance between
the data point and a point on the model curve is simply $(X_d - X_c)^2 + 
(Y_d - Y_c)^2$, where $(X_d,Y_d)$ are the data point coordinates and
$(X_c,Y_c)$ are the coordinates of a point on the model curve.  Because the
curve is a smooth curve, it can be approximated by a straight line in an 
interval that is sufficiently small.  The interval of interest is one
centered on the point on the curve closest to the data point.  Varying the 
$X_d$ of the data point, while holding $Y_d$ constant,
or vice versa, then varies the coordinates of nearest point on
the curve. The variations of these coordinates, $(\Delta X_c, \Delta Y_c)$, 
are related to the data point coordinate variations by
\begin{mathletters}
\begin{eqnarray}
\Delta X_d &=& {\Delta X_c^2 + \Delta Y_c^2\over\Delta X_c}
\label{ape13a}\\
\noalign{\noindent for constant $\Delta Y_d$ and}
\Delta Y_d &=& {\Delta X_c^2 + \Delta Y_c^2\over\Delta Y_c}
\label{ape13b}
\end{eqnarray}
\end{mathletters}
for constant $\Delta X_d$.  The $\Delta X_d, \Delta Y_d$ are the data point
coordinate variations corresponding to variations $\Delta X_c, 
\Delta Y_c$.  Equations (\ref{ape13a}) and (\ref{ape13b}) were derived with 
simple trigonometry.  Using equations (\ref{ape12a}) and (\ref{ape12b}),
equations (\ref{ape13a}) and (\ref{ape13b}) become
\begin{mathletters}
\begin{eqnarray}
(\Delta x_1)_d &=& {(\Delta x_1)_c^2\,\sigma^2(r) + (\Delta r)_c^2\,\sigma^2(x_1)
\over(\Delta x_1)_c\,\sigma^2(r)}
\label{ape14a}\\
\noalign{\noindent and}
(\Delta r)_d &=& {(\Delta x_1)_c^2\,\sigma^2(r) + (\Delta r)_c^2\,\sigma^2(x_1)
\over(\Delta r_1)_c\,\sigma^2(x_1)} \qquad .
\label{ape14b}
\end{eqnarray}
\end{mathletters}
The derivatives $\partial R/\partial x_1$ and $\partial R/\partial r$ use the 
approximations
\begin{mathletters}
\begin{eqnarray}
{\partial R\over\partial x_1} &\simeq& {\Delta R\over(\Delta x_1)_d}
\label{ape15a}\\
\noalign{\noindent and}
{\partial R\over\partial r} &\simeq& {\Delta R\over(\Delta r)_d} \qquad .
\label{ape15b}
\end{eqnarray}
\end{mathletters}
These derivatives are then computed using equations (\ref{ape15a}) and 
(\ref{ape15b}) combined with (\ref{ape14a}) and (\ref{ape14b}): a particular
value of $(\Delta x_1)_c$ is added to the $(x_1)_c$ of the point on the model 
curve that is closest to the data point.  This implies a particular
$(\Delta r)_c$, which, in turn, implies a particular $\Delta R$.  These
are placed in equations (\ref{ape14a}) and (\ref{ape14b}) to compute the 
variations induced at the data point, $(\Delta x_1)_d$ and $(\Delta r)_d$.
These, in turn, are used in equations (\ref{ape15a}) and (\ref{ape15b}) to
compute the desired derivatives.  In the special cases where $(\Delta x_1)_c=0$
and $(\Delta r)_c=0$, $\partial R/\partial x_1 = 0$ and $\partial R/
\partial r = 0$, respectively, which can be verified geometrically and do
{\it not\/} come from equations (\ref{ape14a}) and (\ref{ape14b}).

   Generalizing the above to the two-component LVG model is trivial.  The
definitions of the $\{a_i\}$ are now the seven parameters
of the two-component models, instead of the three parameters shown in equations 
(\ref{ape4a}), (\ref{ape4b}), and (\ref{ape4c}).

\bigskip
\bigskip

\section{Uncertainty Analysis of the Dust-Derived Column Density for the 
Two-Component Models\label{appf}}

As stated in Section~\ref{ssec33}, the column density of molecular gas derived from the
continuum observations is derived from expression~(\ref{mr39}), which is written here as
\begin{equation}
Y = C(\{a_i\}; x_1, x_2, x_3)\ X(x_1,x_2) \qquad ,
\label{apf1}
\end{equation}
where $Y$ is the observed column density, $\NHd$, $C$ is the factor multiplying
$\NHdo$, and $X$ is $\NHdo$.  The $\{x_i\}$ are defined as Appendix~\ref{appe}, except that
$x_2$ is now $\Tdc$.  The $\{a_i\}$ are the seven parameters that specify each 
model curve for the two-component models (see Section~\ref{ssec33}).  The function $C$ is 
analogous to the $R$ in Appendix~\ref{appe}.  Consequently, the uncertainty derivation is 
very similar to that in Appendix~\ref{appe}.  The only other expression 
needed is that for $X$.  This can be in the Wien limit for simplicity, as was done 
in Section~\ref{ssec21} and Appendix~\ref{appd}.  Specifically, equation~(\ref{apd1}) is now
written as
\begin{equation}
X(x_1,x_2) = x_2\, D\, exp\left({F\over x_1}\right)\qquad .
\label{apf2}
\end{equation}
As in Appendix~\ref{appe}, taking partial derivatives of $Y$ with respect
to the $\{a_i\}$ and $\{x_i\}$ is necessary.  Consequently, we use the following:
\begin{mathletters}
\begin{eqnarray}
{\partial X\over\partial x_1} &=& X \left({-F\over x_1}\right)
\label{apf3a}\\
\noalign{\noindent and}
{\partial X\over\partial x_2} &=& {X\over x_2}\qquad .
\label{apf3b}
\end{eqnarray}
\end{mathletters}
Taking the partial derivatives of $Y$ with respect to the $\{a_i\}$ and 
$\{x_i\}$ and adding them as in equation~(\ref{ape6}) gives
\begin{eqnarray}
\sigma^2(Y) &=&\left[\sum_{i=1}^7 \left({\partial C\over\partial a_i}\right)^2
\,\sigma^2(a_i)\right]X^2
\ +\ \left[\left({\partial C\over\partial x_1}\right)^2\,\sigma^2(x_1)
\ +\ \left({\partial C\over\partial r}\right)^2 \sigma^2(r)\right]X^2
\nonumber\\
&&\ +\ 2\left[-F\,{\partial C\over\partial x_1}\,{\sigma^2(x_1)\over x_1^2}
\ +\ r\,{\partial C\over\partial r}\,{\sigma^2(x_2)\over x_2^2}\right]X^2
\ +\ C^2\,\sigma^2(X)\qquad ,
\label{apf4}
\end{eqnarray}
where $r$ is defined as in Appendix~\ref{appe}.  The partial derivatives of $C$ were
determined using the same approach as for the partial derivatives of $R$ in
Appendix~\ref{appe}.  Expression~(\ref{apf4}) was used to give the uncertainties
in the derived $\NHd$ values for the two-component models.




\clearpage


\clearpage
 
\clearpage 

\clearpage
 
\clearpage 

\clearpage 

\clearpage 

\clearpage 

\clearpage 

\clearpage 

\clearpage 

\clearpage 

\clearpage 

\clearpage 

\clearpage 

\clearpage 

\clearpage 

\clearpage 

\clearpage

\clearpage 

\clearpage 

\begin{figure}
\epsscale{0.7}
\plotone{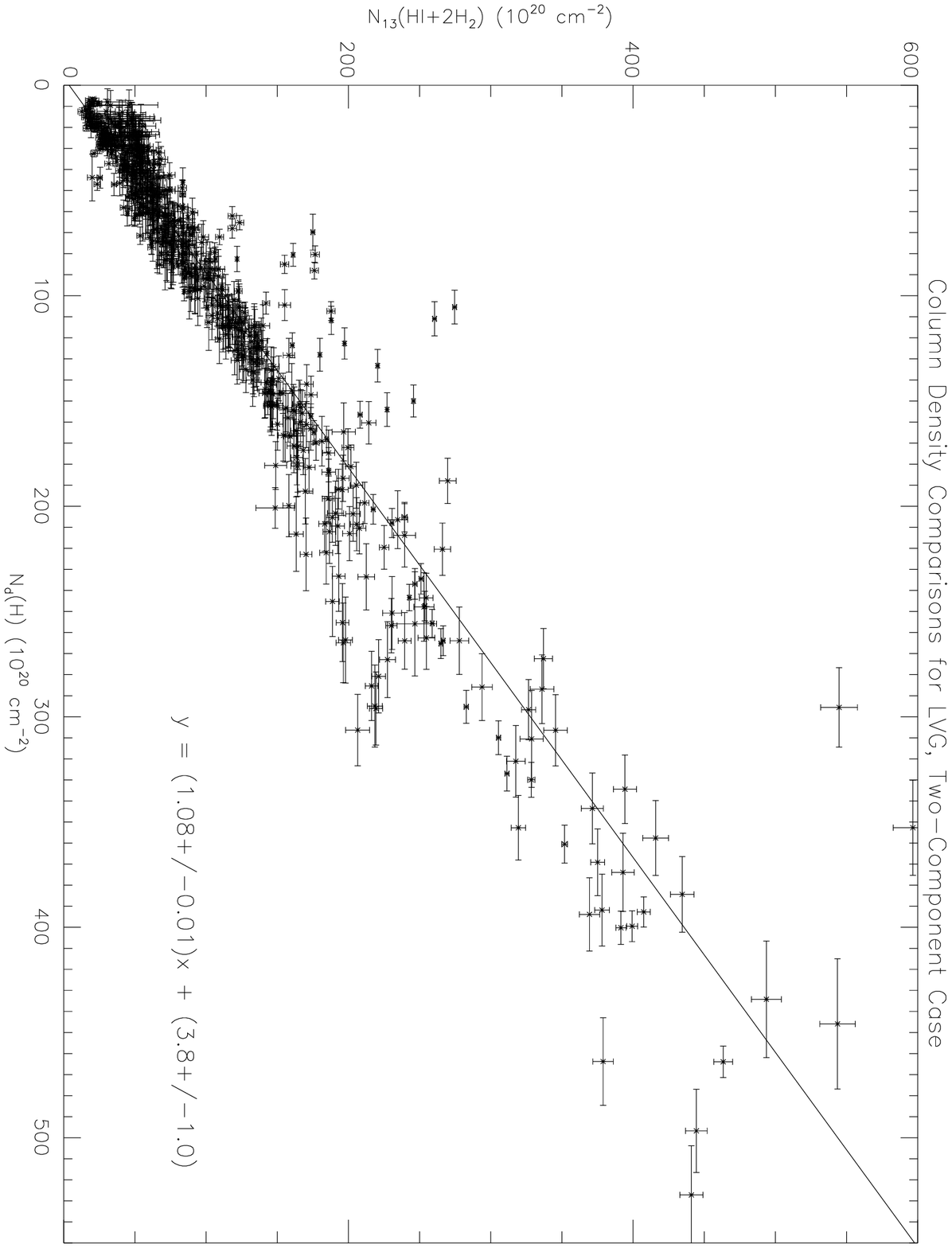}
\caption{The plot above compares the gas column densities as derived
from gas tracers and those derived from dust continuum emission for
the case of non-LTE gas emission and two components.  The plot 
shows the gas column density as derived from $\cOone$ using the 
results from the LVG method versus the gas column density as derived 
from the 140$\um$ and 240$\um$ continuum emission.  The points in 
this plot represent the positions where $\Ia$, $\Ib$, and $\Ic$ are 
at more than 5-$\sigma$, as in Figure~\ref{fig6}.  
\label{fig28}}
\end{figure}

\clearpage 

\begin{figure}
\epsscale{0.8}
\plotone{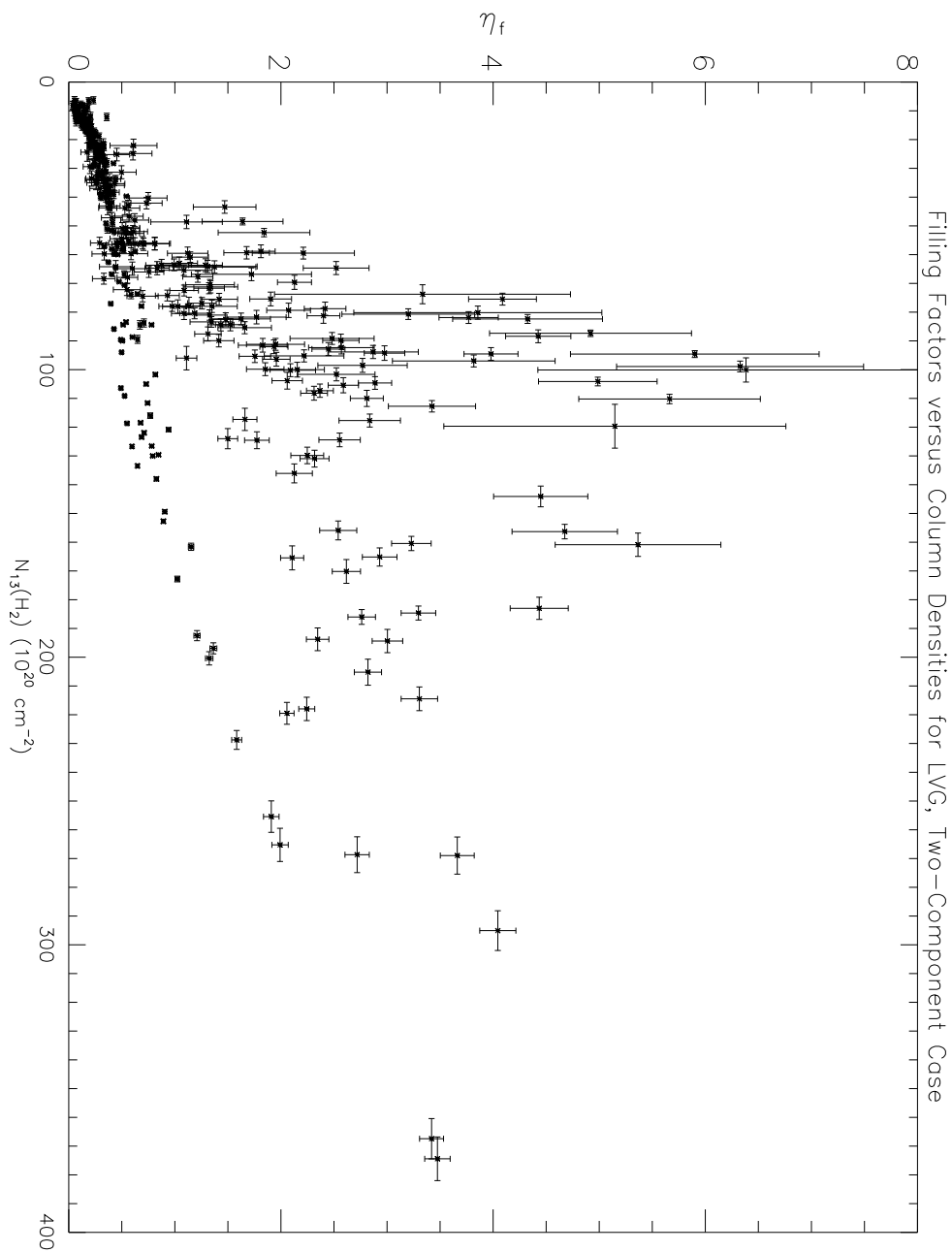}
\caption{The area filling factors of substructures in the Orion clouds are 
plotted against the molecular gas column densities as computed in the LVG, 
two-component case.  The points in this plot represent the positions 
where $\Ia$, $\Ib$, and $\Ic$ are at more than 5-$\sigma$ and where the 
peak radiation temperature of $\COone$ is at more than 3-$\sigma$.  This 
is a total of 372 points. 
\label{fig29}}
\end{figure}

\clearpage 

\begin{figure}
\epsscale{0.7}
\plotone{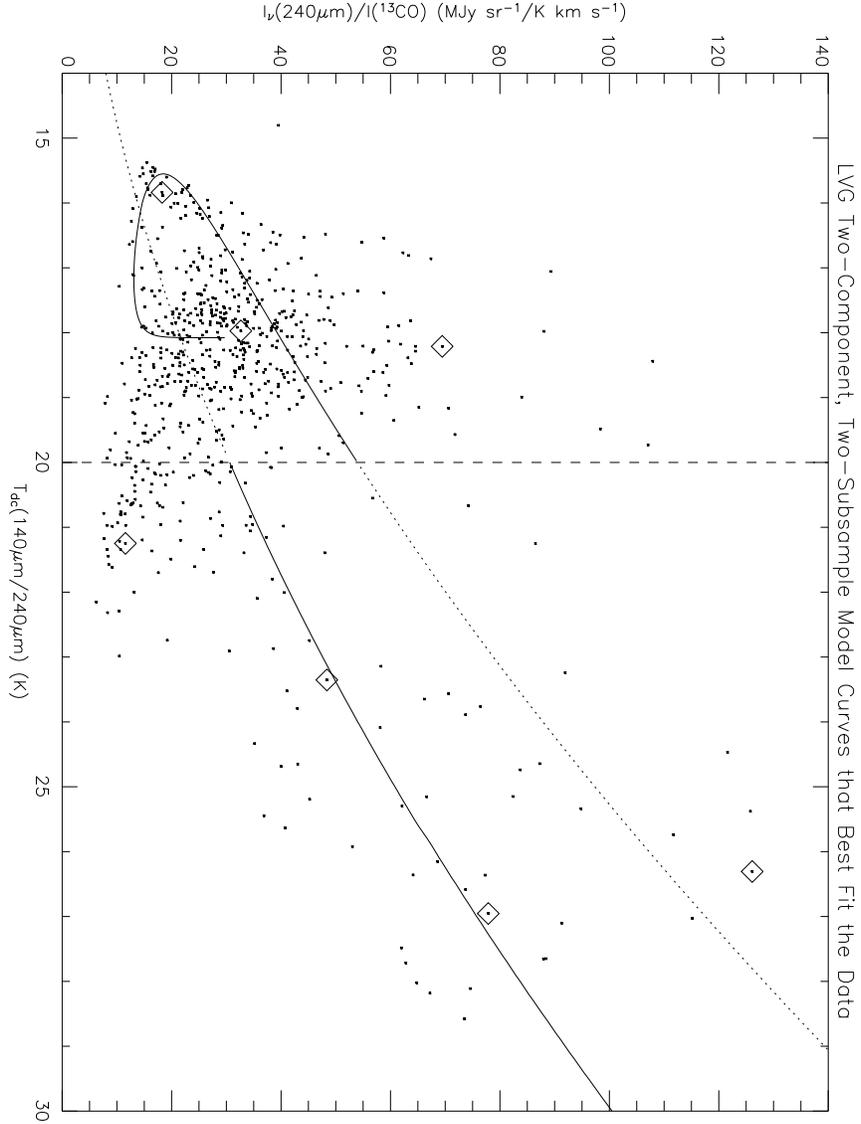}
\caption{The model curves of $\rd$ versus $\Td$ that best fit the data are 
shown for the non-LTE, two-component, two-subsample case.  The two solid 
curves represent fits to the $\Tdc<20\,$K and $\Tdc\geq 20\,$K subsamples.
The fit to the former is the curve from a LVG, two-component model with $c_0$
restricted to be $\geq 1$.  The fit to the latter is an LVG, one component
model with no such restriction.  The dotted curves represent the extensions
of the model curves across the $\Tdc=20\,$K boundary, itself represented
as a dashed line.  The plot includes the same data points as in 
Figure~\ref{fig6} with error bars excluded for clarity.  The parameters and 
their values are listed in Table~\ref{tbl-2}.  The diamonds in this figure 
are the fiducial points discussed in Section~\ref{ssec35} and listed 
in Table~\ref{tbl-2x}.
\label{fig30}}
\end{figure}

\clearpage 

\begin{figure}
\epsscale{0.7}
\plotone{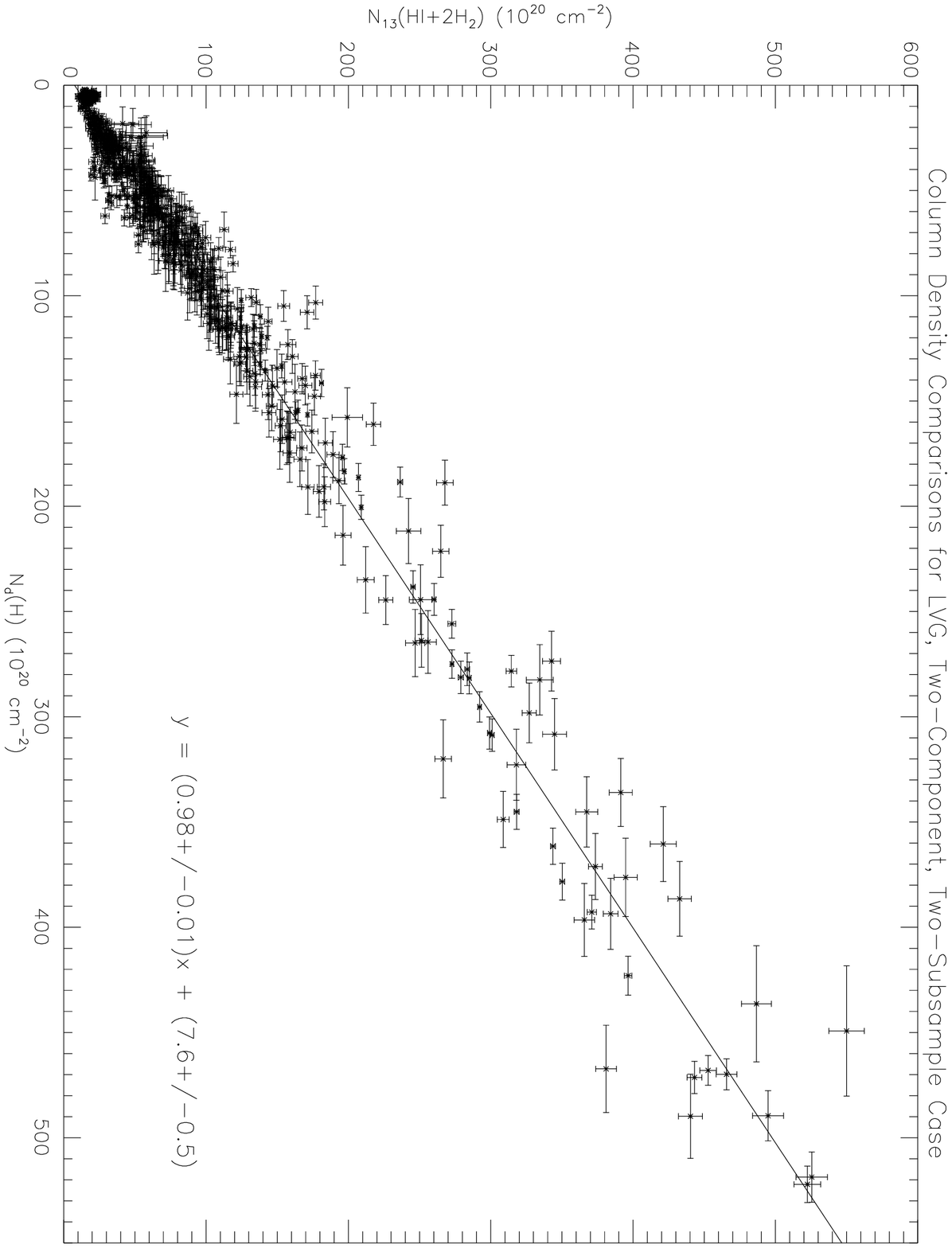}
\caption{The plot above compares the gas column densities as derived
from gas tracers and those derived from dust continuum emission for
the case of non-LTE gas emission, two components, and two subsamples.  
The plot shows the gas column density as derived from the $\cOone$ using 
the results from the LVG method versus the gas column density as derived 
from the 140$\um$ and 240$\um$ continuum emission.  The points in 
this plot represent the positions where $\Ia$, $\Ib$, and $\Ic$ are 
at more than 5-$\sigma$, as in Figure~\ref{fig6}.  
\label{fig31}}
\end{figure}

\clearpage 

\begin{figure}
\epsscale{0.8}
\plotone{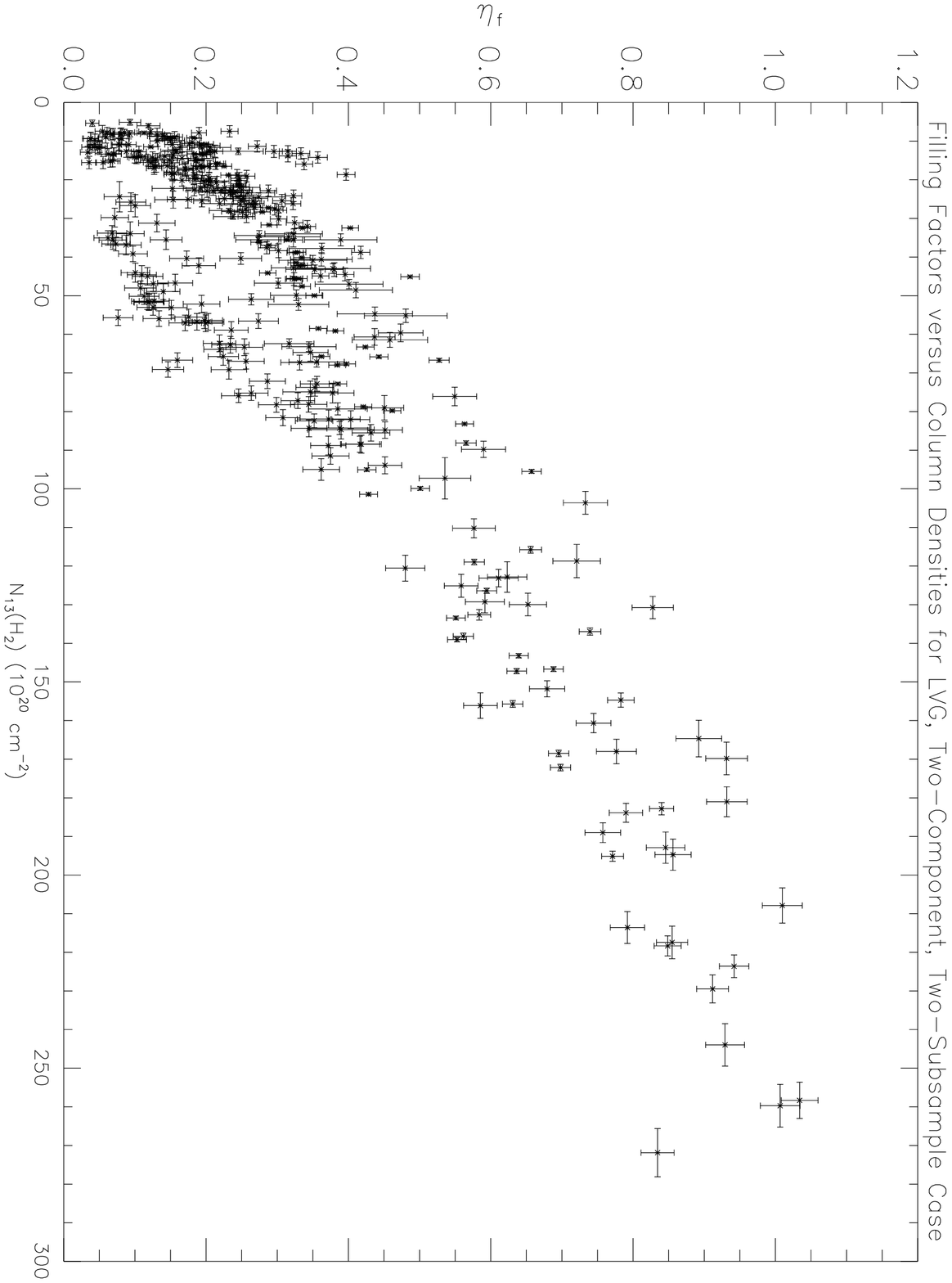}
\caption{The area filling factors of substructures in the Orion clouds are 
plotted against the molecular gas column densities as computed in the LVG, 
two-component, two-subsample case.  The points in this plot represent the 
positions where $\Ia$, $\Ib$, and $\Ic$ are at more than 5-$\sigma$ and where 
the peak radiation temperature of $\COone$ is at more than 3-$\sigma$.  This 
is a total of 372 points. 
\label{fig32}}
\end{figure}

\clearpage 

\begin{figure}
\epsscale{0.73}
\plotone{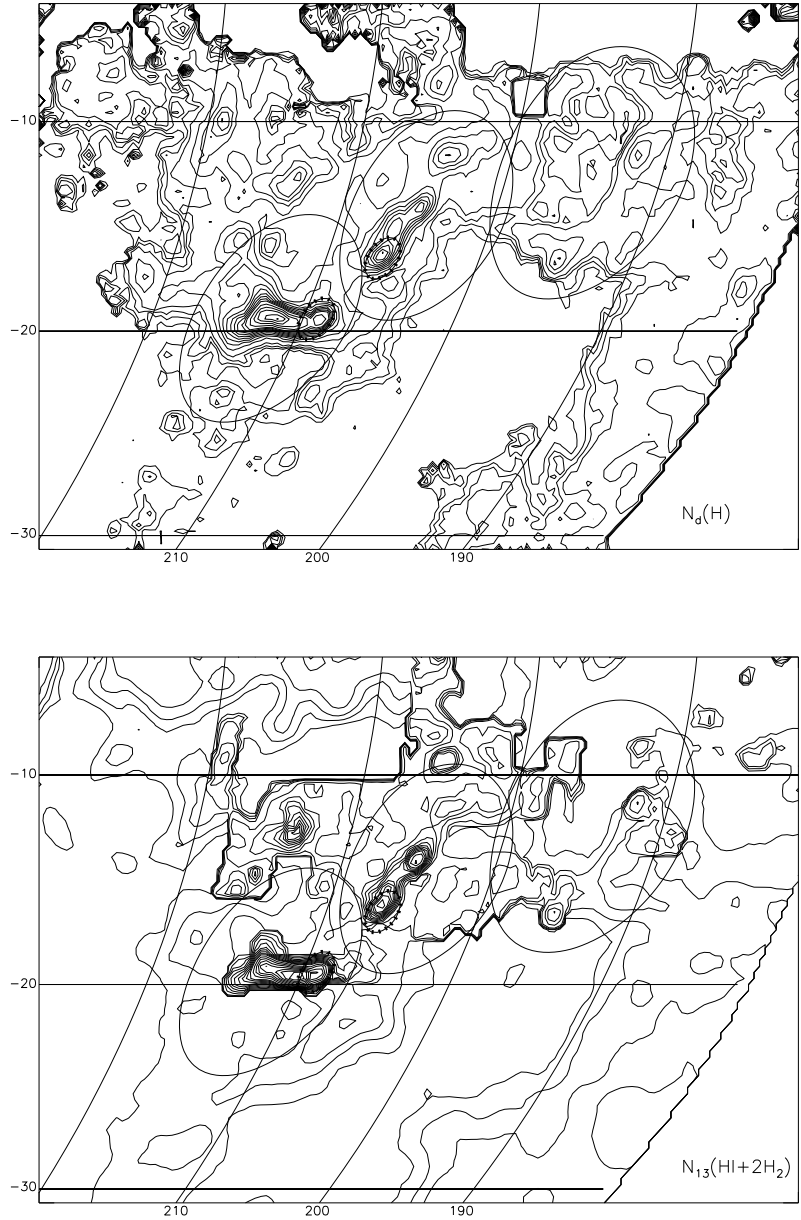}
\caption{Contour maps of the column densities for the LVG, one-component
case are shown above.  The upper panel shows the map for the column densities
derived from the 140$\um$ and 240$\um$ continuum observations.  The contour
levels are 3, 5, 7, 10, 20, 30,..., 100, 120, 140,..., 200 in units of 
$10^{20}\ H\ nuclei\cdot\unit cm^{-2}$.  The lower panel shows the map for the 
column densities derived from the $\cOone$ and H$\,$I observations.  The 
contour levels are 3, 5, 7, 10, 20, 30,..., 100, 120, 140,..., 240 in units 
of $10^{20}\ H\ nuclei\cdot\unit cm^{-2}$.
\label{fig33}}
\end{figure}

\clearpage 

\begin{figure}
\epsscale{0.73}
\plotone{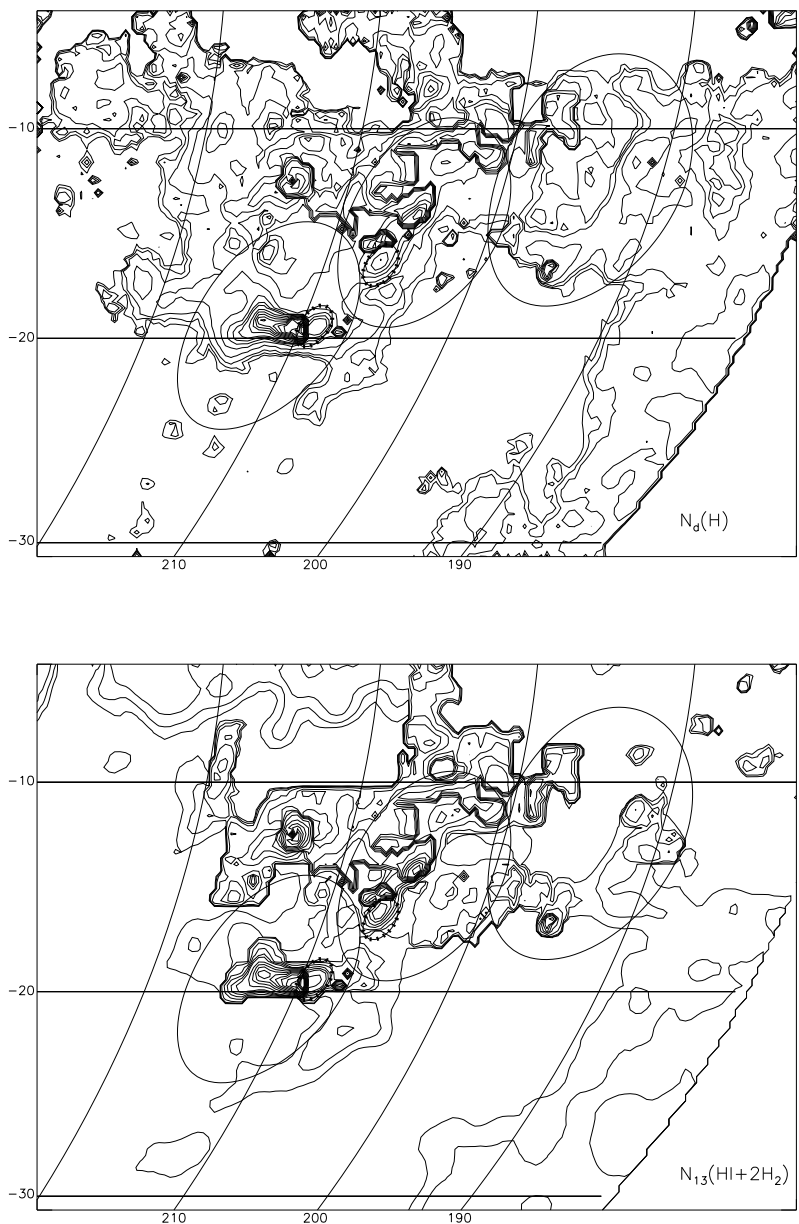}
\caption{Contour maps of the column densities for the LVG, two-component,
two-subsample case are shown above.  The upper panel shows the map for the 
column densities derived from the 140$\um$ and 240$\um$ continuum observations.  
The lower panel shows the map for the column densities derived from the $\cOone$ 
and H$\,$I observations.  The contour levels for both panels are 5, 7, 10, 20, 
40, 60, 80, 100, 150, 200,..., 650 in units of $10^{20}\ H\ nuclei\cdot\unit cm^{-2}$.
\label{fig34}}
\end{figure}

\clearpage 

\begin{figure}
\epsscale{0.7}
\plotone{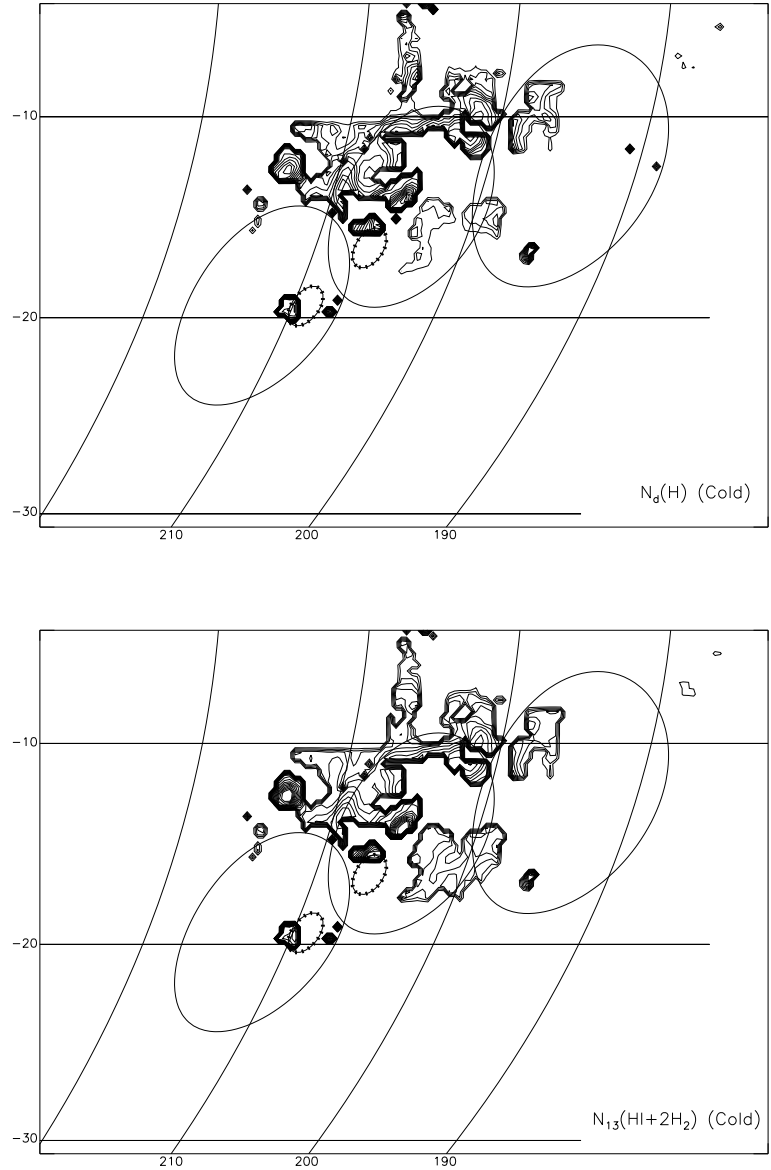}
\caption{Contour maps of the column densities for the cold ($\Tdo=\Tkone\leq 10\,$K)
dust and gas of component~1 in the LVG, two-component, two-subsample case are shown 
above.  The upper panel shows the map for the column densities of the cold gas 
derived from the 140$\um$ and 240$\um$ continuum observations.  The lower panel shows 
the map for the column densities of the cold gas derived from the $\cOone$ observations.
Along those lines of sight with this cold gas the H$\,$I column densities were also
added.  The contour levels for both panels are 10, 20, 30,..., 100, 120, 140,..., 200,
300, 400, 500 in units of $10^{20}\ H\ nuclei\unit cm^{-2}$.
\label{fig35}}
\end{figure}

\clearpage






\clearpage

\begin{deluxetable}{cc}
\tablecaption{Two-Component Model Parameter Values. \label{tbl-1}}
\tablewidth{0pt}
\tablehead{
\colhead{Parameter} & \colhead{Best-Fit Value} 
}
\startdata
$\DT$ & 0$\,$K \\
\noalign{\bigskip}
$c_0$ & 0.04 \\
\noalign{\medskip}
$\Tdz$ & 18$\,$K \\
\noalign{\medskip}
$\nvtcz$ & $5\times 10^{16}\ckms$ \\
\noalign{\medskip}
$n_{c0}$ & $1.8\times 10^4\unit cm^{-3}$\\
\noalign{\bigskip}
$\nvtco$ & $8\times 10^{15}\ckms$ \\
\noalign{\medskip}
$n_{c1}$ & $1\times 10^3\unit cm^{-3}$\\
\noalign{\bigskip}
$\chi_\nu^2$ & 5.69 \\
\noalign{\medskip}
$\nu$ & 667 \\
\enddata



\tablecomments{The {\it formal\/} uncertainties of $\DT$ and $\Tdz$ are both 
$\rm\le 3\times 10^{-4}\,K$. The {\it formal\/} relative uncertainties of the 
other parameters are $\le 3\times 10^{-4}$.}

\end{deluxetable}

\clearpage

\begin{deluxetable}{ccc}
\tablecaption{Two-Component, Two-Subsample Model Parameter Values. \label{tbl-2}}
\tablewidth{0pt}
\tablehead{
\colhead{Parameter} & \colhead{Best-Fit Parameter Values}\span\omit\\
\colhead{} & \colhead{$\Tdc<20\,K$ Subsample} & \colhead{$\Tdc\geq 20\,K$ Subsample}
}
\startdata
$\DT$ & 0$\,$K & $-3\pm 4\,$K \\
\noalign{\bigskip}
$c_0$ & 1.0 & --- \\
\noalign{\medskip}
$\Tdz$ & 18$\,$K & --- \\
\noalign{\medskip}
$\nvtcz$ & $5\times 10^{15}\ckms$ & --- \\
\noalign{\medskip}
$n_{c0}$ & $1.0\times 10^5\unit cm^{-3}$ & --- \\
\noalign{\bigskip}
$\nvtco$ & $2\times 10^{16}\ckms$ & $(3.2\pm {1.8\atop ?})\times 10^{15}\ckms$ \\
\noalign{\medskip}
$n_{c1}$ & $1\times 10^5\unit cm^{-3}$ & $(5.6\pm {4.4\atop 2.4})\times 10^3\unit cm^{-3}$ \\
\noalign{\bigskip}
$\chi_\nu^2$ & 4.60 & 9.98 \\
\noalign{\medskip}
$\nu$ & 525 & 139 \\
\enddata



\tablecomments{The {\it formal\/} uncertainties of $\DT$ and $\Tdz$ for the $\Tdc<20\,K$
subsample are both $\rm\le 2\times 10^{-4}\, K$. The {\it formal\/} relative uncertainties 
of the other parameters for this subsample are $\le 2\times 10^{-4}$.}

\end{deluxetable}

\clearpage

\begin{deluxetable}{cccccc}
\tablecaption{Inferred Physical Conditions at the Fiducial Points\tablenotemark{a}\label{tbl-2x}}
\tablewidth{0pt}
\tablehead{
\colhead{Coordinates\tablenotemark{a}} & \colhead{Parameter\tablenotemark{b}} 
& \colhead{Parameter Values\tablenotemark{c}}\span\omit\span\omit\span\omit\\
\colhead{} & \colhead{} & \colhead{Case 1\tablenotemark{d}} & \colhead{Case 2\tablenotemark{d}} 
& \colhead{Case 3\tablenotemark{e}} & \colhead{Case 4\tablenotemark{e}}\\
}
\startdata
15.84, 18.26 & $\DT$ & $-4$ & $-1$ & 0 & 0 \\
	     & $\NDv$ & --- & $3.2\times 10^{15}$ & $(2\times 10^{15})$ $8\times 10^{15}$ 
	     & $(2\times 10^{14})$ $5\times 10^{15}$ \\
	     & $\nH$ & --- & $1\times 10^5$ & $(1.8\times 10^4)$ $1\times 10^3$ 
	     & $(1\times 10^4)$ $5.6\times 10^3$\\
	     & Distance & 1.17 & 1.78 & 1.67 & 1.92 \\
\noalign{\medskip}	     
17.97, 32.65 & Distance & 0.72 & 1.06 & 0.31 & 0.93 \\
\noalign{\medskip}
18.21, 69.46 & Distance & 3.29 & 3.43 & 0.40 & 2.45 \\

\noalign{\bigskip}
\noalign{\bigskip}

21.25, 11.56 & $\DT$ & +9 & $-3$ & 0 & $-3$ \\
	     & $\NDv$ & --- & $3.2\times 10^{15}$ & $(2\times 10^{15})$ $8\times 10^{15}$ 
	     & $3.2\times 10^{15}$ \\
	     & $\nH$ & --- & $5.6\times 10^3$ & $(1.8\times 10^4)$ $1\times 10^3$ 
	     & $5.6\times 10^3$\\
	     & Distance & 3.56 & 3.67 & 2.45 & 3.67 \\
\noalign{\medskip}
23.35, 48.37 & Distance & 0.99 & 0.47 & 5.52 & 0.47 \\
\noalign{\medskip}
26.31, 126.1 & Distance & 4.82 & 6.48 & 5.29 & 5.03 \\
\noalign{\medskip}
26.95, 77.87 & Distance & 0.88 & 0.45 & 1.71 & 1.05 \\

\enddata

\tablenotetext{a}{As depicted in the $\rd$ versus $\Td$ plots of Figures~\ref{fig6}, 
\ref{fig7}, \ref{fig8}, \ref{fig12}, \ref{fig13}, \ref{fig14}, \ref{fig15}, 
\ref{fig26}, and \ref{fig30}.}
\bigskip
\tablenotetext{b}{$\DT$ is in units of Kelvins, $\NDv$ is in units of $\ckms$, and 
$\nH$ is in units of $\rm cm^{-3}$.  ``Distance" is the orthogonal distance of the 
fiducial point from the theoretical curve in units of sigmas.  The parenthesized
quantities in Cases~3 and 4 are the parameter values for component~0.  In particular,
the parenthesized values for $\NDv$ are really the values for the product $c_{_0}*\nvtcz$.}
\upskip
\tablenotetext{c}{The parameter values for $\DT$, $\NDv$, and $\nH$ are not explicitly
listed for some fiducial points.  The parameter values for these points are the same as
those of the points with the listed values [e.g. point (15.48, 18.26) or (21.25, 11.56)]
that lie above the points without the listed values.}
\upskip
\tablenotetext{d}{Case~1: LTE, one-component models.\qquad Case~2: LVG, one-component models.}
\upskip
\tablenotetext{e}{Case~3: LVG, two-component models.\ \ Case~4: LVG, two-component, 
two-subsample models.}


\end{deluxetable}

\clearpage

\begin{deluxetable}{lrrrrr}
\tablecaption{Gas Masses Sampled by $\cO$, $\CO$, \& H$\,$I (M$_\odot$)\label{tbl-3}}
\tablewidth{0pt}
\tablehead{
\colhead{Field\tablenotemark{a}} & \colhead{Case 1\tablenotemark{b}} &  
\colhead{Case 2\tablenotemark{c}} & \colhead{Case 3\tablenotemark{d}} &
\colhead{Case 4\tablenotemark{e}} & \colhead{$\CO$ \& H$\,$I\tablenotemark{f}}
}
\startdata
Orion Nebula & 21000 & 18000 & 35000 & 31000 & 20000 \\
\noalign{\medskip}
Orion$\,$A & 88000 & 85000 & 146000 & 129000 & 122000 \\
\noalign{\bigskip}
NGC$\,$2024 & 10000 & 9100 & 18000 & 9200 & 17000 \\
\noalign{\medskip}
Orion$\,$B & 93000 & 88000 & 248000 & 189000 & 104000 \\
\noalign{\bigskip}
$\lambda\,$Orionis & 38000 & 36000 & 92000 & 57000 & 45000 \\
\noalign{\medskip}
\noalign{\hrule}
\noalign{\medskip}
TOTAL\tablenotemark{g} & 220000 & 209000 & 486000 & 376000 & 271000 \\
\enddata


\tablenotetext{a}{See W96 for definitions of the fields.}
\tablenotetext{b}{Using  LTE, one-component models and observations of $\cO$ 
and H$\,$I.}
\tablenotetext{c}{Using LVG, one-component models and observations of $\cO$ 
and H$\,$I.}
\tablenotetext{d}{Using LVG, two-component models and observations of $\cO$ 
and H$\,$I.}
\tablenotetext{e}{Using LVG, two-component, two-subsample models and 
observations of $\cO$ and H$\,$I.}
\tablenotetext{f}{Adopting an X-factor of $2.6\times 10^{20}\cKkms$.}
\tablenotetext{g}{Sum of Orion$\,$A, Orion$\,$B, $\lambda\,$Orionis fields.}


\end{deluxetable}

\clearpage

\begin{deluxetable}{lrrr}
\tablecaption{Gas Masses Inferred from 140, 240$\um$ Continuum Emission 
(M$_\odot$)\label{tbl-4}}
\tablewidth{0pt}
\tablehead{
\colhead{Field\tablenotemark{a}} & \colhead{Cases 1 \& 2\tablenotemark{b}} &  
\colhead{Case 3\tablenotemark{c}} & \colhead{Case 4\tablenotemark{d}}
}
\startdata
Orion Nebula & 14000 & 29000 & 27000 \\
\noalign{\medskip}
Orion$\,$A & 107000 & 169000 & 151000 \\
\noalign{\bigskip}
NGC$\,$2024 & 9700 & 14000 & 9700 \\
\noalign{\medskip}
Orion$\,$B & 81000 & 225000 & 179000 \\
\noalign{\bigskip}
$\lambda\,$Orionis & 63000 & 122000 & 82000 \\
\noalign{\medskip}
\noalign{\hrule}
\noalign{\medskip}
TOTAL\tablenotemark{e} & 251000 & 516000 & 411000 \\
\enddata


\tablenotetext{a}{See W96 for definitions of the fields.}
\tablenotetext{b}{Using one-component models and continuum observations.}
\tablenotetext{c}{Using two-component models and continuum observations.}
\tablenotetext{d}{Using two-component, two-subsample models and continuum
observations.}
\tablenotetext{e}{Sum of Orion$\,$A, Orion$\,$B, $\lambda\,$Orionis fields.}


\end{deluxetable}

\clearpage

\begin{deluxetable}{lrr}
\tablecaption{Best Estimates of Total\tablenotemark{a}\ \ Gas Masses 
(M$_\odot$)\label{tbl-5}}
\tablewidth{0pt}
\tablehead{
\colhead{Tracer} & \colhead{One-Component Case\tablenotemark{b}} &  
\colhead{Two-Component Case\tablenotemark{c}}
}
\startdata
Derived from Continuum & 251000 & 411000 \\
\noalign{\medskip}
Derived from $\cOone$\tablenotemark{d} & 263000 & 429000 \\
\enddata


\tablenotetext{a}{The total {\it hydrogen\/} mass from the 3 big fields (i.e., 
sum of Orion$\,$A, Orion$\,$B, $\lambda\,$Orionis fields.}
\tablenotetext{b}{Using LVG, one-component models.}
\tablenotetext{c}{Using LVG, two-component, two-subsample models.}
\tablenotetext{d}{After correcting for insufficient coverage.}


\end{deluxetable}

\clearpage

\begin{deluxetable}{lrr}
\tablecaption{Masses\tablenotemark{a}\ \ of Gas in a Cold 
($\Tk\leq 10\,$K) Component (M$_\odot$)\label{tbl-6}}
\tablewidth{0pt}
\tablehead{
\colhead{Field\tablenotemark{b}} & \colhead{Case 3\tablenotemark{c}} &  
\colhead{Case 4\tablenotemark{d}}
}
\startdata
Orion Nebula & 14000 & 14000 \\
\noalign{\medskip}
Orion$\,$A & 43000 & 28000 \\
\noalign{\bigskip}
NGC$\,$2024 & 5700 & 6600 \\
\noalign{\medskip}
Orion$\,$B & 155000 & 126000 \\
\noalign{\bigskip}
$\lambda\,$Orionis & 54000 & 28000 \\
\noalign{\medskip}
\noalign{\hrule}
\noalign{\medskip}
TOTAL\tablenotemark{e} & 252000 & 181000 \\
\enddata


\tablenotetext{a}{Inferred from $\cO$ and H$\,$I observations.}
\tablenotetext{b}{See W96 for definitions of the fields.}
\tablenotetext{c}{Using LVG, two-component models and observations of $\cO$ 
and H$\,$I.}
\tablenotetext{d}{Using LVG, two-component, two-subsample models and 
observations of $\cO$ and H$\,$I.}
\tablenotetext{e}{Sum of Orion$\,$A, Orion$\,$B, $\lambda\,$Orionis fields.}


\end{deluxetable}

\end{document}